\documentclass{aa}  

\usepackage{graphicx}
\usepackage{orcidlink}
\usepackage{subcaption}
\usepackage{multirow}
\usepackage{enumitem}
\usepackage{threeparttable}

\usepackage{txfonts}
\hypersetup{
    colorlinks=true,     
    linkcolor=blue,     
    citecolor=blue,     
    urlcolor=blue       
}
\definecolor{green_new}{rgb}{0.0, 0.5, 0.0}
\definecolor{green_new}{rgb}{0.0, 0.0, 0.0}
\begin{document}

   \title{VINTERGATAN-GM: Long-lived satellite planes induced by a massive Gaia-Sausage-Enceladus-like merger}
   \authorrunning{R. Rodríguez-Cardoso et al.}
   \titlerunning{Long-lived satellite planes in VINTERGATAN-GM}
  \author{R.~Rodríguez-Cardoso\inst{1, 2}\orcidlink{0000-0002-9158-195X},
    S.~Roca-Fàbrega\inst{3}\orcidlink{0000-0002-6299-152X},
    Oscar~Agertz\inst{3}\orcidlink{0000-0002-4287-1088},
    Jesus~Gallego\inst{1, 2}\orcidlink{0000-0003-1439-7697},
    Justin~Read\inst{4}\orcidlink{0000-0002-1164-9302},
    Andrew~Pontzen\inst{5}\orcidlink{0000-0001-9546-3849},
    Martin~P.~Rey\inst{6}\orcidlink{0000-0002-1515-995X},
    I.~Santos-Santos\inst{5}\orcidlink{0000-0001-6054-2897},
    M.~Gámez-Marín\inst{7}\orcidlink{0000-0002-1979-0941},
    Jess~Kocher\inst{3, 8}\orcidlink{0009-0005-2049-3847}}

   \institute{Departamento de Física de la Tierra y Astrofísica, Fac. CC. Físicas, Universidad Complutense de Madrid, Plaza Ciencias, 1, 28040 Madrid, Spain\\
              \email{ramorodr@ucm.es}
        \and Instituto de Física de Partículas y del Cosmos (IPARCOS), Fac. CC. Físicas, Universidad Complutense de Madrid,
Plaza Ciencias,~1,~28040 Madrid, Spain
        \and Lund Observatory, Division of Astrophysics, Department of Physics, Lund University, SE-221 00 Lund, Sweden
        \and Department of Physics, University of Surrey, Guildford GU2 7XH, UK
        \and  Institute for Computational Cosmology, Department of Physics, Durham University, Lower Mountjoy, South Road, Durham DH1 3LE, UK
        \and Department of Physics, University of Bath, Claverton Down, Bath BA2 7AY, UK 
        \and  Departamento de Física Teórica, Universidad Autónoma de Madrid, E-28049 Cantoblanco, Madrid, Spain 
        \and Materials Science and Applied Mathematics, Malmö University, SE-205 06 Malmö, Sweden
        }

   \date{Received 20 March 2026 / Accepted }

 
  \abstract
       {Satellite galaxies in the Local Group exhibit prominent spatially flattened configurations around their hosts, with a significant fraction also sharing coherent orbital motion. However, galaxy formation simulations in the Lambda Cold Dark Matter model have historically struggled to produce similarly planar, kinematically coherent, and stable structures, giving rise to the so-called “planes of satellites problem”. 
       }
       {We investigate the formation of planes of satellites and assess whether their emergence depends on the mass of a major merger at  $z\sim2$, comparable to the Gaia-Sausage-Enceladus (GSE) merger in the Milky Way.}
       {We employed the VINTERGATAN-GM suite of five high-resolution cosmological zoom-in realizations of the same Milky Way-mass halo, generated through targeted genetic modifications of a GSE progenitor. The total mass ratio of this $z\sim2$ merger was incrementally increased  from 1:10 to 1:2.1, while fixing the galaxy's final dynamical mass and large-scale environment. We examined whether these simulations produce planes of satellites comparable to those observed in the Local Group and how their properties depend on the GSE progenitor mass.}
       {We find a consistent trend that the more massive the GSE progenitor, the more planar and kinematically coherent the satellite population. Simulations with a GSE progenitor with a total mass ratio larger than 1:6 develop long-lived kinematically coherent planes with properties comparable to the one observed in the Milky Way. These structures arise because sufficiently massive mergers, infalling along the direction of maximum compression of the lagrangian volume, produce flattened host halos with anisotropic velocity dispersions. As a result, the minor axis of the resultant host halo aligns with the merger direction and the direction of flattening of the local cosmic web and planes of satellites emerge through two complementary processes: (i) satellites preferentially infall along the stable host’s equatorial plane, and (ii) anisotropic dynamical friction in the nonspherical halo gradually reshape their orbits toward this plane, generating coherent and long-lived planar configurations.}
       
  {}


   \maketitle
%

\section{Introduction}
\label{sec:intro}
The so-called small-scale problems of the Lambda cold dark matter ($\Lambda$CDM) model refer to discrepancies between observations and theoretical predictions on dwarf-galaxy scales in the standard cosmological framework (see \citealt{Bullock_Boylan_2017, Del_popolo_2017, Sales_2022} for reviews). Among these, the planes of satellites problem has long been regarded as one of the most challenging and debated issues (see \citealt{Pawloski_2018} for a comprehensive review).

Satellite galaxies around the Milky Way (MW) are observed to be preferentially distributed within a thin, extended plane oriented nearly perpendicular to the Galactic disk (e.g., \citealt{Lynden_Bell_1976,Kunkel_1976,Kroupa_Theis_Boily_2005,Pawlowski_Altenburg_Kroupa_2012}). This planar arrangement of satellites is commonly referred to as the Vast POlar Structure (VPOS). In addition to this pronounced spatial anisotropy, proper-motion measurements have revealed that a substantial fraction of MW satellites present orbital angular momentum vectors aligned with the normal to the VPOS \citep{Metz_2008,Pawloski_Kroupa_2013,Fritz_2018_coorb,Pawlowski_Kroupa_2020,Hammer_2021}. Recent Gaia-based proper-motion measurements show that approximately $40\%$ of MW satellites are co-orbiting\footnote{Throughout this work, the term ``co-orbitation'' refers to kinematic coherence irrespective of the sense of rotation, considering all satellites within an angular aperture of $\alpha_{\rm co-orbit} = 36.87^\circ$ as commonly adopted in the literature \citep{Fritz_2018_coorb, Santos_santos_sim_2020, Santos_santos_MW_2020, santos_santos_2023,Taibi_portrait_2024,Martinez_Garcia_delpino_2025, Gamez_Marin_TNG_2025}} around the VPOS normal \citep{Fritz_2018_coorb,Santos_santos_MW_2020,Taibi_portrait_2024}, underscoring the high level of kinematic coherence of this structure.

Similar planar satellite configurations have also been reported around Andromeda (M31; \citealt{Conn_2013,Ibata_2013}), Centaurus~A \citep{Muller_2018}, NGC~253 \citep{Martinez-Delgado_2021}, and NGC~4490 \citep{Pawlowski_2024, Karachentsev_2024}, among others. However, the three-dimensional spatial and kinematic coherence of satellite planes beyond the MW remains less well constrained due to projection effects, distance uncertainties, and the general lack of proper-motion measurements.

The cosmological significance of satellite planes remains debated, with no clear consensus on how frequently they arise or how long they persist. Numerous studies have searched for analogous satellite configurations in $\Lambda$CDM simulations and attempted to quantify their frequency. While spatially thin positional planes are occasionally found, they are generally rare and often transient \citep{Libeskind_2005,Libeskind_2009,Lovell_OrbitSubAccretion_2011,Bahl_2014,Cautun_2015,Buck_CoherenceM31_2016,Ahmed_CoherenceM31_2017,Santos_santos_sim_2020}. Some works have identified satellite planes that also exhibit kinematic coherence (e.g., \citealt{Shao_2019,Samuel_2021,santos_santos_2023,GamezMarin_2024,Gamez_Marin_TNG_2025}); however, most of these planar structures appear to be unstable and short-lived \citep{Bahl_2014,Gillet_2015,Buck_CoherenceM31_2016,Xu_2023,Zhao_2023}. Observationally, despite the uncertainties associated with backward orbit integration, studies of the MW VPOS likewise reach different conclusions regarding whether it is transient \citep{Lipnicky_2017,Maji_OrbitIntegration_2017,Sawala_Nature_2023} or long-lived \citep{Martinez_Garcia_delpino_2025}. A recent overview of satellite-plane lifetimes is presented by \citet{Sawala_TransietPlanes_2025}.

Building on this, \cite{santos_santos_2023} and \cite{Gamez_Marin_TNG_2025} identified the so-called kinematic persistent planes (KPPs) in hydrodynamical cosmological simulations. These structures are defined as groups of satellites whose orbital angular-momentum poles remain clustered around a common axis over extended timescales, forming a long-lived kinematic skeleton whose instantaneous positional planarity can be enhanced by additional transient satellites \citep{Shao_2019,santos_santos_2023,Sawala_Nature_2023,GamezMarin_2024}.

Despite growing evidence that such kinematically coherent satellite structures can arise in $\Lambda$CDM simulations, their physical origin and broader significance remain uncertain. Several mechanisms have been proposed to explain the clustering of satellite orbital poles, including anisotropic accretion along large-scale filaments \citep{Zentner_2005,Lovell_OrbitSubAccretion_2011,Libeskind_2011,Libeskind_Subhaloaccretion_2014,Wang_SatelliteAligment_2014,Libeskind_2015,Wang_SatelliteFilament_2020,Dupuy_Accretion_2022}; flattening of the local cosmic web (CW; \citealt{GamezMarin_2024}); the accretion of satellite groups \citep{Lynden_Bell_1995,Donghia_Lake_groupaccretion_2008,Li_Helmi_2008,Smith_GroupAccretion_2016};  the formation of satellites as tidal dwarf galaxies in gas-rich mergers \citep{Kroupa_2010,Fouquet_2012,Hammer_2013,Kroupa_2015, Akib_2025}; or the action of torques induced by an aspherical host halo, which can channel satellite orbits toward the host’s equatorial plane \citep{Shao_2019,Shao_2021}. 

The host mass-assembly history may therefore play an important role in shaping kinematically coherent satellite structures. In the MW, while the possible connection between the recent accretion of the LMC group and the VPOS has been extensively investigated \citep{Li_Helmi_2008,Nichols_LMCsatellites_2011,Garavito-Camargo_2021,santos_santos_2023,Vasiliev_2024,Martinez_Garcia_delpino_2025}, little attention has been paid to the impact of the last major merger experienced by the MW: the Gaia--Sausage--Enceladus (GSE) event, corresponding to the accretion of a massive dwarf galaxy at $z\sim2$ by the proto-MW \citep{Helmi_ProtoMW_2018, Belokurov_ProtoMW_2018}. Observationally, both the satellite population associated with the GSE progenitor and its total mass remain poorly constrained, with stellar and halo mass estimates spanning more than an order of magnitude ($M_\star \sim 10^8$--$10^9\,\mathrm{M}_\odot$, $M_{\rm halo} \sim 10^{10}$--$10^{11}\,\mathrm{M}_\odot$; e.g., \citealt{Bignone_2019,Bonaca_2020,Feuillet_2020,Kruijssen_2020,Mackereth_Bovy_2020,Naidu_2020,Naidu_2021,Limberg_2022,Lane_2023,Carrillo_2024}).

Cosmological simulations of galaxy formation provide a powerful framework for exploring the impact of a GSE-like merger on the formation of satellite planes in the MW. Ideally, this would require large cosmological volumes capable of sampling a representative range of merger histories, while simultaneously achieving sufficient resolution to resolve dwarf satellite populations, avoid numerical satellite disruption \citep{van_den_bosch_2018, Errani_2024, Santos_Santos_2025} and model the host galaxy disk, which is thought to play an important role in the anisotropic disruption of satellites \citep{Santos_santos_sim_2020,santos_santos_2023, Samuel_2021}. Large volume simulations with such a high resolution remain computationally prohibitive. High-resolution zoom-in simulations offer an alternative approach, but conventional suites generally comprise galaxies with independent formation histories shaped by different cosmological initial conditions (ICs). Consequently, the causal effect of a specific merger is difficult to disentangle from other differences in their large-scale environment, host mass, and accretion history.

The \textsc{Vintergatan-GM} suite of simulations \citep{Rey_VintergatanGM_2023} circumvents these limitations by employing the \textsc{genetIC} modification code \citep{Roth_Pontzen_2016_GM,Rey_Pontzen_2018,Stopyra_Pontzen_2021_GM}. This method allows for targeted modifications of the ICs to systematically vary the mass ratio of a specified GSE-like merger at $z \sim 2$, while preserving the same $z=0$ halo mass, large-scale environment, and surrounding structures. As a result, the impact of the GSE-like merger mass ratio can be isolated while controlling for most other factors affecting the evolution of the central galaxy and its satellite system.  Previous studies have shown that varying the merger mass ratio in this suite leads to pronounced differences in the morphology of the central galaxy at $z=0$ \citep{Rey_VintergatanGM_2023} and in the satellite stellar mass function several Gyr after the merger \citep{Joshi_VintergatanGM_2024}.

In this work, we investigate whether MW-mass halos in the \textsc{Vintergatan-GM} simulation suite develop satellite planes comparable to those observed around the MW and M31, and whether their properties depend on the mass of the GSE-like progenitor. We describe the simulations and analysis methods in Section~\ref{sec:methods}. The presence of high-quality positional planes is examined in Section~\ref{sec:positional_planes}, while the kinematic coherence of satellite systems is analyzed in Section~\ref{sec:kpp}. In Section~\ref{sec:origin}, we explore the physical mechanisms driving the differences observed across the simulations. In Section~\ref{sec:discussion}, we compare our findings with those of previous studies. Our conclusions are summarized in Section~\ref{sec:conlcusions}.

\section{Methods}
\label{sec:methods}
\subsection{VINTERGATAN-GM simulations}
This paper analyzes the \textsc{Vintergatan-GM} suite of simulations presented in \cite{Rey_VintergatanGM_2023}. The suite comprises five zoom-in cosmological hydrodynamical simulations of MW-mass halos. All simulations assume a flat $\Lambda$CDM cosmology with $H_0 = 67.27 \,{\rm km\,s^{-1}\,Mpc^{-1}}$, $\Omega_{\rm m,0} = 0.3139$, $\Omega_{\rm b,0} = 0.04916$, $\sigma_{8} = 0.8440$, and $n_{\rm s} = 0.9645$ \citep{Planck_2015}, and evolve from $z = 99$ to $z = 0$. It includes a fiducial MW–mass system and four additional realizations in which the cosmological ICs were genetically modified (GM) using the \textsc{genetIC} code \citep{Roth_Pontzen_2016_GM, Rey_Pontzen_2018, Stopyra_Pontzen_2021_GM} to alter the mass ratio of a major merger occurring at $z\sim2$ while keeping the $z=0$ dynamical mass and the surrounding large-scale structure fixed. This suite of GM ICs was originally introduced in \cite{Rey_Starkenburg_2022} (see their `Milky-Way-like' family), where multiple dark matter-only (DMO) zoom-in simulations of two MW-mass hosts were performed. The fiducial \textsc{Vintergatan-GM} simulation targets a dark matter (DM) halo with a total dynamical mass $M_{\rm vir} \sim 10^{12}\rm{M_\odot}$, consistent with current observational estimates of the MW halo mass (see \citealt{Wang_MWmass_2020} for a review), and with no other massive neighbors within $5\,\rm{R_{vir}}$\footnote{Throughout this paper, the virial radius $R_{\rm vir}$ is defined following the standard definition of \citet{Bryan_Norman_1998} as the radius enclosing a virial overdensity $\Delta_c$ that depends on redshift according to $\Delta_c = 18\pi^2+60x-32x^2$, where $x=\frac{\Omega_0(1+z)^3}{E^2(z)}$ and $E(z)\equiv H(z)/H_0$ is the dimensionless Hubble parameter. The virial mass $M_{\rm vir}$ is then defined as the total mass enclosed within $R_{\rm vir}$.}, and which experiences an early major merger with a radially biased orbit (infall at $z\sim2$, with a $M_{\rm vir}$ ratio of 1:6 and a radial-to-tangential velocity ratio between the progenitors $\rm{v_r/v_\theta = 16}$). This merger resembles a GSE-like event, and the halo's assembly history was selected to broadly resemble the inferred properties of the progenitors of the proto-MW \citep{Belokurov_ProtoMW_2018, Helmi_ProtoMW_2018}. The GM realizations systematically vary the mass of this $z\sim2$ GSE-like progenitor, denoted as Merger B, by modifying the local overdensity field to 90, 95, 110, and 120 percent of the fiducial value. These adjustments result in total (stellar) mass ratios spanning 1:10 (1:24) to 1:2.1 (1:2.1). Throughout this work, we refer to these GM simulations as Smallest, Smaller, Larger, and Largest. The detailed properties of the modified $z\sim2$ progenitor are listed in Table~\ref{tab:table_progenitor_B}. These simulations also experience an earlier major merger, labeled Merger~A, at $z\sim3.3$.  The mass of this early merger varies slightly across the suite, ranging from total (stellar) mass ratios spanning from 1:0.7 (1:1.8) to 1:1.9 (1:2.4) (please see Table~1 in \citealt{Rey_VintergatanGM_2023}), although both this variation and its impact on the subsequent evolution are small compared to those of Merger~B.
\begin{table*}
\caption{Properties of the GSE-like progenitor (Merger~B).}
\label{tab:table_progenitor_B}
\setlength{\tabcolsep}{10pt} 
\centering
\begin{tabular}{lccccc}
\hline
Property & Smallest & Smaller & Fiducial & Larger & Largest \\
\hline

$z_{\rm infall}$
& $1.94$
& $2.03$
& $1.99$
& $1.99$
& $2.24$ \\[2pt]

$M_{\rm vir}^{\rm infall}\,[{\rm M}_{\odot}]$
& $2.2\times10^{10}$
& $2.3\times10^{10}$
& $3.8\times10^{10}$
& $7.4\times10^{10}$
& $1.1\times10^{11}$ \\[2pt]

$M_{\star}^{\rm infall}\,[{\rm M}_{\odot}]$
& $2.2\times10^{8}$
& $3.1\times10^{8}$
& $6.0\times10^{8}$
& $1.2\times10^{9}$
& $2.3\times10^{9}$ \\[2pt]

$M_{\rm vir}^{\rm prog}/M_{\rm vir}^{\rm host}$
& $1{:}10$
& $1{:}9.8$
& $1{:}6.0$
& $1{:}2.9$
& $1{:}2.1$ \\[2pt]

$M_{\star}^{\rm prog}/M_{\star}^{\rm host}$
& $1{:}24$
& $1{:}15$
& $1{:}8.1$
& $1{:}4.3$
& $1{:}2.1$ \\[2pt]

$(v_r/v_\theta)_{\rm infall}$
& $9.5$
& $9.1$
& $16$
& $8.8$
& $5.5$ \\[2pt]

$z_{\rm pericenter}$
& $1.70$
& $1.76$
& $1.79$
& $1.80$
& $1.98$ \\[2pt]

$z_{\rm coalescence}$
& $1.40$
& $1.45$
& $1.52$
& $1.47$
& $1.65$ \\[2pt]

\hline
\end{tabular}

\tablefoot{
$z_{\rm infall}$ is defined as the redshift of the first snapshot in which the GSE-like progenitor crosses the virial radius of the central halo.  At this time, 
we report the progenitor virial and stellar masses, merger mass ratios, and radial to tangential velocity $v_r/v_\theta$.  The quantities $z_{\rm pericenter}$ and $z_{\rm coalescence}$ denote the redshifts of the first pericentric passage and final coalescence, respectively.
}
\end{table*}

Each simulation has a mass resolution of $m_{\rm DM} = 2\times10^5\,\rm{M}_\odot$ and a minimum initial gas cell mass of $m_{\rm gas}=3.6\times10^4\,\rm{M}_\odot$. A detailed description of the numerical set-up and galaxy formation physics is provided in \cite{Rey_VintergatanGM_2023}; here we summarize the main elements. \textsc{Ramses} \citep{Teyssier_Ramses_2002} is an adaptive mesh refinement (AMR) hydrodynamics code that solves Poisson’s equation via a particle–mesh gravity solver and models gas dynamics using an HLLC (Harten--Lax--van Leer--Contact) Riemann solver, assuming an ideal gas with $\gamma = 5/3$. The AMR scheme enables spatial resolutions of order $\sim 20$ pc in the interstellar medium. The simulations adopt the \textsc{Vintergatan} galaxy formation model \citep{Agertz_VG_2021}, which includes prescriptions for star formation, stellar feedback from SNeIa and SNeII, and winds from massive and AGB stars. The simulations follow equilibrium cooling of a metal-enriched plasma \citep{Courty_Alimi_2004, Rosdahl_2013}, including heating and photoionization from a spatially uniform, time-dependent ultraviolet background implemented in \textsc{RAMSES} following \cite{Haardt_Madau_96}. Gas with $n_{\rm H} \ge 0.01\ \mathrm{cm}^{-3}$ self-shields from this radiation field \citep{Aubert_2010, Rosdahl_Blaizot_2012}, allowing it to cool to $T \le 100\ \mathrm{K}$ and condense to densities $\rho \ge 100\ \mathrm{cm}^{-3}$, where star formation is modeled via a Schmidt-law prescription \citep{Schmidt_1959, Rasera_Teyssier_2006, Agertz_2013}, producing stellar particles of initial mass $10^{4}\ {\rm M}_\odot$ sampled from a \cite{Chabrier_2003} initial mass function (IMF). Feedback energy is deposited thermally when the cooling radius is resolved by at least six AMR cells, and otherwise injected as momentum \citep{Kim_Ostriker_2015, Martizzi_2015, Agertz_VG_2021}. The \textsc{Vintergatan} model has been shown to successfully reproduce key structural, kinematic, and chemical properties of MW-like galaxies, as well as their merger-driven evolution \citep{Agertz_VG_2021,Renaud_2021,Renaud_2021b,Segovia_Otero_2022}. The same GM initial conditions are also studied within the \textsc{Paradigm} project \citep{Joshi_Paradigm_2025}, which compares the \textsc{Vintergatan} and IllustrisTNG physical models. For the satellite populations, \citet{Joshi_Paradigm2_2025} found no systematic dependence of disruption fractions or timescales on merger mass ratio or physical model, although \textsc{Vintergatan} produces more luminous satellites because of its higher star-formation efficiency in low-mass halos.

\label{sec:VG_simulations}

\subsection{Subhalo finding and tracking}
\label{sec:halofinder}

To analyze the formation of planes of satellites in our simulations, it is essential to obtain an accurate tracking of the orbital evolution of satellite galaxies. This requires reliable identification of subhalos throughout their infall into the host halo. However, the detection of subhalos during their interaction with the host remains a nuanced challenge in numerical simulations. While halo finders generally agree on the positions and properties of isolated halos \citep{Knebe_2012_galaxycomparison}, the detection of subhalos is considerably more difficult due to their tendency to blend into the variable background density of their larger host (see \citealt{Symfind_Mansfield_2024} for a recent overview). Even widely used algorithms such as \textsc{Rockstar} \citep{Rockstar_behroozi_2012} and \textsc{SubFind} \citep{Subfind_springel_2001} struggle to identify subhalos that undergo substantial mass loss from strong tidal stripping \citep{Onions_2012_subhalofindercomparison, Diemer_sparta_2024, Forouhar_Moreno_2025}.

To ensure robust subhalo tracking in these regimes, we adopted the method described in detail by \citet{Rodirguez_Cardoso_Agora_2025}. We summarize the main steps here. Our approach begins with the \textsc{Rockstar} halo finder, which identifies halos and subhalos in individual snapshots by locating overdensities in the 6D phase-space distribution. We then apply the merger tree code \textsc{Consistent-Trees} \citep{Consistent_behroozi_2012} to establish temporal connections between halos and subhalos. For convenience, we refer to this combination of \textsc{Rockstar} and \textsc{Consistent-Trees} as “\textsc{RCT}.”

\textsc{RCT} catalogs are known to suffer from premature subhalo loss, especially during pericentric passages, where subhalos are frequently misclassified as merged \citep{Diemer_sparta_2024, Symfind_Mansfield_2024, Rodirguez_Cardoso_Agora_2025, Forouhar_Moreno_2025, Barrow_2026}. For this reason, we adopt a particle-based post-processing approach similar to those implemented in \textsc{Symfind} \citep{Symfind_Mansfield_2024}, \textsc{Sparta} \citep{Diemer_sparta_2024} and \textsc{Bloodhound} \citep{Kong_Bloodhound_2026}. Specifically, we adopt the \textsc{RCT} catalogs for isolated halos, but once a halo crosses the virial radius of a larger host and becomes a subhalo, we track its evolution exclusively through its associated DM particles identified prior to infall. In subsequent snapshots, the subhalo’s position is determined from its most bound particles, while unbound particles are iteratively removed. Structural properties are computed from the remaining bound material, and subhalo tracking is terminated once the number of bound particles falls below a minimum threshold (${N_{\rm thresh}} = 20$) or the subhalo becomes indistinguishable from the host core for several consecutive snapshots.

This algorithm has been tested in the AGORA \texttt{CosmoRun} simulation suite by \citet{Rodirguez_Cardoso_Agora_2025} across multiple numerical codes, including \textsc{RAMSES}, demonstrating excellent performance in tracking subhalos down to their actual physical disruption. In Appendix~\ref{app:rct_part_tracking}, we present the results of applying this method to the \textsc{Vintergatan-GM} simulation suite and assess its performance by directly comparing the particle-tracking results with the standard RCT output, highlighting the improvement in tracing subhalos after many orbits toward the inner regions of the host halo (see also Figure~\ref{fig:CDF_sats_rad}).

\subsection{Satellite selection}
\label{sec:star_assigment}
During the construction of the corrected merger trees described in the previous section, stellar particles are assigned to halos and subhalos. For isolated central halos, we adopt the same iterative stellar assignment procedure as in \citet{Samuel_stellarassigment_2020} and \citet{Jung_Agora_2024}, which we briefly summarize here. For each halo, we initially selected all stellar particles within a radius of $0.8\,R_{\rm vir}$. From this sample, we retain only particles whose velocities relative to the halo are smaller than twice the halo’s maximum circular velocity. Using this subset, we compute the radius enclosing $90\%$ of the stellar mass, $r_{90}^{\rm star}$, together with the stellar velocity dispersion, $\sigma_{\rm vel}^{\rm star}$. The selection is then refined by imposing two additional constraints: stellar particles must lie within $1.5\,r_{90}^{\rm star}$ of the center of mass of the halo and the stellar particles, and their relative velocities must be below $2\,\sigma_{\rm vel}^{\rm star}$. The values of $r_{90}^{\rm star}$ and $\sigma_{\rm vel}^{\rm star}$ are recalculated for the updated particle set, and this procedure is iterated until the values converge at the $99\%$ level between successive iterations.

Assigning stellar particles to subhalos embedded within a more massive host halo is more challenging, as it becomes difficult to distinguish particles genuinely bound to the subhalo from those belonging to the central stellar halo or disk, particularly during pericentric passages when the two populations may overlap in six-dimensional phase space. To address this, stellar particles are assigned to subhalos using temporal information during subhalo tracking, in a manner analogous to the identification of DM particles that remain bound to the subhalo.
Specifically, stellar particles associated with each subhalo are first identified in the snapshot immediately prior to infall, using the same iterative criteria adopted for isolated halos. Unlike DM particles, subhalos may continue to form stellar particles after infall. Therefore, at each subsequent snapshot, we add to the pre-infall set of stellar particles the newly formed stellar particles that satisfy the iterative selection criteria and are associated with the subhalo. At every post-infall snapshot, a gravitational unbinding procedure is applied to the combined set of pre-infall and newly formed stellar particles, removing those that have become unbound due to tidal stripping.

As in the treatment of DM particles during subhalo-tracking described in \citet{Rodirguez_Cardoso_Agora_2025}, we assume that subhalos do not significantly accrete stellar particles from the host during infall, and that any stellar mass growth is driven exclusively by in situ star formation from the subhalo’s own cold gas. While this assumption may break down in rare cases, such as during major mergers, the number of host stellar particles temporarily captured by subhalos is expected to be negligible compared to the subhalo’s initial stellar mass \citep{Behroozi_2015,Diemer_sparta_2024}. Even when a host stellar particle becomes temporarily bound to a subhalo, it is typically lost again on short timescales and has little impact on the long-term evolution of the subhalo’s stellar mass \citep{Han_2012}. This simplification significantly reduces computational cost. As an exception, if a subhalo contains its own (sub)satellite population prior to crossing the host’s virial radius, we allow the stellar particles associated with these systems to be accreted by the subhalo during infall, in direct analogy with the treatment of DM particles described in \citet{Rodirguez_Cardoso_Agora_2025}.

The performance of this method for associating stellar particles with isolated halos and subhalos is illustrated in the bottom row of Figure~\ref{fig:comparison_rct}, which shows both the smooth evolution of the stellar mass and the stripping of stellar particles during subhalo infall.

Finally, we define a (satellite) galaxy as any (subhalo) halo with a stellar mass $M_\star \geq 10^{5}\,M_\odot$, corresponding to systems resolved with more than $\sim$10 stellar particles. Observationally, this threshold approximately selects the classical MW satellites, corresponding to about 15 systems in the MW and 32 in M31 \citep[e.g.,][]{Pace_2025}, while the faintest known VPOS members fall below it. However, the \textsc{Vintergatan} model forms stars more efficiently in low-mass halos than several other galaxy-formation models \citep{Joshi_Paradigm2_2025}, yielding a larger number of satellites above this stellar-mass threshold. Our simulated samples contain $\sim40$ -- 45 satellites, comparable in size to observed samples extending down to $M_\star\sim10^{3.5}$--$10^{4}\,{\rm M_\odot}$, such as the observational samples used in recent works about planes of satellites (e.g., \citealt{Santos_santos_MW_2020, Taibi_portrait_2024, Martinez_Garcia_delpino_2025}). Some satellites in our simulations may therefore correspond observationally to ultra-faint dwarfs with lower stellar masses than implied by our nominal threshold. Importantly, this likely overestimate of stellar mass in low-mass halos in our simulations affects only the size and observational interpretation of the selected satellite sample. Stellar mass enters our analysis solely as a selection criterion and does not otherwise influence the plane-finding methodology or the orbital evolution of the selected subhalos, which is governed primarily by their total, dark-matter-dominated masses, the host halo mass, and the surrounding large-scale structure.

\section{High-quality positional planes}
\label{sec:positional_planes}
To systematically search, snapshot by snapshot, for positional planar configurations of satellites in our simulations, we employed the 4-galaxy-normal density plot (4GND plot) method, originally introduced by \cite{Pawlowski_2013_4GND} and later extended by \cite{Santos_santos_sim_2020, Santos_santos_MW_2020}. This technique enables us to identify whether a subsample of the total population of ${N_{\rm tot}}$ satellites defines a planar arrangement, based on the outputs of a fitting procedure that utilizes the tensor of inertia (ToI; \citealt{Metz_ToI_2007, Pawlowski_2013_4GND}) of the satellite population. 

\subsection{Four-galaxy-normal density plots}
The 4GND plot method is applied to each timestep of the simulations; here we briefly summarize the main steps of the method below (see \citealt{Pawlowski_2013_4GND} for a detailed description):

\begin{enumerate}
    \item A plane is fitted, using the ToI method \citep{Metz_ToI_2007, Pawlowski_2013_4GND}, to the positions of every combination of four\footnote{Three points always define a plane, but do not allow any quantification of plane thickness. Therefore, four satellites are the minimum required to enable thickness estimation while still producing a sufficiently large number of combinations for a robust signal, see e.g., \citep{Pawlowski_2013_4GND}.} distinct satellites. The resulting plane normal vector (i.e., the 4-galaxy-normal) is stored along with the principal-axis lengths $a$, $b$, and $c$, corresponding to the major, intermediate, and minor axes, respectively.
    \item All 4-galaxy-normal vectors are transformed into a galactocentric coordinate system aligned to a fixed reference direction\footnote{We use the host halo minor axis at $z=0$ in the Largest simulation as the reference direction, which allows consistent comparison across snapshots and simulations.} and projected onto a regularly binned sphere to create a density map. As opposite normal vectors indicate the same plane, density maps in this study are shown using Aitoff spherical projection diagrams in Galactic coordinates (longitude $l$, latitude $b$) within the interval $l \in [-90^{\circ}, +90^{\circ}]$. Following \cite{Pawlowski_2013_4GND}, each normal vector is weighted by $w = \log\left(\frac{a + b}{c}\right)$ to emphasize planar satellite configurations over filamentary or spheroidal ones. In these plots, overdensities correspond to groups of satellites contributing to the same dominant planar spatial configuration. 
    \item Density peaks are identified and isolated by ordering all bins according to their density values. The primary overdensity is defined around the bin with the highest density. Subsequent overdensities are selected in decreasing order of density, with the requirement that each new peak is separated by more than $15^{\circ}$ from the center of all previously defined overdensities. In this way, distinct overdensity regions are systematically differentiated. For each region, the coordinates of the density peak are given by the center of its highest-density bin.
    \item We quantify the contribution of a given satellite $s$ to a density peak $p$, denoted $C_{p,s}$. We define an aperture of $15^{\circ}$ 
    
    around the peak and select all 4-galaxy-normals that fall within this region. For each normal, the four contributing satellites are identified. If satellite $s$ is among them, its contribution to peak $p$ is counted as the weight of that 4-galaxy-normal. The final contribution $C_{p,s}$ is obtained by summing the weights of all 4-galaxy-normals within the aperture that include satellite $s$. This value is then normalized by $C_{N,{\rm all}}$, the total weighted number of 4-galaxy-normals (including those outside the aperture), ensuring consistent comparison across samples with different satellite population sizes $N_{\rm tot}$.    
\end{enumerate}

\subsection{Extension to the 4GND plot method}
\label{sec:extension}
To analyze the relevance of density peaks at each timestep and to compare simulations with observations, we apply the extension described in \cite{Santos_santos_sim_2020, Santos_santos_MW_2020}, summarized in the following additional steps:

\begin{enumerate}[start=5]
    \item  To each peak we assign a value $C_p$, the peak strength, defined as the normalized number (or percentage) of 4-galaxy normals within $15^\circ$ of the peak center. By construction, $C_p = \sum_s C_{p,s}$, where the contribution number $C_{p,s}$ for satellite $s$ is defined in step (4) above. 
    \item For each peak, a collection of planes is constructed for different choices of satellite number $N_{\rm sat}$. Satellites are ranked by their contribution to the $p$th peak, and the ToI technique is applied to the top $N_{\rm sat}$ contributors to obtain the best-fitting plane. Therefore, for each peak, a set of planes is derived from progressively larger subsets of satellites. 
\end{enumerate}

Following the approach adopted by \citet{Santos_santos_sim_2020,Santos_santos_MW_2020}, we use this methodology to identify high-quality positional planes, defined as those that are both well populated and highly flattened, as quantified by $N_{\rm sat}$ and $c/a$. We then compare the highest-quality positional planes across the different simulations.

\subsection{Analysis of 4GND metrics}

\begin{figure*}
   \centering
   \includegraphics[width=0.99\linewidth]{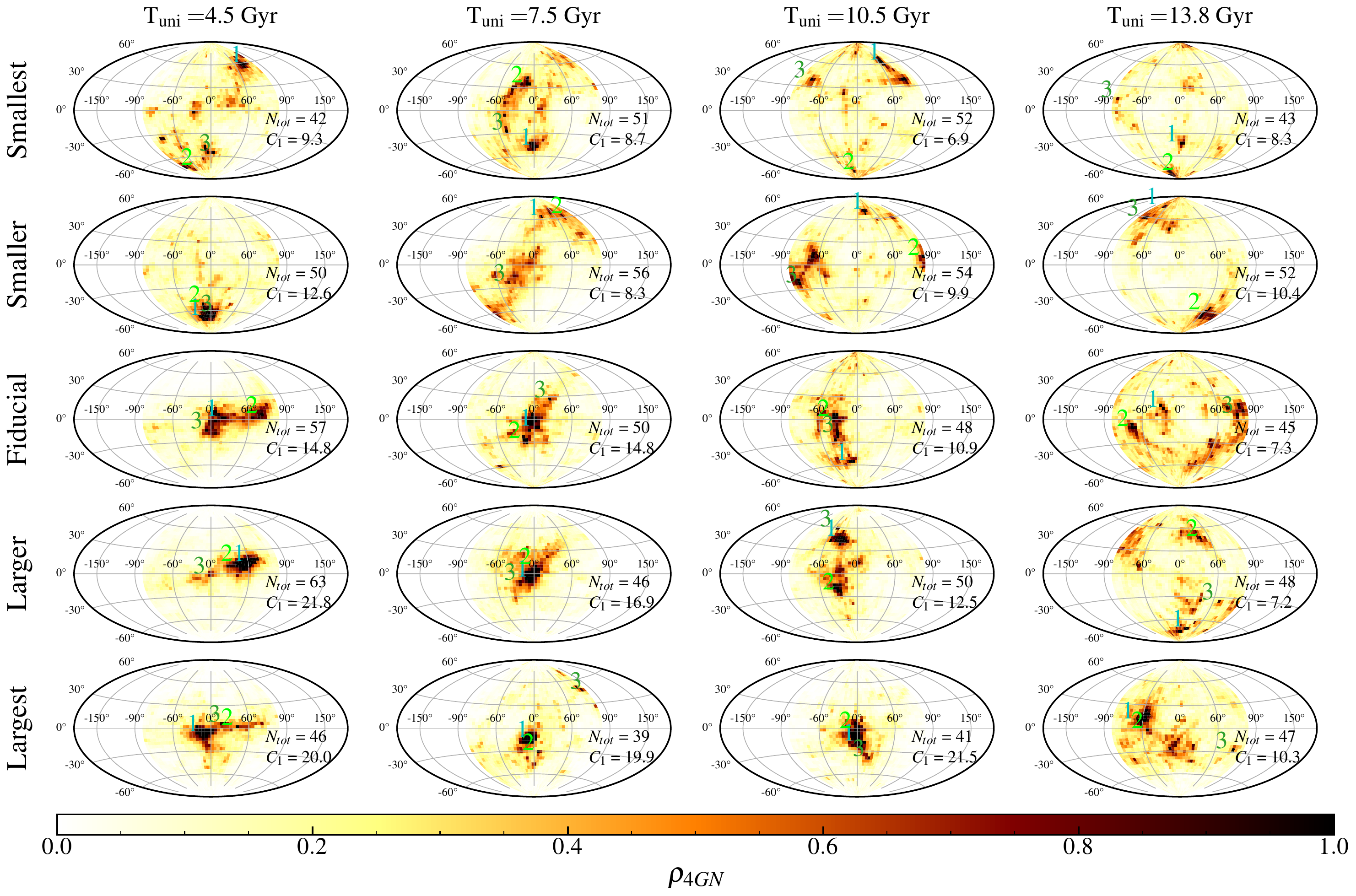}
   \caption{Examples of four-galaxy-normal-density plots (4GND plots) at four different timesteps (4.5, 7.5, 10.5, and 13.8 Gyr) for each of the simulations. Each column represents a different timestep and each row a different simulation. The main density peaks are labeled with numbers ordered by decreasing central bin density. The legend in each panel reports the total number of satellites at that timestep, $N_{\rm tot}$, and the peak strength of the main peak, $C_1$. The colorbar values are proportional to the normalized bin density of four-galaxy-normals. Overdensities in these plots indicate groups of satellites contributing to the same planar configuration and thus reflect an anisotropic satellite distribution, in contrast to the uniform density expected for an isotropic distribution.}
    \label{fig:4GNDPlot}%
\end{figure*}

Figure~\ref{fig:4GNDPlot} shows examples of 4GND plots for all the \textsc{Vintergatan-GM} simulations at four representative timesteps, $T_{\rm uni}=$ 4.5, 7.5, 10.5, and 13.8~Gyr\footnote{Throughout this paper, $T_{\rm uni}$ denotes cosmic time measured from the Big Bang, with $T_{\rm uni}=13.8$~Gyr corresponding to the present day.}. These approximately equally spaced timesteps were selected to illustrate the evolution of satellite spatial anisotropy since the GSE-like merger to $z=0$. The three primary density peaks are labeled in decreasing order of their central-bin density. Because peaks are identified independently at each timestep, a given peak number does not necessarily trace the same planar configuration through time. This figure may be compared with the corresponding 4GND plots for the observed satellites in the MW and M31 presented in \cite{Santos_santos_MW_2020}, as well as with results from other cosmological zoom-in simulations shown in \cite{Santos_santos_sim_2020}.

The 4GND plots reveal that, in several simulations and at specific timesteps, the distribution of 4-galaxy-normal vectors is highly anisotropic, in sharp contrast to the uniform density expected for a random isotropic satellite configuration. At $T_{\rm uni} = $~4.5, 7.5 and 10.5 Gyr, a general trend across the simulations is that the overdensities are more pronounced in the Larger and Largest runs than in the Smallest and Smaller ones, with the Fiducial simulation lying between these two groups. This behavior reflects a systematic increase in the anisotropy of the satellite spatial distribution from the Smallest to the Largest simulations at these epochs. The same trend is captured by the strength of the main peak, $C_1$, which increases steadily from $\rm{\sim8}\%$ in the Smallest run to $\rm{\sim20}\%$ in the Largest case. By contrast, at the 13.8 Gyr timestep, these differences largely disappear, as the Fiducial, Larger, and Largest simulations exhibit a reduced anisotropy in their 4-galaxy-normal distributions. This reduced anisotropy in the final snapshots is mainly driven by the recent accretion of satellites that are not aligned with the plane, together with the temporary misalignment of transient satellites, both of which weaken the 4GND overdensities (see Appendix~\ref{app:persvstrans} for more details).

\begin{figure*}
   \centering
   \includegraphics[width=0.99\linewidth]{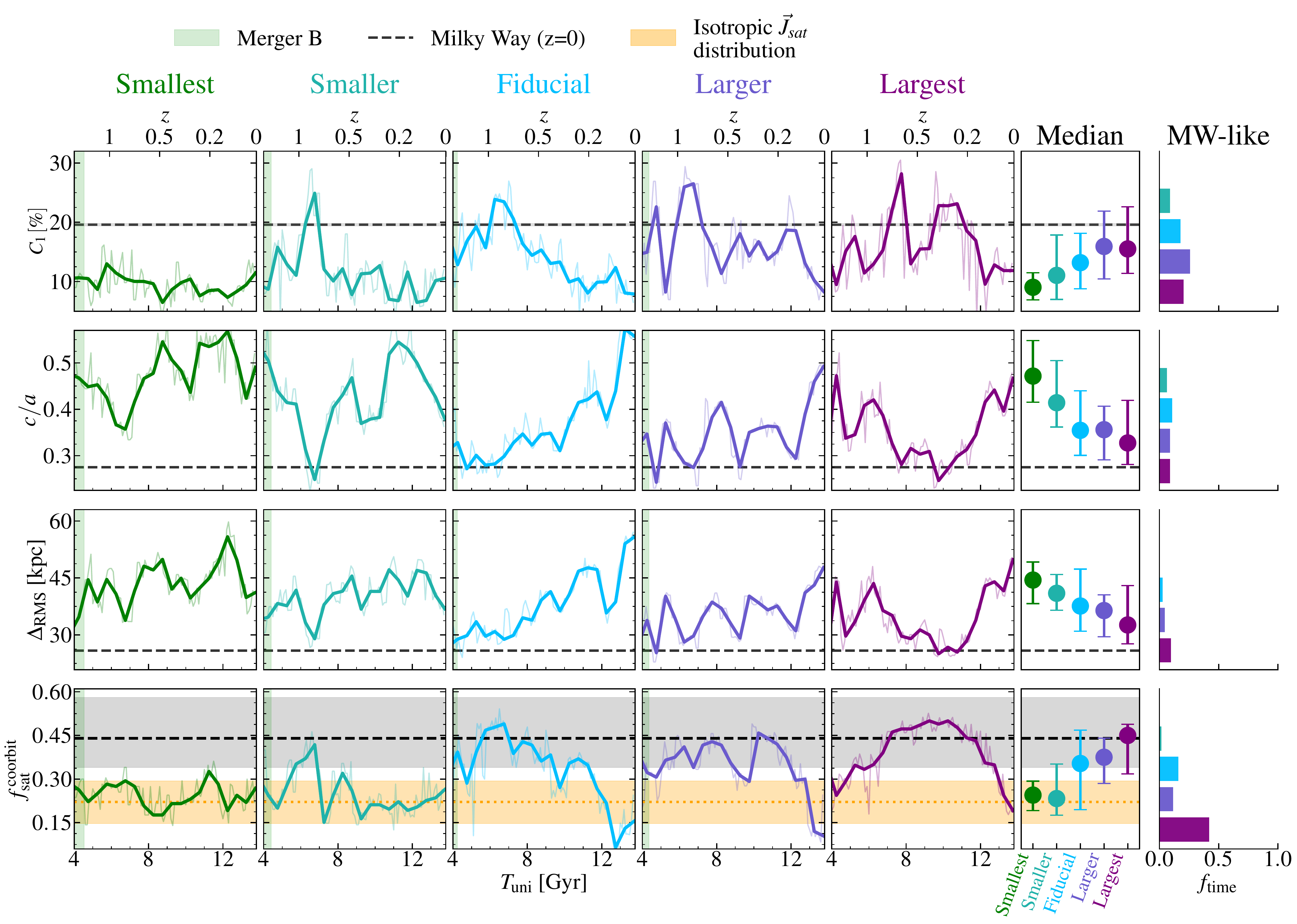}
   \caption{4GND metrics evolution over cosmic time for each simulation. 
   The first five columns show the temporal evolution of each metric for the five simulations, with green shaded regions indicating the time interval of Merger~B. Thin lines show the full time evolution, while thick lines represent the median values computed in 0.5~Gyr bins. \textbf{First row:} Evolution of the peak strength ($C_1$), which quantifies the degree of anisotropy in the distribution of 4-galaxy-normal vectors. \textbf{Second row:} Evolution of the short-to-long axis ratio ($c/a$) of the plane associated with the main peak, computed using $f_{\rm sat} = 90\%$. 
   \textbf{Third row:} Evolution of the root-mean-square thickness ($\Delta_{\rm RMS}$) of the plane associated with the main peak, computed considering $f_{\rm sat} = 90\%$. \textbf{Fourth row:} Evolution of the fraction of co-orbiting satellites relative to the normal of the plane used in the second and third rows. The sixth column shows, for each metric and simulation, the median value over the time evolution since Merger~B, with vertical bars indicating the central 16th–84th percentile range.
   Black dashed lines denote the present-day ($z=0$) values measured for the MW. In the fourth row, the gray band shows the observed MW value and its uncertainty derived from proper motions uncertainties, while the yellow band corresponds to the expectation for an isotropic distribution of satellite angular momentum directions. The rightmost column displays the fraction of time ($f_{\rm time}$), computed since Merger~B, during which each simulation reaches or exceeds the levels of planarity measured for the MW, according to each metric. A clear trend emerges across the suite, with more massive GSE-like mergers producing thinner and more coherent satellite planes throughout the post-merger evolution.
   }
              \label{fig:4GND_metrics_evolution}%
\end{figure*}

To better analyze the positional planes in our simulations, we examine the time evolution of the 4GND metrics associated with the main overdensities in the 4GND plots, rather than focusing on individual snapshots. The first five columns of Figure \ref{fig:4GND_metrics_evolution} show the temporal evolution of the different 4GND metrics for each simulation.  The properties $c/a$, $\Delta_{\rm RMS}$, and $f_{\rm sat}^{\rm coorbit}$ are computed at each timestep for the plane associated with the main peak and considering a satellite fraction of $f_{\rm sat} = 90\%$\footnote{As described in step (6) of Section~\ref{sec:extension}, for each peak a collection of planes is constructed for different choices of the number of satellites, $N_{\rm sat}$, or equivalently $f_{\rm sat} = N_{\rm sat}/N_{\rm tot}$.}. We adopt $f_{\rm sat}=90\%$ to assess the planarity of nearly the entire satellite population and enable a stringent comparison with the MW, whose satellite distribution remains highly flattened when almost all known satellites, including recently discovered faint dwarfs, are considered (e.g., \citealt{Santos_santos_MW_2020}). The temporal evolution of these metrics for different satellite fractions ($f_{\rm sat}=70\%$ and $50\%$) is shown in Figure \ref{fig:f_sat_70_50}. The observed value for the MW plane when considering a sample of 46 satellites is shown as a black dashed line using the values listed in Table~\ref{tab:4gnd_metrics_mw_m31}, extracted from \citet{Santos_santos_MW_2020}.

Throughout this work, the fraction of co-orbiting satellites with respect to the plane normal, $f_{\rm sat}^{\rm coorbit}$ is defined as the fraction of satellites whose orbital poles lie within an angular aperture corresponding to 10 \% of the sphere ($\alpha_{\rm co\text{-}orbit}=36.87^\circ$) from the plane normal, consistent with the standard definition in the literature \citep{Fritz_2018_coorb,Santos_santos_MW_2020,Taibi_portrait_2024,Martinez_Garcia_delpino_2025}. This choice enables a direct comparison between our results and those observed for the MW (e.g., \citealt{Santos_santos_MW_2020}), as well as with predictions from previous cosmological simulations \citep{Santos_santos_sim_2020,Gamez_Marin_TNG_2025}. We have additionally verified that moderate variations in the adopted $\alpha_{\rm co\text{-}orbit}$ aperture do not lead to systematic differences in our conclusions.

\begin{table}
\centering
\caption{4GND metrics for the observed planes in MW when considering 46 satellites. 
}
\begin{tabular}{lc}
\hline
& MW46 \\
\hline
$C_1$
& $19.59\pm0.19$ 
  \\[2pt]

${c/a}^*$
& $0.27\pm0.002$  
 \\[2pt]

${\Delta_{\rm RMS}\, [\rm{kpc}]}^*$
& $24.84\pm0.6$  
 \\[2pt]

${f_{\rm sat}^{\rm \,coorbit}}^*$
& $0.44\pm0.11$  
\\[2pt]
\hline
\end{tabular}
\vspace{2pt}
\par\footnotesize
\textit{Note.} Properties marked with (*) were computed using the plane associated with the main peak and corresponding to $f_{\rm sat} = 90\%$.
\label{tab:4gnd_metrics_mw_m31}
\end{table}

The temporal evolution of the 4GND planarity metrics exhibits strong variations, reflecting the continuous evolution of the satellite population through the accretion and disruption of satellites, as well as temporary alignments that produce transient planar structures. To identify systematic trends across the simulation suite, the sixth column of Figure \ref{fig:4GND_metrics_evolution} shows the median value of each metric for each simulation, computed over the time evolution since the GSE-like merger. In addition, to facilitate comparison with the observed MW system, the rightmost column displays the fraction of time ($f_{\rm time}$), measured since Merger B, during which each simulation reaches or exceeds the levels of planarity observed in the MW for each metric.

The evolution of the main peak strength, $C_1$, reveals systematic differences across the genetically modified simulations, particularly evident when considering the median values. The Smallest and Smaller runs fluctuate around the observed $\sim10\%$ for most of the time and remain well below the MW value ($\sim20\%$), except for a brief peak in the Smaller run. In contrast, the Larger and Largest simulations show higher $C_1$ values that fluctuate around MW levels, frequently exceeding the MW threshold. Overall, the median values display a continuous increase of $C_1$ with the mass of the GSE-like merger, indicating a progressively stronger clustering of the 4-galaxy-normal vectors.

When analyzing the quality of these planes by considering the $c/a$ and $\Delta_{\rm RMS}$ evolution, we found that they are closely linked to the behavior of $C_1$, since thin planes characterized by low $c/a$ and $\Delta_{RMS}$ correspond to strongly anisotropic distributions of 4-galaxy-normal vectors. Consequently, a similar trend across the simulation suite is recovered. The median values decrease systematically from $c/a\sim0.5$ and $\Delta_{\rm RMS}\sim45$ kpc in the Smallest run to $c/a\sim0.3$ and $\Delta_{\rm RMS}\sim30$ kpc in the Largest run. When compared with the observed MW values, the MW plane appears particularly extreme and is only matched by the Larger and Largest simulations for approximately 5 \% and 10 \% of the evolution time, respectively.

Consistent with the trends identified above, the fourth row of Figure~\ref{fig:4GND_metrics_evolution} reveals a corresponding progression in the degree of co-orbitational coherence around the plane normal. The Smallest and Smaller runs show little to no co-orbitation, with $f_{\rm sat}^{\rm coorbit}$ values consistent with an isotropic distribution of angular-momentum poles and far below those observed for the MW. At the opposite extreme, the Largest simulation exhibits a level of co-orbitation comparable to that of the MW over an extended period of approximately 6~Gyr. The Fiducial and Larger simulations represent intermediate regimes, showing progressively increasing degrees of co-orbitation with the plane. A more detailed analysis of the kinematic coherence and co-orbitation of the satellite systems is presented in Section~\ref{sec:kpp}.

Overall, the positional analysis reveals a systematic dependence on the mass of the GSE-like merger: simulations with more massive progenitors develop more anisotropic satellite distributions, thinner planes, and higher fractions of co-orbiting satellites, thereby more closely resembling the VPOS. Only the Larger and Largest runs simultaneously reach MW-like values across the different planarity metrics, for approximately 5\% and 10\% of their post-merger evolution, respectively. These results suggest that sufficiently massive GSE-like mergers favor the emergence of VPOS-like satellite configurations.

Finally, despite these systematic trends, the structural properties of high-quality positional planes can vary substantially on relatively short timescales across the simulation suite, as seen in Figure~\ref{fig:4GND_metrics_evolution}. Such variations are consistent with previous cosmological simulations interpreting extreme positional planes as largely transient features \citep{Gillet_2015, Buck_CoherenceM31_2016, Ahmed_CoherenceM31_2017, Santos_santos_sim_2020, Sawala_TransietPlanes_2025}, as well as with orbit-integration studies showing that the highly flattened configuration defined by the classical MW satellites is transient \citep{Maji_OrbitIntegration_2017,Sawala_Nature_2023}. To further investigate this issue, in Appendix~\ref{app:persvstrans} we analyze how transient are satellite members in these positional planes. For that, we select the thinnest positional plane in each simulation and distinguish between the satellite members whose orbital planes remain persistently aligned with it and transient members that only temporarily contribute to the configuration. The persistent fraction increases overall from $13.3\%$ in the Smallest run to $51.3\%$ in the Largest run, with the Smaller, Fiducial, and Larger simulations showing intermediate values. This indicates that exceptionally thin positional planes -- such as that found in the Largest simulation at $T_{\rm uni}\sim9.5$~Gyr -- are generally built from a combination of a stable persistent population and transient satellites that temporarily enhance their instantaneous planarity. Consequently, even when an underlying planar structure remains stable over several Gyr, its instantaneous thinness can vary substantially as transient satellites become misaligned or as newly accreted satellites enter along different directions (see Appendix~\ref{app:persvstrans} and Figure~\ref{fig:aligment_perstrans}).

\section{Kinematically persistent planes (KPPs)}
\label{sec:kpp}

As shown at the end of the previous section (and in Appendix~\ref{app:persvstrans}), the transient high-quality positional planes appear to be constructed around a more stable kinematic ``skeleton'' that provides long-term support to the planar configuration. This is consistent with previous studies showing that a significant fraction of satellites in both the VPOS (e.g., \citealt{Taibi_portrait_2024,Martinez_Garcia_delpino_2025}) and cosmological simulations exhibit coherent co-orbiting motions \citep{Shao_2019,santos_santos_2023,Gamez_Marin_TNG_2025}. In this section, we investigate the kinematic coherence of the satellite populations in greater detail. In particular, we search for KPPs, defined as groups of satellites whose orbital angular-momentum poles remain clustered around a common axis over extended timescales.

\subsection{Identifying directions of maximum satellite co-orbitation}
\label{sec:kpp_sub}
\begin{figure*}
    \centering
    \includegraphics[width=0.999\linewidth]{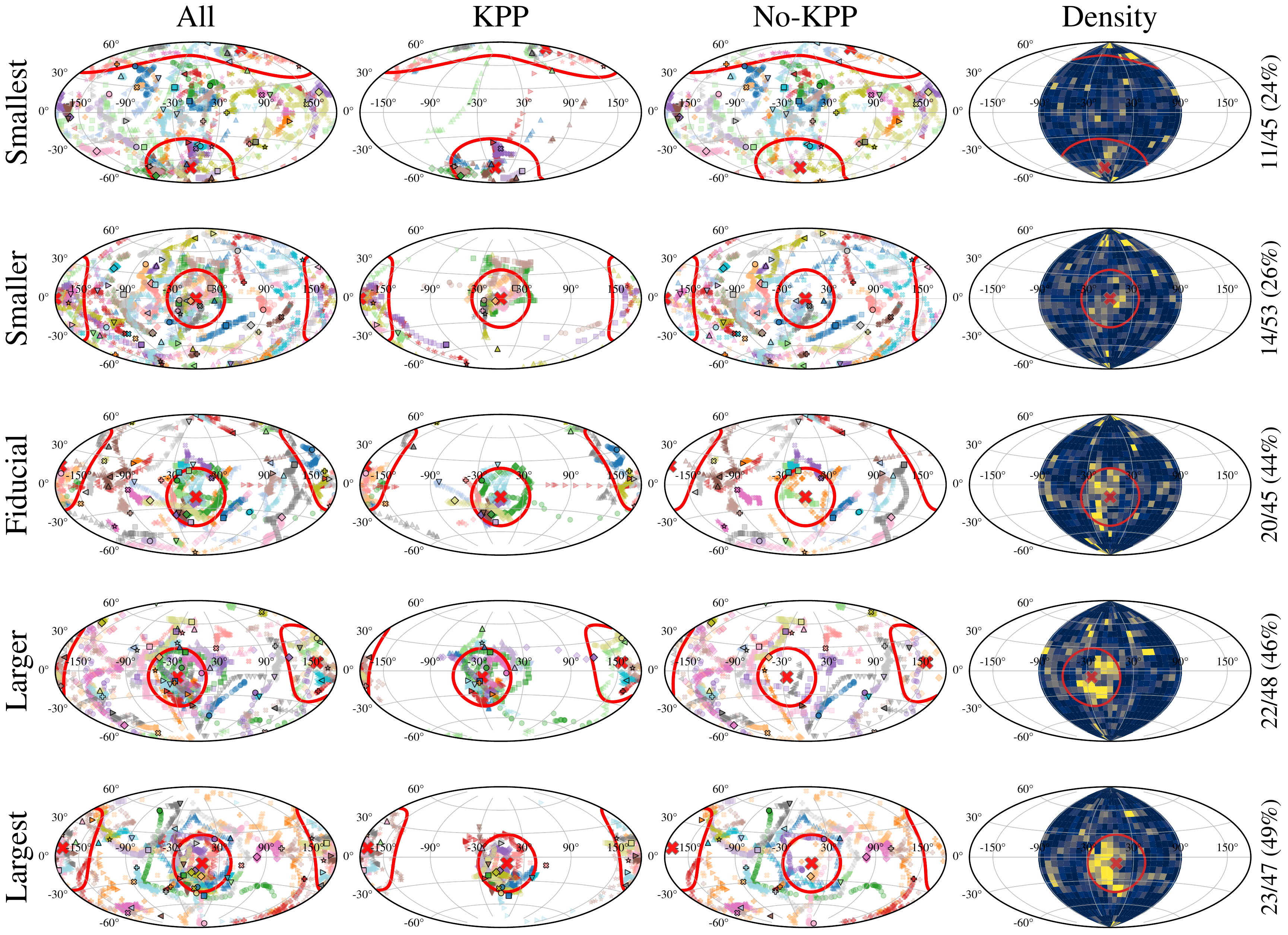}
    \caption{Aitoff projections showing the evolution of the orbital poles of satellites along their orbits. Satellite identities are indicated by marker and color, and different points correspond to different output times between $z_{\mathrm{infall}}$ and $z=0$, with the final snapshot highlighted by markers with black edges. Each row represents a different simulation. Red crosses mark the (axial) direction of maximum satellite co-orbitation, $\hat{n}_{\mathrm{KPP}}$, and the red lines indicate the angular aperture $\alpha_{\rm coorbit}=36.87^\circ$ around these directions. \textbf{First column:} all surviving satellites at $z=0$. \textbf{Second column:} only the satellites that belong to the KPP. \textbf{Third column:} satellites that are not members of the KPP. \textbf{Fourth column:}~binned density map of the distribution of satellite orbital poles. Each satellite contributes at every output time step, and the colorbar intensity is proportional to the bin density normalized by the total number of counts, allowing comparison across simulations with different satellite populations. In this column, orbital poles are treated as axial vectors to enhance co-orbiting structures irrespective of the sense of rotation; therefore, the density map is restricted to one hemisphere (lon $\in [-90^\circ, 90^\circ]$). On the far right, the corresponding statistics indicate the population fraction of KPP members in each simulation. More massive GSE-like progenitors produce stronger clustering of satellite angular momentum poles and larger KPP populations across the simulation suite.
    }
   \label{fig:aitoff_angular}
\end{figure*}

To identify the axis of maximum satellite co-orbitation, we analyze overdensities in the temporal evolution of projected satellite angular momentum poles, following a procedure based on the ``Scanning of Stacked Orbital Poles'' method introduced by \citet{santos_santos_2023}. For each satellite surviving at $z=0$, we computed the time evolution of its angular momentum vector $\vec{J}_{\rm sat}$ and project its direction onto an Aitoff diagram over the interval from $z_{\rm infall}$ to $z=0$. Since simulation snapshots are not uniformly spaced in time, we restrict the analysis to outputs separated by at least $100\,\mathrm{Myr}$ to avoid overweighting specific epochs. The resulting Aitoff projections are shown in the left column of Figure~\ref{fig:aitoff_angular}, where different satellites are color-coded and each point corresponds to a different simulation snapshot.

The direction of maximum co-orbitation is identified by scanning the Aitoff projection with a fixed angular aperture of $\Delta_{\rm scan} = 36.87^\circ$ and searching for the region with the highest overdensity of satellite angular momentum poles. To further enhance the sensitivity to co-orbiting structures, we treat $\vec{J}_{\rm sat}$ as an axial vector during the stacking procedure, ignoring the sense of rotation. This choice increases the local density of directions and facilitates the identification of preferred co-orbitation axes. The resulting density map of satellite orbital poles, treated as axial vectors, is shown in the fourth column of Figure~\ref{fig:aitoff_angular}. The identified axis of maximum co-orbitation, $\hat{n}_{\rm KPP}$\footnote{Throughout this paper we use $\hat{n}_{\rm KPP}$ to refer to the axis of maximum co-orbitation identified in this section, which corresponds to $\vec{J}_{\rm stack}$ in the notation of \cite{santos_santos_2023, GamezMarin_2024, Gamez_Marin_TNG_2025}.}, for each simulation is indicated by red crosses. Although the sense of rotation is ignored when identifying the direction of maximum co-orbitation, we retain both hemispheres in first three columns of Figure~\ref{fig:aitoff_angular} in order to visually distinguish corotating and counter-rotating satellites.

Once $\hat{n}_{\rm KPP}$ has been determined, satellites are ranked according to the number of timesteps, $N_{\rm co\text{-}orb}$, during which their orbital poles lie within the same angular aperture, $\alpha_{\rm co\text{-}orbit}=36.87^\circ$, around this axis. Satellites are classified as KPP members if their orbital poles remain within this aperture for at least $50\%$ of the analyzed timesteps. In contrast to \citet{santos_santos_2023}, we do not exclude satellites whose angular momentum poles vary by more than $40^\circ$. This choice allows us to retain satellites that experience significant torques, provided they ultimately spend more than half of the analyzed time within the co-orbiting aperture around $\hat{n}_{\rm KPP}$. The second and third columns of Figure~\ref{fig:aitoff_angular} show the KPP members and nonmembers, respectively.

The stacked orbital-pole distributions reveal a systematic increase in kinematic coherence across the suite. The fraction of satellites classified as KPP members grows from $24\%$ in the Smallest simulation -- consistent with an approximately isotropic distribution of angular momentum poles $\rm{\sim20}\%$ -- to $49\%$ in the Largest simulation. The Smaller, Fiducial, and Larger runs exhibit intermediate and progressively increasing KPP fractions of $26\%$, $44\%$, and $46\%$, respectively. Thus, the Fiducial, Larger, and Largest simulations develop well-populated KPPs containing $\gtrsim40\%$ of their surviving satellites, comparable to the $\sim40$--$50\%$ co-orbiting fraction inferred for satellites in the MW \citep{Fritz_2018_coorb,Taibi_portrait_2024,Martinez_Garcia_delpino_2025}.

\begin{figure}
    \centering
    \begin{subfigure}{0.499\textwidth}
    \includegraphics[width=\textwidth]{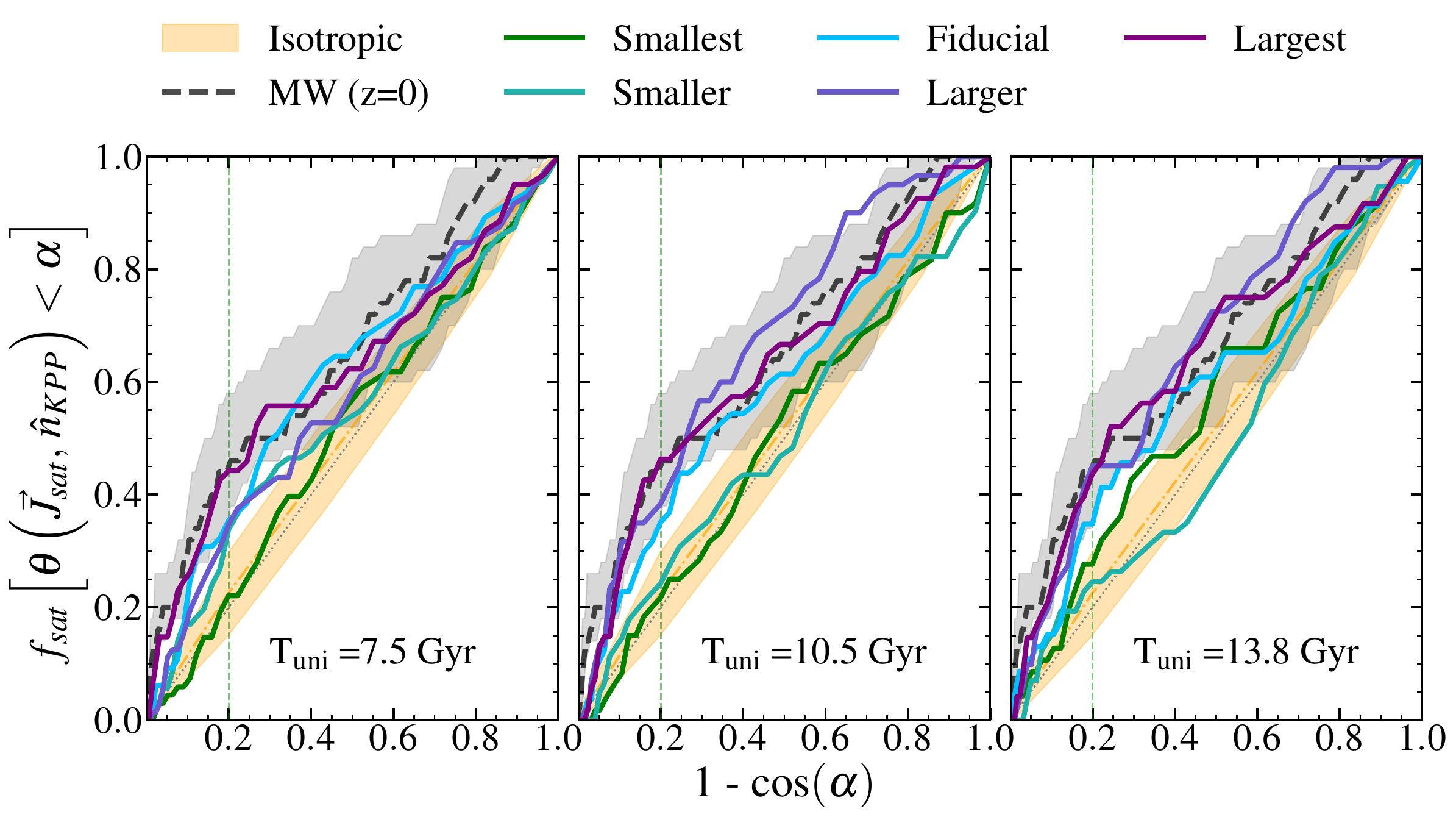}
    \end{subfigure}
    \begin{subfigure}{0.49\textwidth}
    \includegraphics[width=\textwidth]{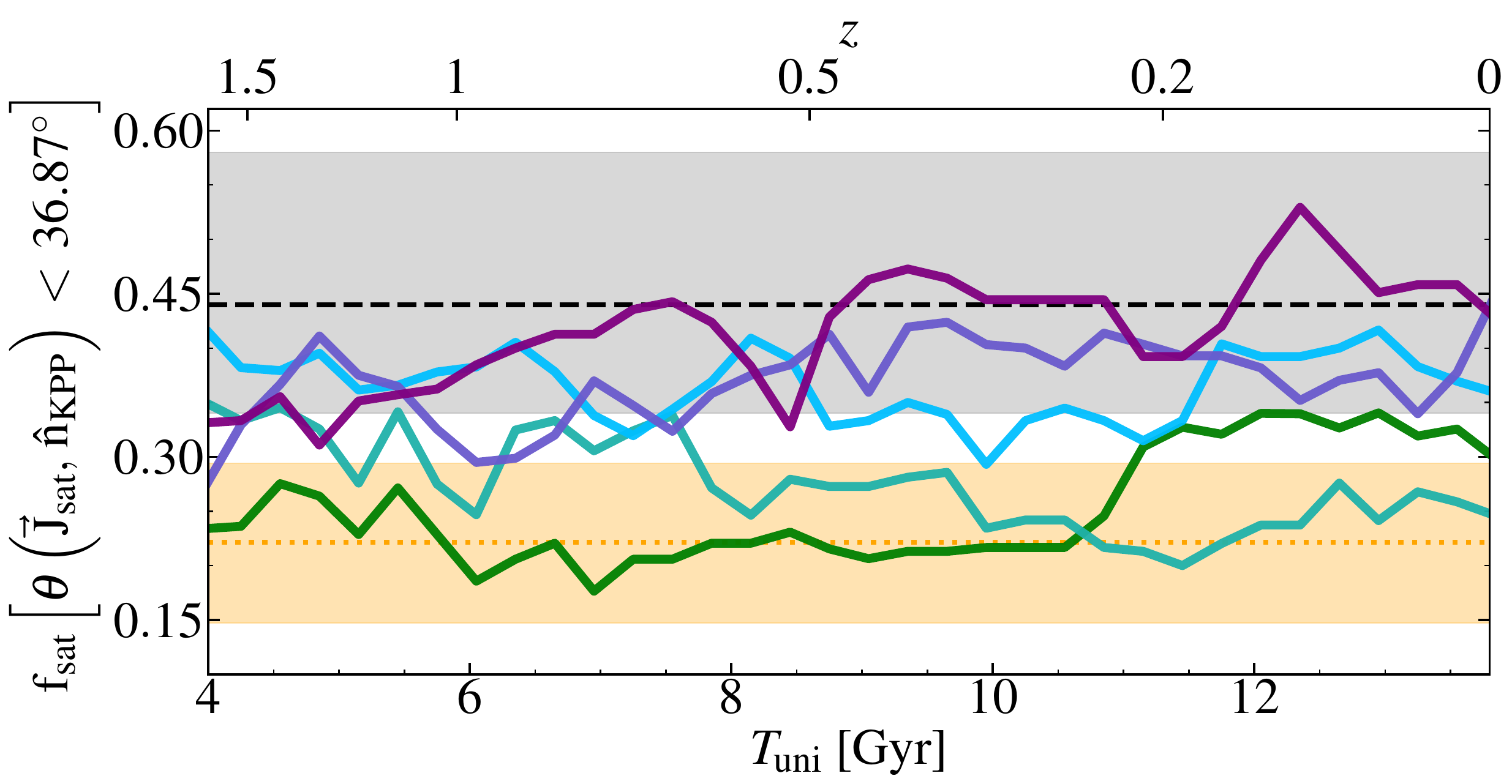}
    \end{subfigure}
    \caption{Evolution of the clustering of satellite angular-momentum poles around the axis of maximum co-orbitation, $\hat{n}_{\rm KPP}$. \textbf{Top panels:} Fraction of satellites whose orbital poles, $\rm{\vec{J}_{sat}}$, lie within an angular distance $\alpha$ from $\rm{\hat{n}_{KPP}}$, at given time steps for each simulation. The vertical green dashed line marks the co-orbitation criterion adopted in this work, $\alpha_{\rm co\mbox{-}orbit} = 36.87^\circ$. \textbf{Bottom panel}: Time evolution of the fraction of satellites co-orbiting around $\hat{n}_{\mathrm{KPP}}$ axes. In both panels, the orange solid lines and shaded regions show the results obtained for isotropized configurations of orbital poles. The black dashed line with gray shaded region indicates the corresponding measurements for the MW satellites, based on the data compilation presented in \cite{Taibi_portrait_2024}. Across the simulation suite, satellite orbital poles transition from nearly isotropic in the Smallest run to strongly clustered around $\hat{n}_{\mathrm{KPP}}$ in the Largest run.}
   \label{fig:f_sat_nkpp}
\end{figure}

Figure~\ref{fig:f_sat_nkpp} provides a complementary snapshot-by-snapshot view of this clustering. The top panels show, at three representative times, the cumulative distribution of angular distances between the satellite orbital poles and $\hat{n}_{\rm KPP}$. For comparison, the black dashed line and gray band show the observed MW distribution relative to the VPOS normal using \citet{Taibi_portrait_2024}, while the orange curves and shaded regions correspond to $N_{\rm random}=5000$ isotropized realizations\footnote{\label{fn:isotropy_curve}We refer the reader to Appendix~C of \citealt{santos_santos_2023} for a detailed description of how the isotropy curve is determined. This curve depends on both the number of satellites considered and the number of timesteps included in the analysis. To adopt a single reference curve for comparing our simulations, we use the median values of $N_{\rm sat}$ and $N_{\rm out}$ across the five runs.}. The CDFs confirm that the differences inferred from the stacked orbital poles are already present at individual epochs, with the most prominent KPPs showing a clear excess of satellites at small angular separations relative to isotropic expectations. The bottom panel of Figure~\ref{fig:f_sat_nkpp} shows how this clustering evolves with time, through the fraction of satellites lying within $\alpha_{\rm co\text{-}orbit}=36.87^\circ$ of $\hat{n}_{\rm KPP}$. The Smallest and Smaller runs fluctuate around $20$--$30\%$, close to isotropic expectations, whereas the Fiducial and Larger runs typically reach $30$--$40\%$. The Largest simulation shows the strongest and most persistent coherence, maintaining MW-like co-orbiting fractions of $\sim45\%$ for almost 8~Gyr.

\subsection{Comparison between kinematic and positional planes}

The positional planes identified in Section~\ref{sec:positional_planes} are determined independently at each snapshot and therefore have time-dependent membership and orientation, $\hat{n}_{\rm pos}$. In contrast, each KPP defines a persistent axis of maximum satellite co-orbitation, $\hat{n}_{\rm KPP}$. We now examine how these two directions are related.

\begin{figure}
    \centering
    \includegraphics[width=0.99\linewidth]{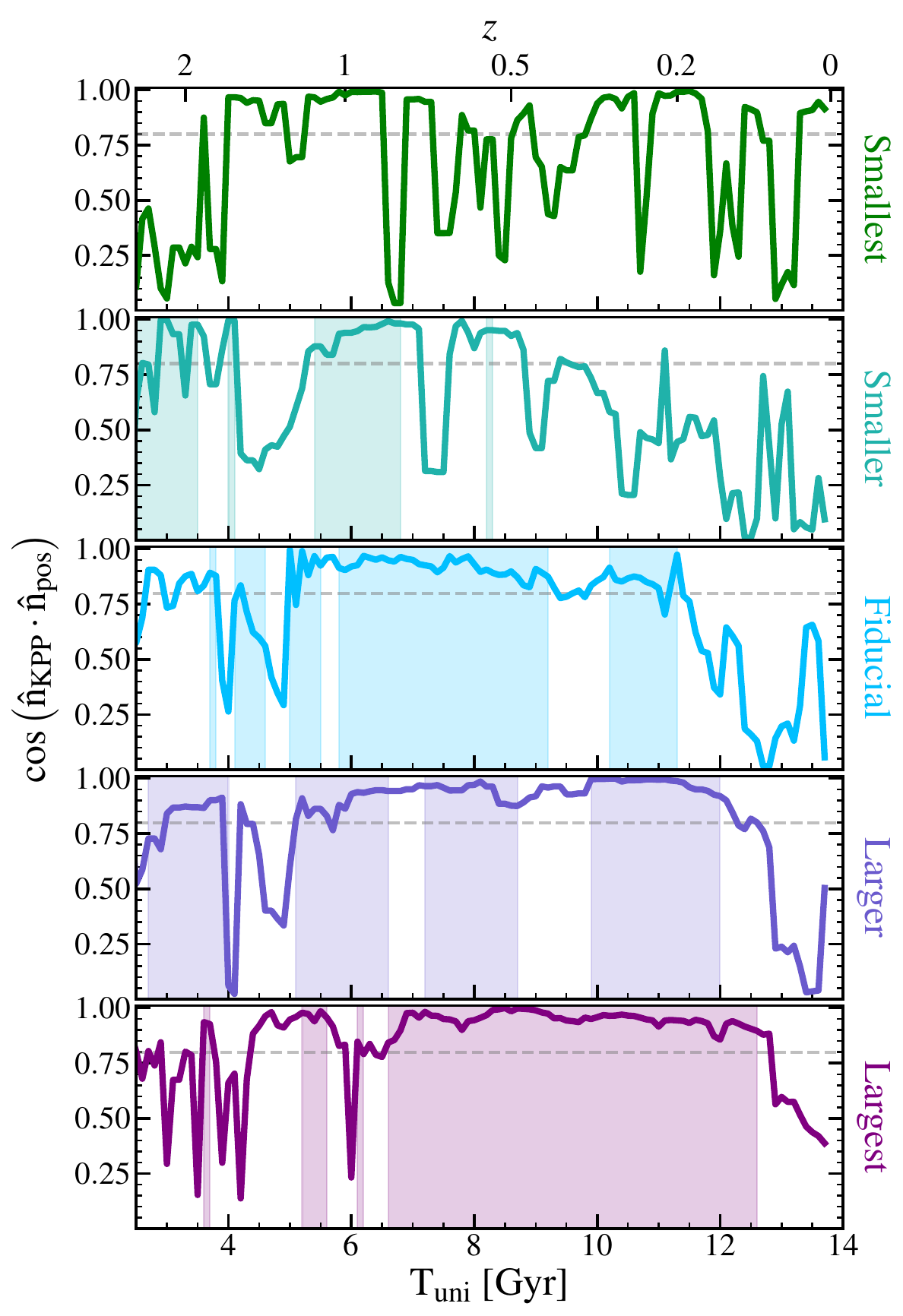}
    \caption{Time evolution of the alignment between the direction of maximum satellite co-orbitation, $\hat{n}_{\rm KPP}$, and the normals to the best-fitting positional planes, $\hat{n}_{\rm pos}$, identified in Section~\ref{sec:positional_planes}. The horizontal gray dashed line marks as a reference $\rm{cos(\alpha_{\rm co-orbit})} = 0.8$. The shaded regions indicate epochs during which the fraction of co-orbiting satellites, $f_{\rm sat}^{\rm coorbit}$, around the positional-plane normal $\hat{n}_{\rm pos}$ is consistent with the values observed in MW (taking into account uncertainties), as identified in the bottom panel of Figure~\ref{fig:4GND_metrics_evolution}. During these epochs, a clear alignment between $\hat{n}_{\rm KPP}$ and $\hat{n}_{\rm pos}$ is observed, highlighting the role of KPPs as long-lived structures that support transient high-quality positional planes.}
    \label{fig:kppvspos}
\end{figure}

In Figure~\ref{fig:kppvspos}, we examine the temporal evolution of the angular separation between the positional-plane normal, $\hat{n}_{\rm pos}$, and the axis of maximum satellite co-orbitation, $\hat{n}_{\rm KPP}$. The colored shaded regions highlight epochs during which the fraction of co-orbiting satellites, $f_{\rm sat}^{\rm coorbit}$, measured with respect to $\hat{n}_{\rm pos}$ is consistent, within uncertainties, with the value observed for the MW, as identified in the bottom panel of Figure~\ref{fig:4GND_metrics_evolution}. These epochs also coincide with the time intervals during which the positional-plane metrics $C_1$, $c/a$, and $\Delta_{\rm RMS}$ shown in Figure~\ref{fig:4GND_metrics_evolution} most closely reproduce the observed MW values. They therefore mark the periods during which the satellite planes in the simulations are both most planar and most kinematically coherent. 

During these epochs, characterized by both well-defined positional planes and a substantial fraction of co-orbiting satellites, a strong alignment between $\hat{n}_{\rm KPP}$ and $\hat{n}_{\rm pos}$ is observed, with angular separations remaining below $36.87^\circ$ (indicated by the gray dashed line) for extended periods, particularly in the Larger and Largest runs.

These results indicate that well-populated KPPs define a preferred orbital plane around which high-quality positional configurations preferentially develop. The instantaneous properties of these positional planes can still vary as satellites outside the persistent kinematic population temporarily contribute to, or become misaligned with, the configuration, as discussed in Appendix~\ref{app:persvstrans}. Thus, the KPP provides a long-lived kinematic backbone, while the quality of the positional plane remains sensitive to the instantaneous distribution of the remaining satellites.

\subsection{Long-lived kinematically coherent planes}
\label{sec:long_lived}

In this section, we now use the previously identified KPPs to investigate the existence and properties of  long-lived ($\gtrsim 4$~Gyr) planar satellite configurations. In particular, we aim to quantify how the different KPP populations affect the ability of each simulation to sustain long-lived planes and to characterize their structural and kinematic properties. We therefore compare the simulations at fixed values of $N_{\rm sat}$, so that different KPP populations should impact the plane quality of long-lived planes.  We identify the most kinematically coherent long-lived plane for a given $N_{\rm sat}$ following the procedure below:
\begin{enumerate}
    \item We selected satellite galaxies with infall times $t_{\rm infall} \leq 10$~Gyr, ensuring that the resulting planes last for at least 4~Gyr, roughly twice the approximate dynamical time of satellite galaxies in MW-like halos.
    \item  For each satellite selected in step (1), we compute the median angular separation, evaluated over all time steps since infall, between the satellite angular-momentum pole, $\vec{J}_{\rm sat}$, and the maximum satellite co-orbitation axis, $\hat{n}_{\rm KPP}$. The satellite galaxies are then ranked according to this median angular separation.
    \item We select the $N_{\rm sat}$ satellites with the smallest median angular separation and track their evolution backward in time, computing the corresponding plane properties at each time step.
\end{enumerate}
Note that if $N_{\rm sat}$ is larger than the number of satellite members of the KPP, non-KPP members with the closest median angular separation will be added to complete the group. This is intentional, as it allows us to quantify how the quality and stability of the plane change once the selected population extends beyond the coherent KPP backbone.

We compare the evolution of the properties of long-lived, kinematically coherent satellite structures identified in our simulations with those of the most kinematically coherent satellite galaxies associated with the MW VPOS. To identify the $N_{\rm sat}$ satellites most kinematically aligned with the VPOS, we adopt the results of \citet{Taibi_portrait_2024}, who determined the orbital angular-momentum poles of 50 confirmed MW satellite galaxies using 5000 Monte Carlo realizations that account for observational uncertainties. For each satellite, they computed the fraction $f_{\rm VPOS}$ of Monte Carlo realizations whose orbital poles lie within an angular aperture of $\alpha_{\rm co\text{-}orbit}=36.87^\circ$ from the VPOS normal, thereby quantifying the degree of alignment between the satellite’s orbital angular momentum and the VPOS normal (see their Table~2).
In this work, we use $f_{\rm VPOS}$ as a ranking metric and select the $N_{\rm sat}$ satellites with the highest values to define the most kinematically coherent satellite sample in the MW VPOS. We refer to this selection as the MW on-plane satellites.

\begin{figure*}
    \centering
    \includegraphics[width=0.99\linewidth]{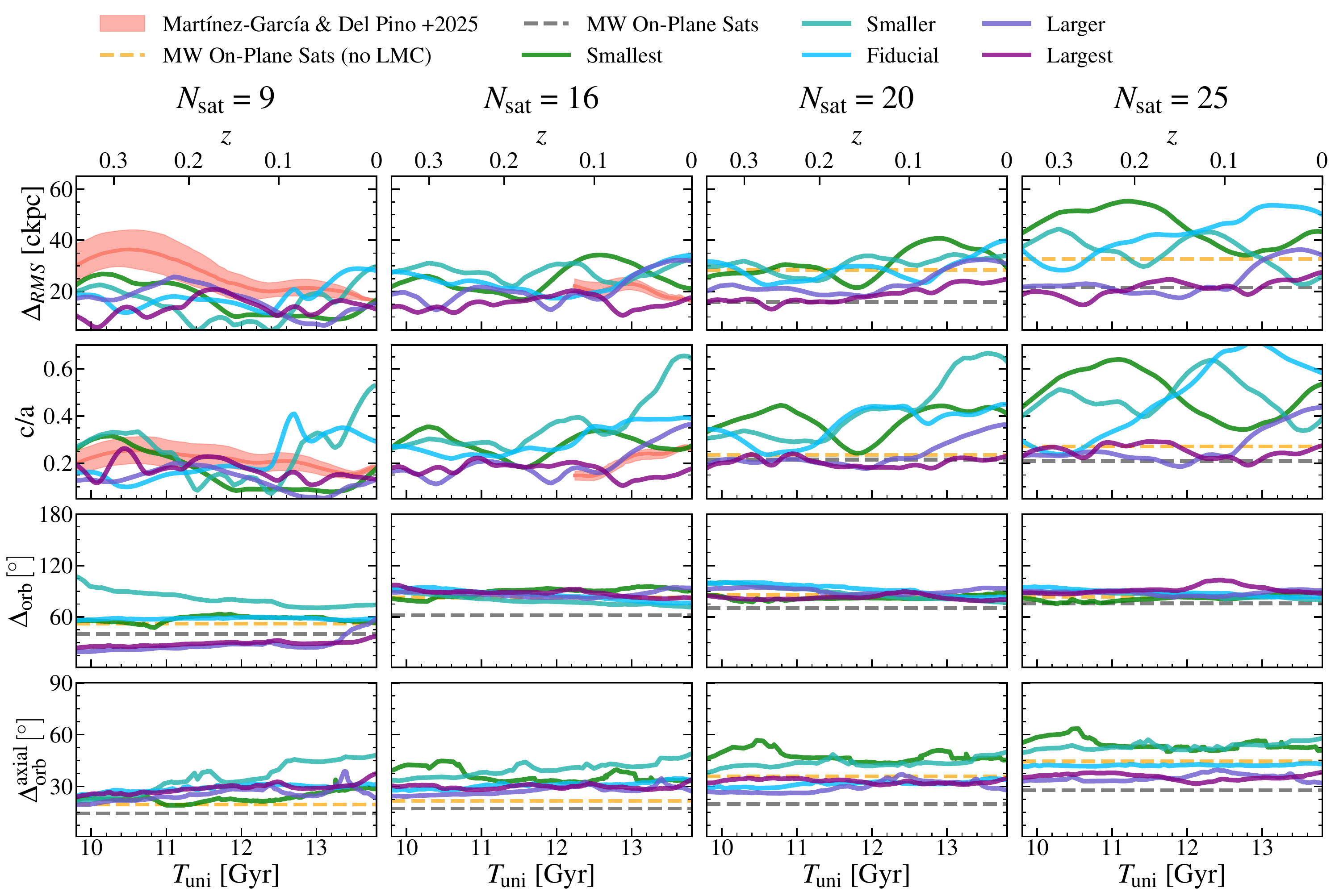}
    \caption{Time evolution of plane properties for long-lived ($\geq 4$ Gyr) planes in each simulation, as a function of the number of satellites involved. Each column corresponds to a different number of satellites in the plane, and each row shows a different plane property. \textbf{First row}: evolution of the RMS height ($\Delta_{RMS}$). \textbf{Second row}: evolution of the minor-to-major axis ratio ($c/a$) of the satellite distribution. \textbf{Third row}: evolution of the orbital pole dispersion ($\Delta_{\mathrm{orb}}$). \textbf{Fourth row}: evolution of the axial orbital pole dispersion ($\Delta_{\rm orb}^{\rm axial}$). For each metric we show the corresponding quantities at $z=0$ computed for the same number ($N_{\rm sat}$) of MW satellites most closely aligned with the VPOS as dashed lines when including (gray) or excluding (orange) the LMC associated satellites. The red solid lines and shaded regions in the first two rows of the first two columns show the predicted evolution of the plane properties for the $N_{\rm sat}=9$ and 16 on-plane VPOS satellite samples from \citealt{Martinez_Garcia_delpino_2025}. Details of the satellite selection procedure in both the MW and the simulations are provided in Section~\ref{sec:long_lived}. A visual representation of the evolution of the planes for $N_{\rm sat}=20$ is shown in Figure~\ref{fig:evolution_plane_multi_plot}.}
   \label{fig:evolution_plane_multi}
\end{figure*}

Figure~\ref{fig:evolution_plane_multi} presents the temporal evolution of the structural and kinematic properties of the long-lived satellite planes identified in each simulation for four representative numbers of satellites $N_{\rm sat} = 9, 16, \,20$ and 25, which roughly correspond to the number of KPP members present in the different simulations. The figure illustrates the time evolution of four relevant metrics:  the plane thickness, quantified by the root-mean-square height $\Delta_{\rm RMS}$; the minor-to-major axis ratio $c/a$ of the satellite spatial distribution; the orbital pole dispersion $\Delta_{\rm orb}$; and the axial orbital pole dispersion $\Delta_{\rm orb}^{\rm axial}$. The orbital pole dispersion $\Delta_{\rm orb}$ measures the root-mean-square angular separation between the satellites’ orbital poles and their mean direction, evaluated over the full $[0^\circ,360^\circ]$ range. The axial orbital pole dispersion $\Delta_{\rm orb}^{\rm axial}$ is defined in an analogous manner but treats orbital poles as axial vectors, such that opposite directions ($\vec{L}$ and $-\vec{L}$) are considered equivalent and angular separations are restricted to the interval $[0^\circ,180^\circ]$. This quantity therefore probes the degree of coherent co-orbitation around the plane normal independently of the sense of rotation. For reference, the same metrics are shown for the MW on-plane satellites most aligned with the VPOS, using the same $N_{\rm sat}$. Because several recent studies have argued that the infall of the LMC group may significantly influence the orbital alignment of MW satellites \citep{Samuel_2021,santos_santos_2023,Taibi_portrait_2024, Vasiliev_2024, Martinez_Garcia_delpino_2025}, we consider two MW selections: one including satellites associated with the LMC\footnote{We follow the selection of satellites adopted by \cite{Taibi_portrait_2024}: Carina II, Carina III, Horologium I, Hydrus I, Phoenix II, Reticulum II, SMC.} (gray dashed lines) and another excluding them (orange dashed lines).  A visual representation of the evolution of the spatial structure and kinematics of these planes for the case $N_{\rm sat}=20$ is shown in Figure~\ref{fig:evolution_plane_multi_plot}.

The spatial properties of the long-lived planes, characterized by $\Delta_{\rm RMS}$ and $c/a$, reveal a clear ordering across the simulation suite. The Larger and Largest simulations consistently produce the thinnest and most flattened configurations, with the Larger and especially the Largest simulations reproducing the extremely low values of $\Delta_{\rm RMS}$ and $c/a$ measured for the MW on-plane satellites for time intervals of about $4$ Gyr. On the other hand, the Smallest and Smaller runs generate substantially thicker structures, and the Fiducial simulation generally lies between these two regimes. The contrast becomes increasingly pronounced as $N_{\rm sat}$ grows, reflecting the limited ability of sparsely populated KPPs to support richer coherent structures.

The behavior of the orbital pole dispersion $\Delta_{\rm orb}$ is markedly different. Across the entire simulation suite, $\Delta_{\rm orb}$ typically remains close to $\sim90^\circ$, indicating a mixture of corotating and counter-rotating satellites. These values closely match those obtained for the MW on-plane satellites when the LMC-associated satellites are excluded. However, including the LMC group substantially lowers the dispersion in the MW, particularly for $N_{\rm sat}=16$, indicating stronger corotation than is reproduced in our simulations. This supports previous suggestions that the recent infall of the LMC group enhances the apparent corotation of MW satellites \citep{santos_santos_2023,Martinez_Garcia_delpino_2025}.

When the sense of rotation is ignored, the axial orbital pole dispersion $\Delta_{\rm orb}^{\rm axial}$ provides a measure of the degree of orbital alignment around the plane normal and therefore of the dynamical coherence of the satellite plane. A systematic trend emerges across the suite: the Larger and Largest simulations show the smallest axial dispersions, the Smallest and Smaller the largest, and the Fiducial run again occupies an intermediate position. The Larger and Largest values are comparable to those of the MW on-plane sample when LMC-associated satellites are excluded, although the full MW sample remains more strongly aligned. This again supports the interpretation that the exceptional orbital alignment observed in the MW is significantly enhanced by the recent infall of the LMC group.

Overall, the differences in KPP richness translate directly into the ability of the satellite systems to sustain long-lived planar configurations. Only the simulations with well-populated KPPs maintain MW-like structural properties when progressively larger satellite populations are included.

Taken together, the positional and kinematic analyses reveal a systematic transition across the \textsc{Vintergatan-GM} suite. The Smallest and Smaller runs produce comparatively thick satellite distributions and show little evidence for a long-lived kinematically coherent plane, whereas the Larger and Largest runs simultaneously develop thin positional configurations and prominent KPPs whose properties approach those observed for the MW satellite system. The Fiducial run represents an intermediate case. This suggests that a fairly massive GSE-like merger in our simulation suite favors the emergence of a VPOS-like satellite configuration, with the transition occurring for merger mass ratios between those sampled by the Fiducial ($1{:}6$) and Larger ($1{:}2.9$) realizations. This points to a GSE progenitor mass of order $7\times10^{10}-10^{11}\,M_\odot$, around that of the Larger and Largest runs in our suite. Such a massive GSE progenitor is consistent with several estimates in the literature (e.g., \citealt{Bignone_2019,Kruijssen_2020,Naidu_2021}), although it lies toward the upper end of, or above, other estimates (e.g., \citealt{Mackereth_Bovy_2020,Lane_2023}).

\section{The origin of satellite planes driven by a massive GSE-like merger event}
\label{sec:origin}
A consistent trend emerges across the simulation suite: the more massive the GSE-like major merger at $z \sim 2$, the higher the kinematical coherence and planarity of the satellite population.  Below, we examine the physical mechanisms associated with increasingly massive GSE-like mergers to investigate whether they drive the emergence and persistence of kinematically coherent satellite planes in our simulations.

\subsection{Large scale structure}
\label{sec:lss}

\begin{figure}
    \centering
    \includegraphics[width=0.95\linewidth]{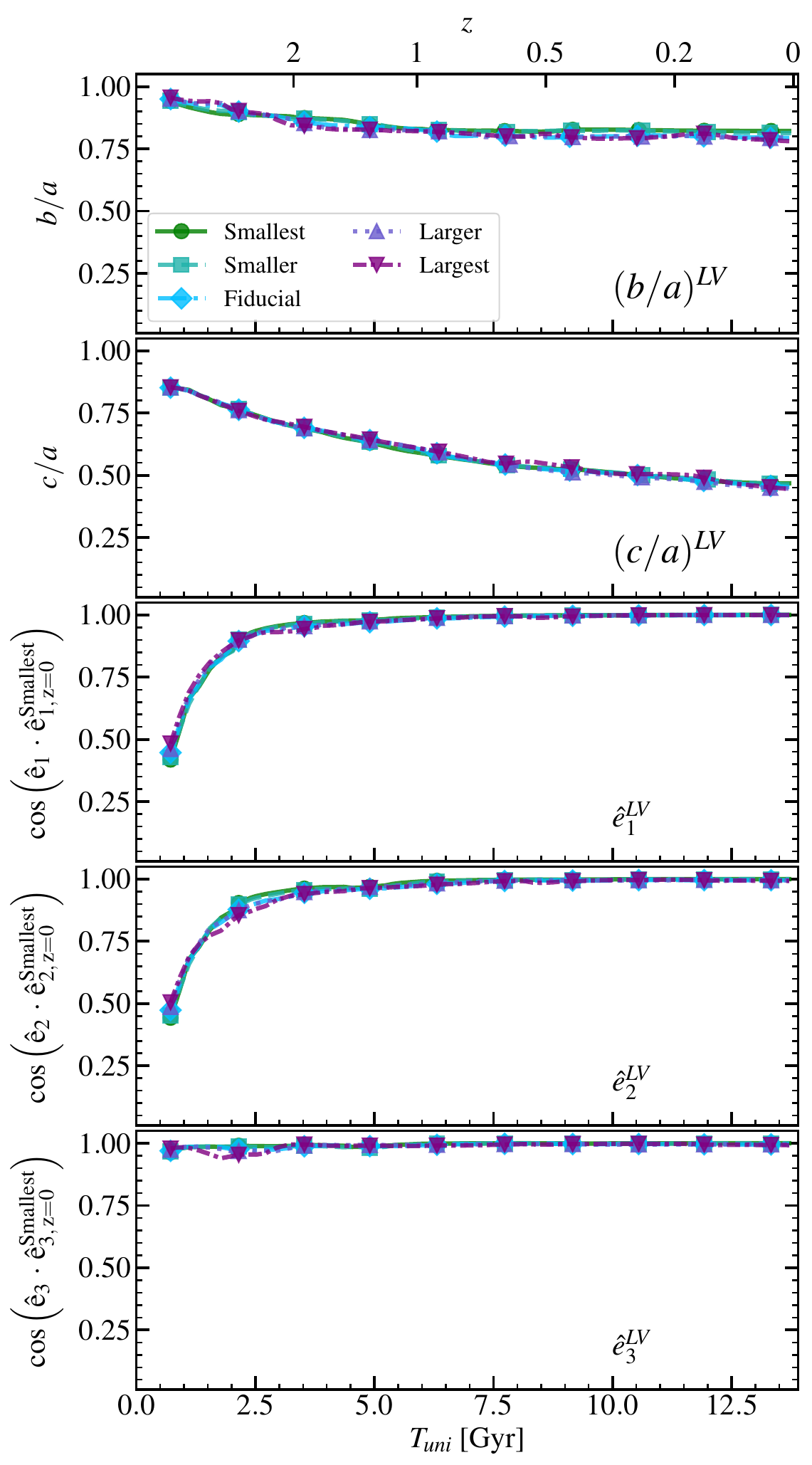}
    \caption{Comparison of the time evolution of the eigenvalues and eigenvectors of the reduced inertia tensor of the LV across the simulation suite. The top two panels show the evolution of the ratios between the eigenvalues $(b/a)^{LV}$ and $(c/a)^{LV}$, respectively. The three bottom panels show the temporal evolution of the cosine of the angle between the LV eigenvectors $\hat{e}_i^{\rm LV}$ of each simulation and the reference eigenvector of the Smallest simulation at $z=0$, $\hat{e}_{i,\,z=0}^{\rm \,Smallest}$. The nearly identical evolution of the LV eigenvalues and eigenvectors across the simulations confirms that the large-scale structure is fixed across the suite, ruling out large-scale anisotropic accretion as the driver of differences in satellite-plane formation.
    }
              \label{fig:evolution_eigenvectors}%
\end{figure}

Several studies have linked the formation of planes of satellites to the anisotropic accretion of subhalos driven by the large-scale structure (e.g., \citealt{Lovell_OrbitSubAccretion_2011, Libeskind_Subhaloaccretion_2014, Dupuy_Accretion_2022}). In this context, \citet{GamezMarin_2024} investigated the relationship between the flattening of the large-scale structure and the formation of KPPs, showing that clustering of satellite orbital poles can naturally arise from the large-scale dynamics predicted by $\Lambda$CDM. In this scenario, specific groups of protosatellites already follow the preferred directions of the anisotropic collapse of the cosmic web (CW) prior to accretion, as they correspond to mass elements embedded within it, leading to an alignment of their angular-momentum vectors with its principal axes.

To characterize the local CW evolution in our simulations, we analyzed the deformation of the lagrangian volume (LV) surrounding the host halo, following an approach almost identical to that of \citet{Robles_LV_2015} and \citet{GamezMarin_2024}. Basically, the CW evolution is characterized by the evolution of the principal directions of the LV and the principal axes of the inertia ellipsoid determined using the reduced inertia tensor of LV particles. We refer the reader to Appendix~\ref{app:LV} for a detailed description of this method and to Figure~\ref{fig:lv_evolution} for a visual representation of LV deformations. We focus here on the evolution of the LV principal directions and axes and compare them across the simulation suite.

Figure~\ref{fig:evolution_eigenvectors} shows the evolution for each simulation of their LV principal directions $\hat{e}_i^{LV}$, as well as the evolution of the ratio between their principal axes $(b/a)^{LV}$ and $(c/a)^{LV}$. In this analysis, $\hat{e}_1^{LV}$, $\hat{e}_2^{LV}$, and $\hat{e}_3^{LV}$ correspond to the principal directions of the major axis $a^{LV}$, intermediate axis $b^{LV}$, and minor axis $c^{LV}$, respectively.

The top three panels display the evolution of the alignment between $\hat{e}_i^{LV}$ and those of the Smallest simulation at $z=0$, chosen as a reference $\hat{e}_{i,\,z=0}^{\rm \,Smallest}$. The alignment of the principal directions is virtually identical across all simulations, reflecting that the large-scale structure remains fixed as imposed by the genetic modifications. In all cases, $\hat{e}_1^{LV}$ and $\hat{e}_2^{LV}$ evolve until $T_{\rm uni} \sim 2$~Gyr and remain fixed thereafter, while $\hat{e}_3^{LV}$ appears stable throughout cosmic evolution. This indicates that the LV establishes its principal directions early, around $T_{\rm uni} \sim 2$~Gyr.

The bottom two panels show the evolution of the axis ratios $(b/a)^{LV}$ and $(c/a)^{LV}$ which quantify the overall LV shape deformation. Both ratios evolve in an identical manner across the simulation suite, confirming the convergence of CW evolution around the host galaxies. Initially, the LV is nearly spherical, with $(b/a)^{LV} \sim (c/a)^{LV} \sim 1$, and gradually becomes more flattened, as $(c/a)^{LV}$ continuously decreases to $\sim 0.4$, while $(b/a)^{LV}$ remains roughly constant around $0.8$ from $T_{\rm uni}\sim2$~Gyr onward. The emergence of this flattened structure, oriented perpendicular to the $\hat{e}_3^{LV}$ principal axis, is also visually evident in Figure~\ref{fig:lv_evolution}. The decline in $(c/a)^{LV}$ reflects the progressive flattening of the LV, driven by anisotropic mass accretion predominantly occurring within the plane perpendicular to $\hat{e}_3^{LV}$.

Overall, the simulation suite exhibits anisotropic mass accretion along the direction of maximum LV compression, a process often invoked as a driver of anisotropic distributions of satellite galaxy positions and angular momentum poles (see \citealt{GamezMarin_2024} and references therein). However, since all simulations share identical large-scale environments and LV evolution, this anisotropic accretion alone cannot explain the differences observed across the suite. Indeed, despite indistinguishable LV flattening and principal directions, only a subset of simulations develop well-populated KPPs and thin satellite planes, while others remain consistent with isotropic expectations. This suggests that while large-scale dynamics may play a role, they do not fully determine the formation of long-lived, kinematically coherent satellite planes, indicating that the trends seen in our simulation suite arise from a different mechanism. The alignment between the principal $\hat{e}_3^{LV}$ direction and the KPPs is examined further in Section~\ref{sec:host_halo_shape} and Figure~\ref{fig:aitoff_directions_all}.

\subsection{Group accretion}
\label{sec:group_accretion}

\begin{figure*}
    \centering
    \includegraphics[width=0.99\linewidth]{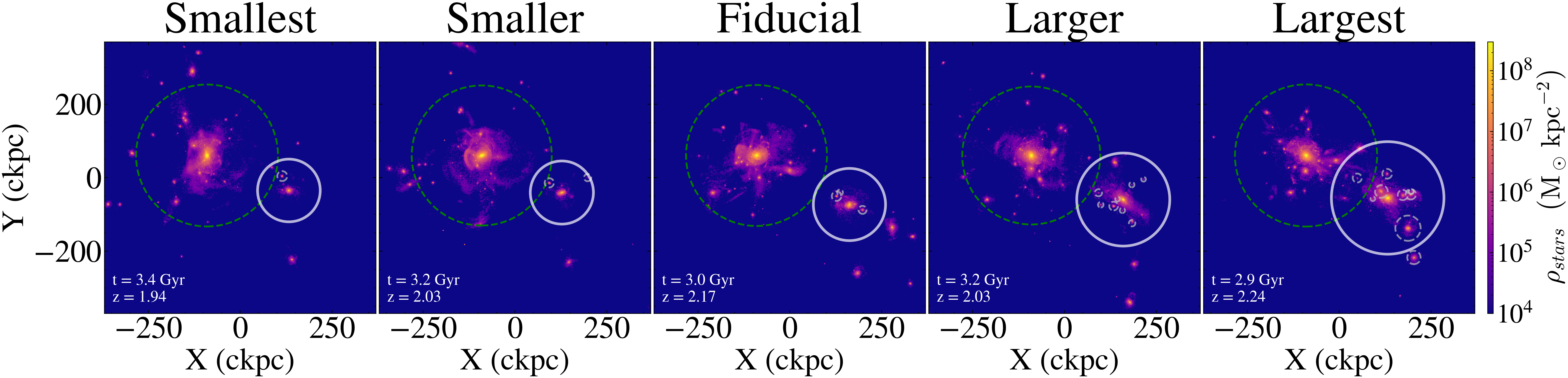}
   \caption{Stellar-density projections shortly before the infall of the GSE-like progenitor at $z\sim2$. Satellites associated with the progenitor are highlighted by gray dashed circles of radius $0.5\,R_{\rm vir}^{\rm sat}$. Green dashed and white solid circles indicate the virial radii of the central host and GSE-like progenitor, respectively. The number of satellites associated with the infalling group increases with progenitor mass.
}
              \label{fig:MM_sats_stellar_density}%
\end{figure*}

\begin{figure*}[h!]
    \centering
    \includegraphics[width=0.99\linewidth]{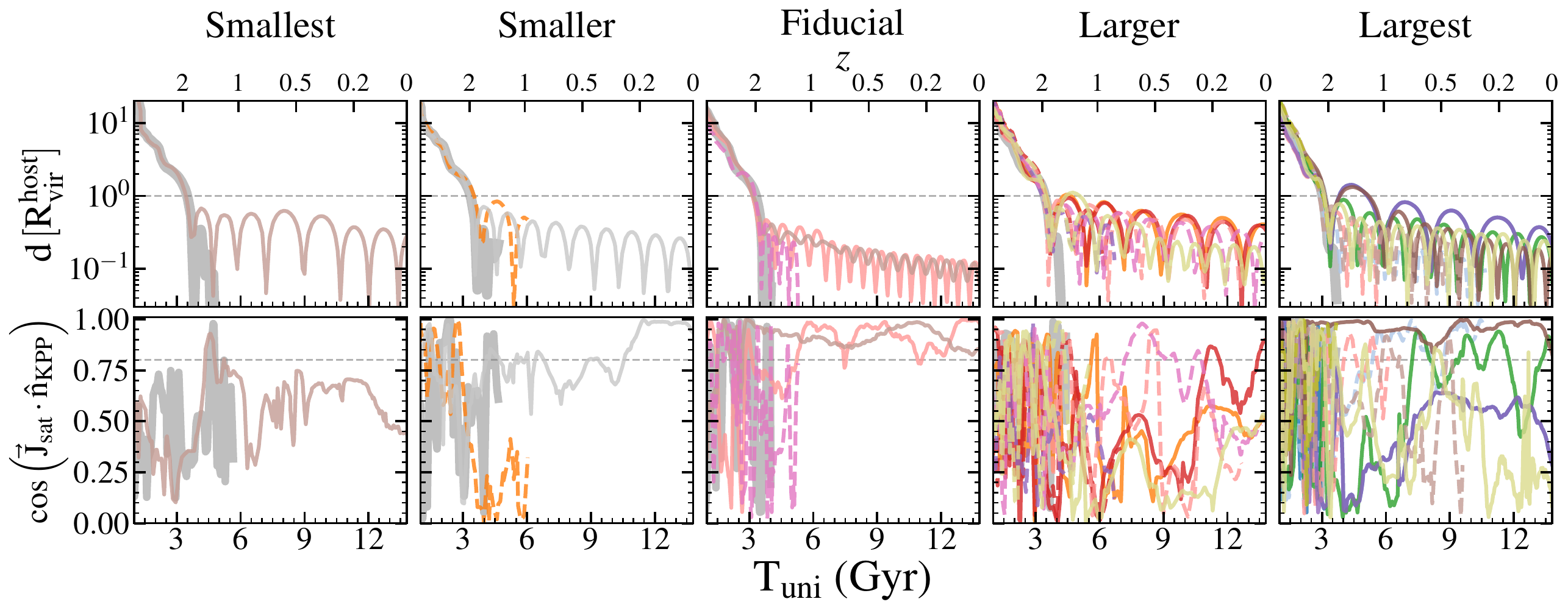}
    \caption{Evolution of GSE-like progenitor (sub)satellites identified in Fig.~\ref{fig:MM_sats_stellar_density}. Each column corresponds to a different simulation.\textbf{ Top row}: satellite trajectories expressed in units of the host virial radius; the host virial radius is indicated by a gray dashed line. \textbf{Bottom row}: evolution of the alignment between each satellite’s angular-momentum vector, $\vec{J}_{\rm sat}$, and the axis of maximum co-orbitation for each simulation, $\hat{n}_{\rm KPP}$, identified in Fig.~\ref{fig:aitoff_angular}. A gray dashed line marks the adopted angular threshold for co-orbitation, $\alpha_{\rm co-orbit} = 36.86^\circ$. Each satellite is shown using a single color within each simulation, and satellites disrupted by $z=0$ are highlighted with dashed lines. Satellites associated with the GSE-like progenitor show no preferential alignment with $\hat{n}_{\rm KPP}$, ruling out GSE-like group accretion as the origin of well populated KPPs.}
              \label{fig:angular_momentum_mm_sats}%
\end{figure*}

The infall of satellite groups has been proposed as a mechanism for producing coherently rotating, planar satellite systems \citep{Lynden_Bell_1976,Li_Helmi_2008,Smith_GroupAccretion_2016}. In this scenario, satellites accreted as part of the same group retain similar orbital planes. The alignment of the LMC and the Magellanic Stream with the VPOS has motivated suggestions that it formed from the accretion of an LMC-dominated dwarf galaxy group \citep{Lynden_Bell_1995,Donghia_Lake_groupaccretion_2008,Li_Helmi_2008,Nichols_LMCsatellites_2011,Vasiliev_2024,Martinez_Garcia_delpino_2025}. A similar scenario has been proposed for the GPoA in M31, driven by a group dominated by NGC 205 \citep{Angus_M31GroupAccretion_2016}.

In our suite of genetically modified simulations, increasing the mass of the $z\sim2$ GSE-like major merger leads to a corresponding increase in the number of (sub)satellites brought into the host halo. As shown in Table~\ref{tab:mm_sats} and Fig.~\ref{fig:MM_sats_stellar_density}, the GSE-like progenitor brings only 1 satellite in the Smallest run, compared to 9 satellites in the Largest case. Such an increase in the number of satellites accreted as part of the same group could favor the formation of KPPs if these satellites conserved a common angular-momentum direction after infall.

To test this hypothesis, we identify satellites associated with the GSE-like progenitor as those galaxies located within its virial radius prior to infall into the host halo. These systems are highlighted in Figure~\ref{fig:MM_sats_stellar_density}. We then follow their orbital evolution and determine whether they become members of the KPP. Figure~\ref{fig:angular_momentum_mm_sats} shows their trajectories and the evolution of the alignment between their angular-momentum vectors and $\hat{n}_{\rm KPP}$.

\begin{table}[h!]
\centering
\caption{Number and $z=0$ fate of satellites associated with the GSE-group accretion in each simulation.}

\begin{tabular}{lcccc}
\hline

& \multirow{2}{*}{$N_{\rm sat}^{\rm sub}$}
& \multirow{2}{*}{Disrupted ($z=0$)}
& \multicolumn{2}{c}{Surviving ($z=0$)} \\[2pt]

&  &  & No-KPP & \textbf{KPP} \\[1pt]
\hline

Smallest  & 1 & 0 & 1 & \textbf{0} \\[2pt]
Smaller   & 2 & 1 & 0 & \textbf{1} \\[2pt]
Fiducial  & 3 & 1 & 0 & \textbf{2} \\[2pt]
Larger    & 8 & 5 & 3 & \textbf{0} \\[2pt]
Largest   & 9 & 5 & 3 & \textbf{1} \\[2pt]

\hline
\end{tabular}
\label{tab:mm_sats}
\end{table}

In all our simulations, the GSE-like progenitor follows a highly radial orbit (see Table~\ref{tab:table_progenitor_B}), similar to the inferred progenitor of the proto–MW \citep{Rey_VintergatanGM_2023}. As a result, the angular momentum direction of the progenitor and its satellites during the first infall is poorly constrained: their position and velocity vectors are nearly antiparallel, and the orbital poles vary strongly during the initial radial passages, stabilizing only after a few orbits. Consequently, satellites accreted as part of the GSE-like group typically exhibit a wide range of orbital orientations and are not preferentially aligned with $\hat{n}_{\rm KPP}$. In the Larger and Largest simulations, zero and one satellites, respectively, are identified as KPP members, corresponding to $0\%$ and $\sim4.3\%$ of the total KPP population. Most satellites infalling within the GSE group in these runs are either disrupted by $z=0$ or remain misaligned with $\hat{n}_{\rm KPP}$. The Smaller and Fiducial simulations constitute mild exceptions, with one and two surviving group satellites identified as KPP members; however, these still represent only a small fraction of the total KPP population, $\sim11.1\%$ and $\sim7.1\%$, respectively. The statistics of these satellites are summarized in Table~\ref{tab:mm_sats}.

Overall, we find that KPPs in our simulations are almost entirely composed of satellites that were not originally associated with the GSE-like progenitor. Specifically, satellites associated with the GSE-like progenitor show no preferential alignment with $\hat{n}_{\rm KPP}$, particularly in the Larger and Largest simulations, where the KPPs are most populated. This indicates that early group accretion at $z\sim2$ does not enhance the kinematic coherence of the satellite population at low redshift. In particular, group accretion does not appear to be the mechanism driving the formation of KPPs in our simulations, nor does it explain the differences observed among the runs. This result is consistent with previous works arguing that spatially extended infalling groups, with characteristic sizes of order $\sim200$~kpc as in the Larger and Largest runs, are insufficiently compact to be the main origin of very thin satellite planes \citep{Metz_AgainstGroupAccretion_2009,Pawloski_2018}. We note, however, that the number of satellites identified in the GSE-like group is limited by resolution, and higher-resolution simulations will be valuable for testing whether richer and more compact infalling groups could enhance the importance of this mechanism.

\subsection{Satellite orbits affected by anisotropies in the host DM halo}
\label{sec:host_halo_shape}

Major mergers are among the primary drivers of the structural evolution of galaxies and their host DM halos, through the redistribution of angular momentum and their impact on galaxy and halo morphology \citep{Bailin_2005,Cox_2006,Zavala_2016,Grand_2017,Pulsoni_MergerDM_2021,Rey_Starkenburg_2022,Yoon_2022,orkney_Edgeshape_2023,Joshi_Paradigm_2025}. Their impact depends on several properties, including the merger mass ratio, orbital configuration, and gas content. In the \textsc{Vintergatan-GM} suite, \citet{Rey_VintergatanGM_2023} showed that increasing the mass of the $z\sim2$ merger produces progressively more compact and bulge-dominated galaxies at $z=0$.

Since the shape of the host DM halo has been proposed to influence satellite galaxy orbits \citep{Binney_1977, Penarrubia_2002, Penarrubia_2004, Shao_2019, Shao_2021}, in this section we investigate how the mass ratio of the GSE-like progenitor at $z\sim2$ affects the subsequent evolution of the host DM halo shape and orientation, and whether these properties correlate with the formation of planes of satellites.

\subsubsection{Anisotropies in host DM halo}
\begin{figure*}
   \centering
    \begin{subfigure}{0.999999\textwidth}
    \includegraphics[width=\textwidth]{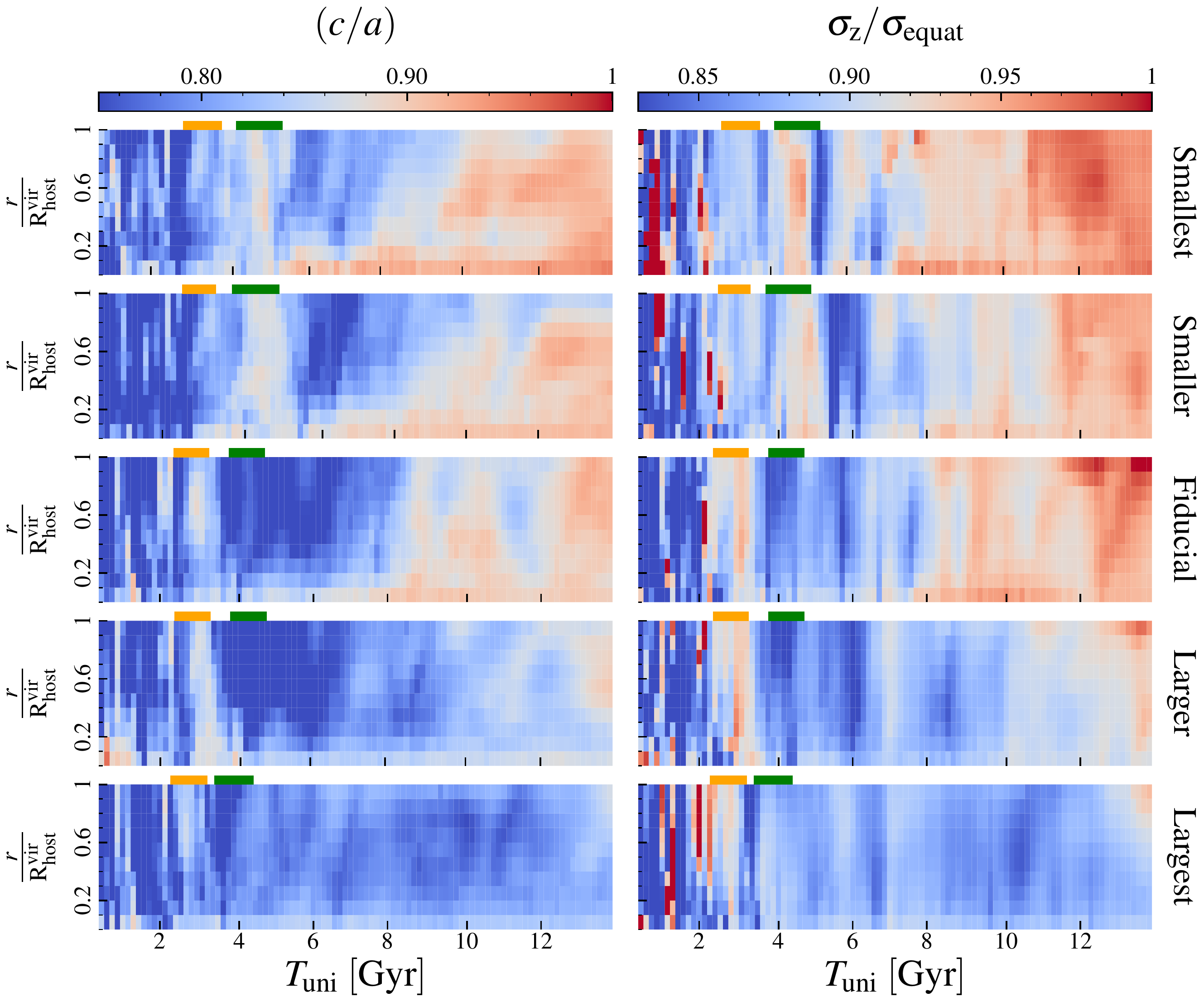}
    \end{subfigure}
   \caption{Evolution over cosmic time of the host DM halo anisotropies in both shape and velocity distribution for each simulation. Both quantities are computed using only host DM particles (excluding those bound to subhalos) within concentric spheres spanning radii from $0.1R_{\rm vir}^{\rm host}$ to $R_{\rm vir}^{\rm host}$, as indicated on the y-axis. \textbf{Left:} evolution of the minor-to-major axis ratio ($c/a$), derived from the eigenvalues of the mass distribution tensor. \textbf{Right:} evolution of the ratio between the velocity dispersion perpendicular to the equatorial plane ($\sigma_z$) and within the equatorial plane ($\sigma_{\rm equat}$). The minor axis obtained from the mass distribution tensor at each timestep and radius defines the equatorial plane. The orange and green horizontal bars above each panel mark the epochs of Merger~A and Merger~B, respectively. Before the mergers, both shape and velocity anisotropies are noisy and similar across the simulation suite. After the mergers, the evolution toward $z=0$ shows a clear dependence on the mass ratio of Merger~B: more massive GSE-like progenitors produce more flattened host halos with a lower velocity dispersion perpendicular to the equatorial plane.}
              \label{fig:host_halo_shape}%
\end{figure*}

In order to characterize the host DM halo shape, we use the mass distribution tensor $\vec{M}_{ij}$ which has been used extensively in halo shape literature \citep{Cole_Lacey_1996, Bailin_Steinmetz_2005, Bett_SpinShape_2007, Bryan_Triaxiality_2013, Mezini_2025}:
\begin{equation}
    M_{ij} = \sum_{n=1}^{N} m_n\, r_{n,i}\, r_{n,j}\,,
    \label{eq:mass_inertia_tensor}
\end{equation}
where $r_{n,i}$ and $r_{n,j}$ are the $i$-th and $j$-th components of the position vector of the $n$-th host particle relative to the host center of mass, $m_n$ is the particle mass, and $N$ is the total number of host particles. 

Rather than considering all particles within $R_{\rm vir}^{\rm host}$ at once, we compute the mass distribution tensor at each analyzed time step within a set of concentric spheres with radii spanning $0.1R_{\rm vir}^{\rm host}$ to $R_{\rm vir}^{\rm host}$, allowing us to probe the radial dependence of the halo shape from the inner to the outer regions. To isolate the smooth host halo potential, we exclude particles associated with resolved subhalos and retain only particles belonging to the main halo within each sphere.

In addition to anisotropies in the spatial distribution of the host DM particles following the major merger, we also investigate possible variations in the velocity distribution. Using the minor-axis direction $\hat{e}_3^{\rm host}$ derived from the $M_{ij}$ tensor, at each snapshot and each radius, we compute the velocity dispersion of the same DM host particles in cylindrical coordinates: along the host minor axis ($\sigma_z$) and within the equatorial plane ($\sigma_R$ and $\sigma_\varphi$), defined as the plane perpendicular to this axis.

Figure~\ref{fig:host_halo_shape} shows the radial and temporal evolution of the host-halo axis ratio, $c/a$, and of the anisotropy in the velocity distribution using the ratio between the velocity dispersion perpendicular ($\sigma_z$) and within the equatorial plane ($\sigma_{\rm equat}$), with the latter being defined as $\sigma_{\rm equat} = \sqrt{(\sigma_R^2 + \sigma_\varphi^2)/2}$. 

At early times ($T_{\rm uni} < 2$ Gyr), before Merger~A and Merger~B, the host DM halo is not yet dynamically settled and shows a noisy evolution. At this stage the halo has a relatively low mass ($\sim7\times10^{10}\,\rm{M}_\odot$) and exhibits relatively low axis ratios, with $c/a \sim 0.7$, consistent with expectations for low-mass halos at high redshift \citep{Butsky_2016,Petit_haloshape_2023}. Following the mergers, the host halo enters a phase of low accretion with no further major merger events. Examining the subsequent evolution until $z=0$, a clear trend emerges with the mass of Merger~B: more massive GSE-like progenitors lead to more flattened host DM halos. In the Smallest simulation, the outer halo ($r>0.2\,R_{\rm vir}^{\rm host}$) briefly becomes slightly flattened during the $\sim3$ Gyr following Merger~B, but this effect gradually disappears after $T_{\rm uni}>8$ Gyr, resulting in a roughly spherical halo with $c/a\sim0.95$ by $T_{\rm uni}>10$ Gyr. At the opposite extreme, the Largest run maintains a consistently flattened halo with $c/a\sim0.8$ across all radii throughout the $\sim10$ Gyr of post-merger evolution. The Smaller, Fiducial, and Larger runs show intermediate behavior, with progressively stronger and longer-lasting flattening as the mass ratio of the GSE-like progenitor increases.

A similar trend emerges when examining the anisotropy of the velocity distribution. After an initially noisy phase preceding the major mergers, the post-merger evolution of the velocity-dispersion ratios shows that simulations with more massive GSE-like progenitors exhibit systematically lower values of $\sigma_z/\sigma_{\rm equat}$ through their evolution to $z=0$. This behavior indicates that the random motions of the host DM particles are comparatively weaker in the direction perpendicular to the equatorial plane. The anisotropy of the velocity distribution closely mirrors the anisotropy in the spatial shape of the host halo: as the halo becomes more flattened, the velocity-dispersion ellipsoid also becomes flattened along the host minor axis.

\begin{figure}
   \centering
    \begin{subfigure}{0.999999\linewidth}
    \includegraphics[width=\linewidth]{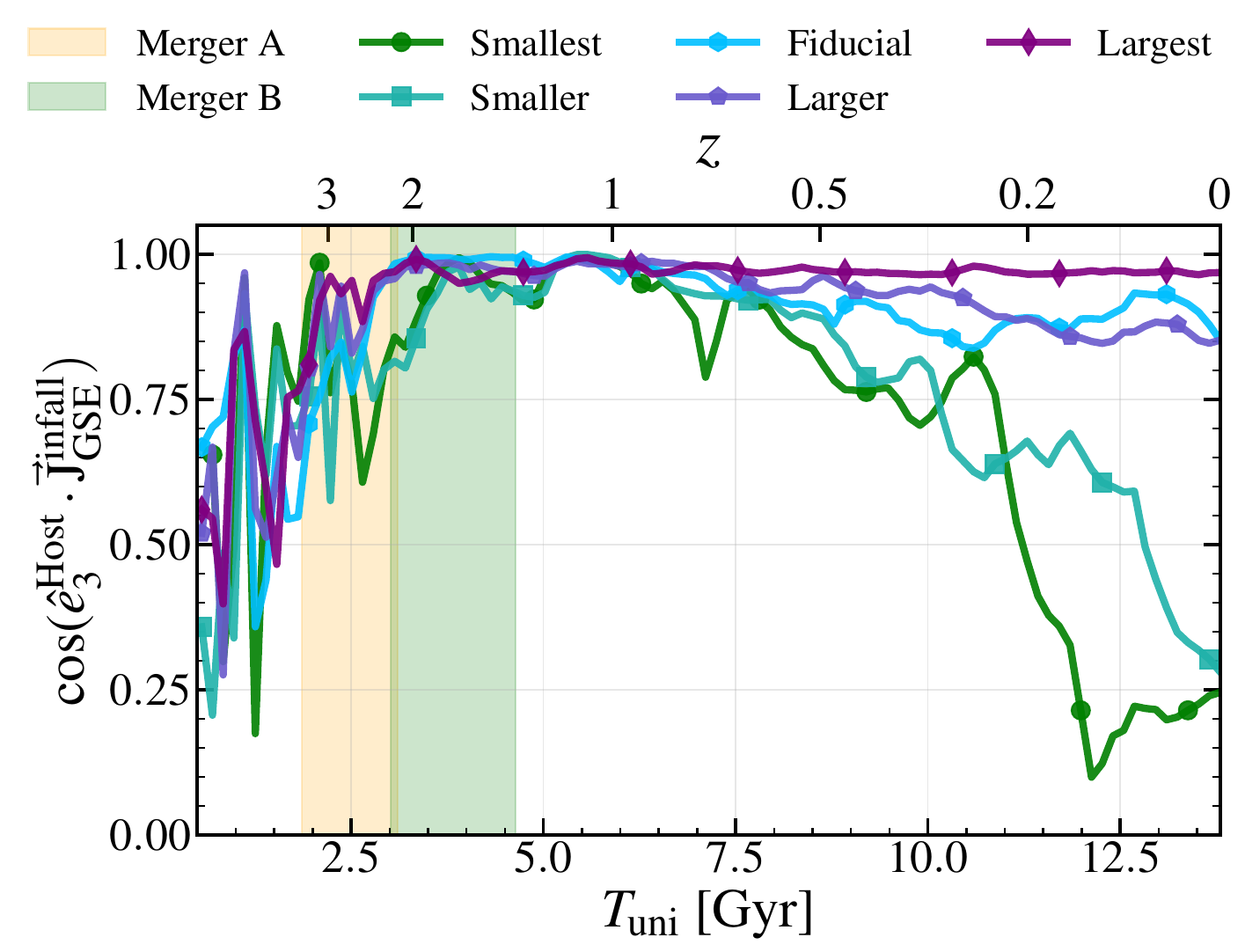}
    \end{subfigure}
   \caption{Evolution of the alignment between the host DM minor axis and the orbital plane of the GSE-like merger across the simulation suite. The host minor-axis direction is defined as the median of the minor-axis orientations measured within the set of concentric spheres used in Figure~\ref{fig:host_halo_shape}. The epochs of Mergers A and B are indicated by the orange and green shaded regions, respectively. Following both consecutive mergers, whose orbital planes are nearly identical, the host DM minor axis becomes strongly aligned with the merger orbital direction, and the duration of this alignment increases with the mass of the GSE-like progenitor.}
              \label{fig:minor_axis_GSE_aligment}%
\end{figure}

To analyze the direction of this flattening, we show in Figure~\ref{fig:minor_axis_GSE_aligment} the alignment between the host minor axis, $\hat{e}_3^{\rm Host}$, and the orbital plane of the GSE-like merger, $\vec{J}_{\rm GSE}^{\rm infall}$. At each timestep, $\hat{e}_3^{\rm Host}$ is defined as the median orientation measured across the radial apertures considered above, while $\vec{J}_{\rm GSE}^{\rm infall}$ is obtained from the median orbital pole over a 1~Gyr interval around GSE infall, reducing the noise associated with its highly radial orbit.

Before the major mergers, $\hat{e}_3^{\rm Host}$ varies substantially, consistent with the unsettled halo structure at these epochs. However, during Merger A the direction of $\hat{e}_3^{\rm Host}$ begins to align with $\vec{J}_{\rm GSE}^{\rm infall}$, since the orbital plane of Merger~A is closely aligned with that of~Merger B, with separation angles of about 10–20$^\circ$ between them across the simulations. This alignment becomes complete for all simulations during Merger~B, highlighting that the resulting post-merger DM halo is aligned with the merger direction. Nevertheless, a clear difference arises among the runs: when the GSE-like progenitor is sufficiently massive, this alignment is preserved until $z=0$ as in the Fiducial, Larger, and Largest runs. In contrast, in the Smallest and Smaller simulations the $\hat{e}_3^{\rm Host}$ begins to deviate from $\vec{J}_{\rm GSE}^{\rm infall}$ roughly $\sim4$ Gyr after Merger~B. This deviation closely correlates with the epochs in which the host DM halo becomes approximately spherical in these runs, implying that the minor axis is intrinsically less well defined.

\subsubsection{Alignment between the host DM halo and kinematic persistent planes}

\begin{figure*}
    \centering
    \includegraphics[width=0.999\linewidth]{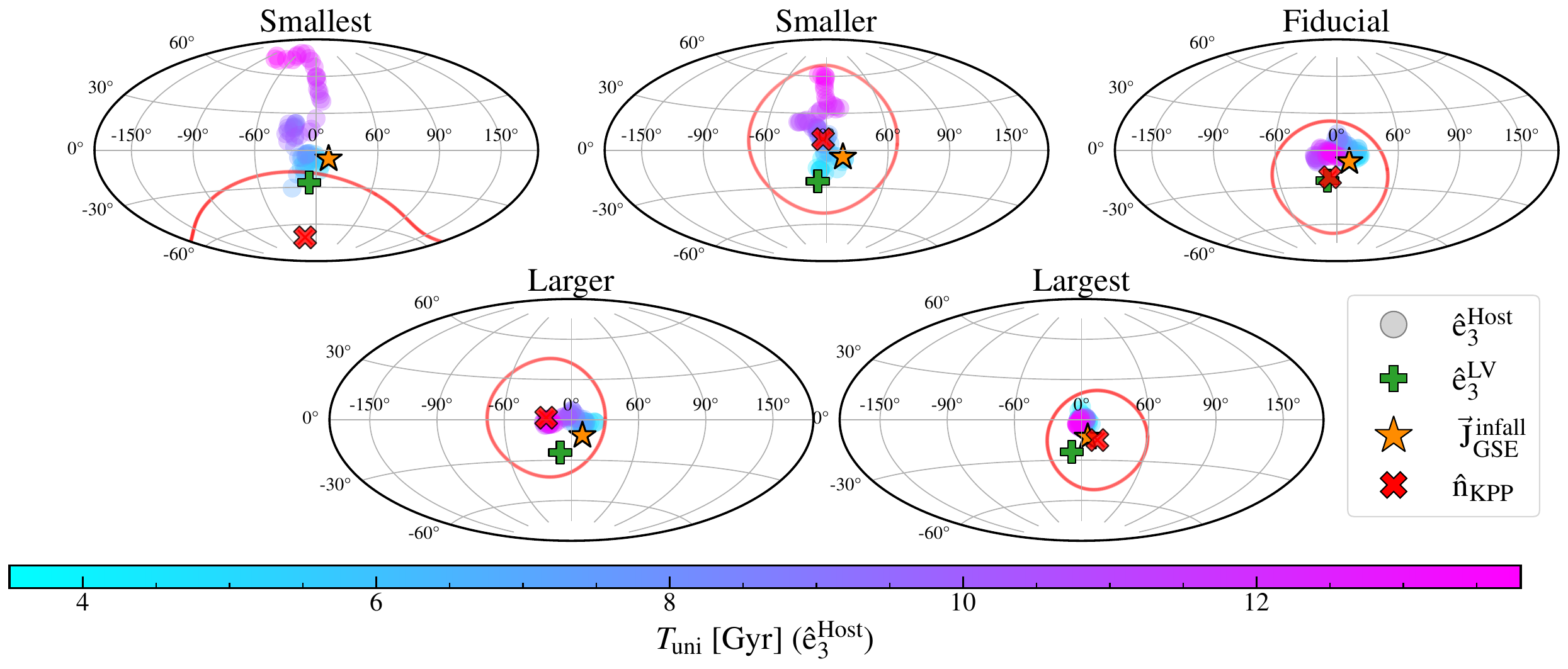}
  \caption{Aitoff projections showing the different relevant directions identified in this work. Colored circles represent the time evolution of the host halo minor axis, $\hat{e}_{3}^{\mathrm{Host}}$, after the GSE-like major merger. The orbital plane of the GSE-like major merger, $\vec{J}_{\rm GSE}^{\rm infall}$, is indicated by an orange star. The axis corresponding to the direction of maximum compression of the LV, $\hat{e}_{3}^{\mathrm{LV}}$, is shown as a green plus symbol. The maximum co-orbitation axis, $\hat{n}_{\mathrm{KPP}}$, is indicated by a red cross, and the red circumference marks the angular distance $\alpha_{50}$ within which $50\%$ of the satellites orbit at $z=0$, providing a measure of the clustering of satellite angular momenta around the co-orbitation axis. In these projections, vectors are treated as axial quantities and are shown only in one hemisphere ($\mathrm{lon} \in [-90^\circ, 90^\circ]$). A clear alignment among all these directions is present in the runs where KPPs are well populated (Fiducial, Larger, and Largest), suggesting that KPP formation may be related to these alignments.}
              \label{fig:aitoff_directions_all}%
\end{figure*}

In order to disentangle the possible connection between the flattening of the host DM halo and the formation of kinematically coherent satellite planes, Figure~\ref{fig:aitoff_directions_all} shows an Aitoff projection in which all the relevant directions identified in this work are overplotted: the post-merger evolution of the host DM halo minor axis ($\hat{e}_{3}^{\mathrm{Host}}$), the orbital plane of the GSE-like major merger ($\vec{J}_{\rm GSE}^{\rm infall}$), the direction of maximum compression of the LV ($\hat{e}_{3}^{\mathrm{LV}}$), and the axis of maximum co-orbitation ($\hat{n}_{\mathrm{KPP}}$) around which KPPs are formed. Several interesting results can be highlighted:

\begin{enumerate}
    \item The orbital plane of the GSE-like major merger, $\vec{J}_{\rm GSE}^{\rm infall}$, is almost identical in all our simulations, as expected a priori in GM simulations, and is closely aligned with the axis of maximum compression of the LV, $\hat{e}_{3}^{\mathrm{LV}}$. This highlights that the major merger is accreted along the direction through which large-scale mass flows are preferentially channeled.
    \item  As shown in Figure~\ref{fig:minor_axis_GSE_aligment}, the Fiducial, Larger, and Largest runs maintain a long-lived alignment between $\hat{e}_3^{\rm Host}$ and the merger direction. Because $\vec{J}_{\rm GSE}^{\rm infall}$ is itself closely aligned with $\hat{e}_3^{\rm LV}$, the host minor axis in these runs also remains aligned with the large-scale compression direction. This alignment is progressively lost in the Smallest and Smaller runs.
    \item Remarkably, in the simulations where we observe a strong clustering of satellite orbital poles and well-populated KPP (Fiducial, Larger, and Largest), the axis of maximum satellite co-orbitation, $\hat{n}_{\mathrm{KPP}}$, around which the KPP is formed, is closely aligned with all the previously identified directions: $\hat{e}_{3}^{\mathrm{Host}}$, $\hat{e}_{3}^{\mathrm{LV}}$ and $\vec{J}_{\rm GSE}^{\rm infall}$.
    \item The angular distance $\alpha_{50}$ from $\hat{n}_{\mathrm{KPP}}$, within which 50\% of satellites orbit at $z=0$, progressively decreases from the Smallest to the Largest simulations. This reflects an increasing degree of orbital clustering around the equatorial plane, mirroring the progressively stronger flattening of the host DM halo.
\end{enumerate}

These alignments suggest that, in this simulation suite, the formation of KPPs is connected to the equatorial plane of the flattened host DM halo. A flattened host halo develops an aspherical potential in which satellite orbits may be progressively reoriented toward the symmetry axes through gravitational tidal torques associated with the higher mass density within the equatorial plane \citep{Shao_2019, Mezini_2025}. In addition, satellite galaxies are affected by dynamical friction, which causes orbital energy loss and gradual orbital decay toward the inner regions of the host DM halo (e.g., \citealt{Binney_Tremaine_1987}). Interestingly, as initially proposed by \cite{Binney_1977} and further investigated by \cite{Penarrubia_2002, Penarrubia_2004}, dynamical friction in flattened halos with anisotropic velocity dispersions is not symmetric and can cause satellites to tilt their orbital planes, progressively aligning them with the equatorial plane of the host DM halo over several orbits. In particular, this anisotropic dynamical friction becomes more effective when the host halo is more flattened \citep{Penarrubia_2002} and when the velocity dispersion perpendicular to the equatorial plane is lower than that within the plane \citep{Binney_1977, Penarrubia_2004}. Both effects therefore predict more efficient orbital-plane alignment in the more strongly flattened and velocity-anisotropic halos.

Moreover, in the Fiducial, Larger, and Largest runs, massive GSE-like major mergers efficiently align the host DM halo minor axis with the direction along which large-scale mass flows are preferentially channeled, $\hat{e}_{3}^{\mathrm{LV}}$. As a result, satellites tend to be accreted on orbits that are already preferentially aligned with the host equatorial plane.

\begin{figure}
    \centering
    \begin{subfigure}{0.999\linewidth}
    \includegraphics[width=\linewidth]{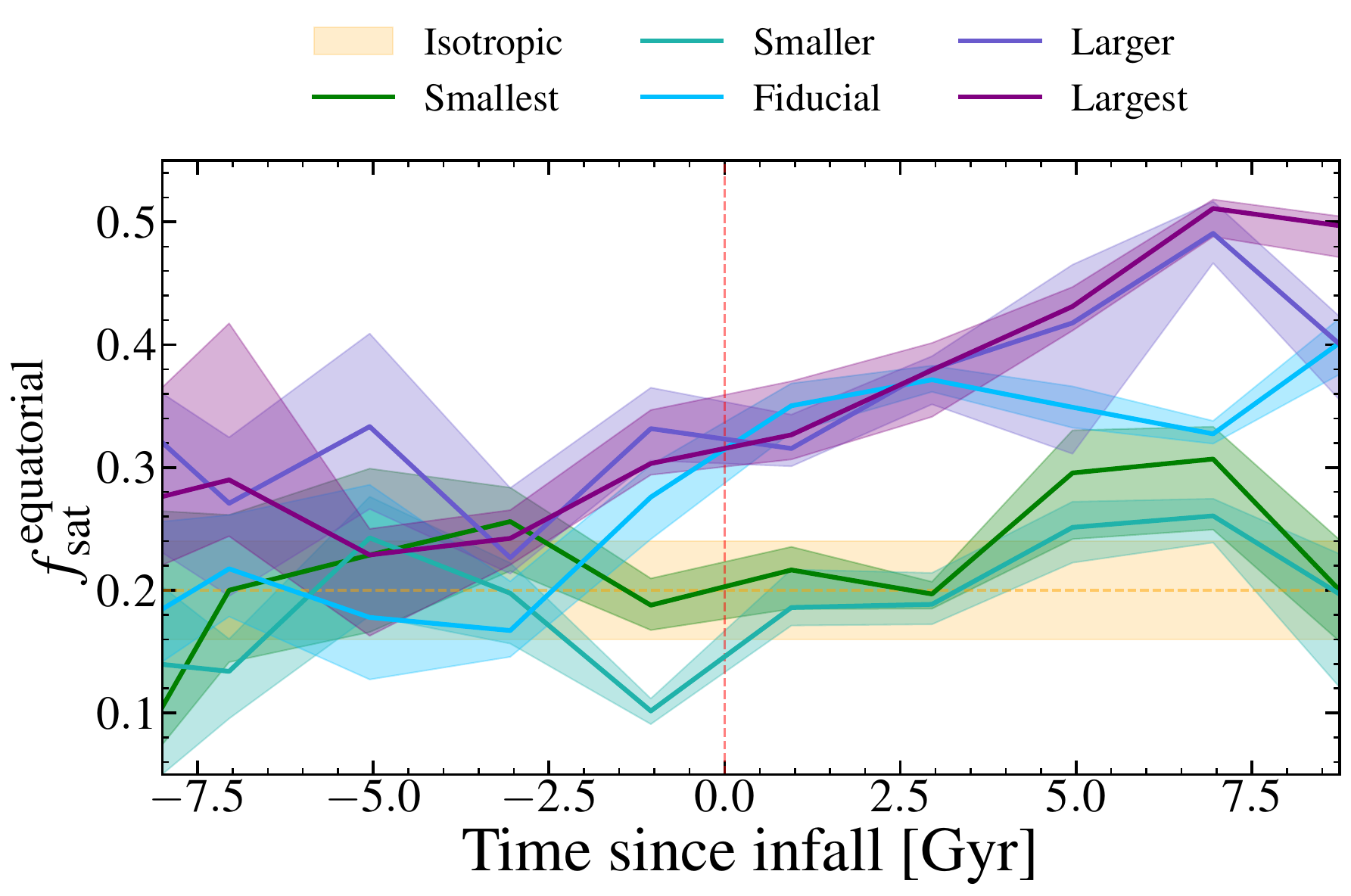}
    \end{subfigure}
    \caption{
    Evolution of the fraction of satellites in equatorial orbits normalized by the infall time of each satellite for each simulation, in order to disentangle the pre- and post-infall behavior. Satellites in equatorial orbits are defined as those with angular momentum vectors lying within an angular distance lower than $\alpha_{\rm co-orbit}= 36.87^\circ$ from the host DM halo minor axis $\hat{e}_{3}^{\mathrm{host}}$ at $z=0$. Lines represent the median fraction of satellites at each time bin and shaded regions mark the median absolute deviation of each bin. Orange dashed line and shaded regions represent the expectations from an isotropic distribution of satellite orbital poles. An enhanced fraction of satellites on equatorial orbits is already present at infall in the Fiducial, Larger, and Largest runs, followed by a secular alignment of satellite orbits toward the host equatorial plane, particularly pronounced in the Larger and Largest runs.}
   \label{fig:equatorial_enhancement}
\end{figure}

To assess whether satellite orbital poles preferentially align with the minor axis of the host DM halo and whether their orbits become confined to the equatorial plane as a consequence of post-infall alignment, we analyze the time evolution of the fraction of satellites on equatorial orbits across the simulation suite. Figure~\ref{fig:equatorial_enhancement} shows the evolution of the fraction of satellites orbiting within the equatorial plane ($f_{\rm sat}^{\rm \, equatorial}$) as a function of each satellite’s time since infall, allowing us to distinguish between pre- and post-infall behavior. Satellites on equatorial orbits are defined as those whose orbital angular momentum poles lie within an angular distance smaller than $\alpha_{\rm co\text{-}orbit}=36.87^\circ$ from the host halo minor axis, $\hat{e}_{3}^{\mathrm{Host}}$, measured at $z=0$.

At times before 2 Gyr of satellite infall, $f_{\rm sat}^{\rm \, equatorial}$ is very noisy, reflecting that these epochs are dominated by satellites at large distances with poorly defined and rapidly varying angular momentum directions. At the infall time, differences between simulations are already present. The Smallest and Smaller runs show $f_{\rm sat}^{\rm \, equatorial}$ consistent with the isotropic expectation ($\sim20\%$), whereas the Fiducial, Larger, and Largest runs exhibit a higher fraction of satellites ($\sim30\%$) that already infall with angular momentum poles aligned with the host DM halo minor axis at $z=0$. This behavior arises because of the more efficient and long-lasting alignment of the host DM halo minor axis and the large-scale direction of preferred mass flows accretion in the Fiducial, Larger, and Largest runs. As a result, in these runs satellite galaxies are preferentially accreted on orbits that are already close to the host equatorial plane.

Regarding the post-infall evolution, the Larger and Largest runs show a clear secular alignment of satellite orbits with the host DM halo equatorial plane. In these runs, $f_{\rm sat}^{\rm \, equatorial}$ increases progressively from $\sim30\%$ at infall to $\sim50\%$ after $\sim 8$ Gyr of post-infall evolution in the Largest run. In contrast, the Smallest and Smaller runs remain consistent with the isotropic expectation, while the Fiducial simulation displays an intermediate behavior between both regimes. This evolution indicates that, in addition to the preferential accretion of satellites already close to the equatorial plane, a fraction of initially misaligned satellites is progressively reoriented after infall in flattened halos.

\begin{figure*}
    \centering
    \includegraphics[width=\linewidth]{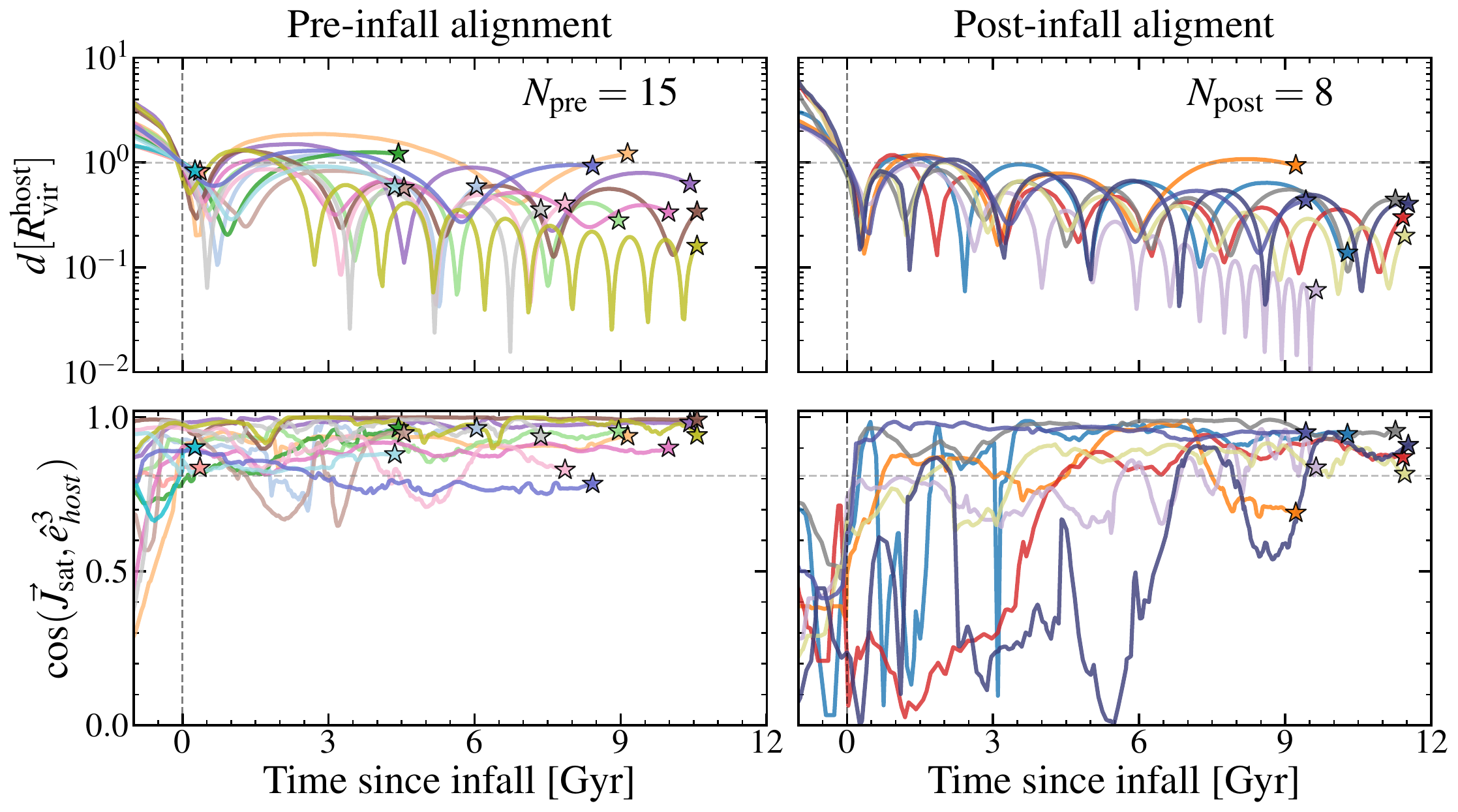}
    \caption{Orbital evolution of KPP members in the Largest simulation, separated according to whether their orbital angular-momentum poles are already aligned with the host equatorial plane at infall (left column) or whether they become aligned only after infall (right column). Satellites are classified using the angular separation between $\vec{J}_{\rm sat}$ and the present-day host minor axis, $\hat{e}_{\rm host}^{3}$, adopting $\alpha_{\rm coorbit}=36.87^\circ$ as the alignment threshold. \textbf{Top panels:} radial trajectories of the satellites, expressed in units of the host virial radius, with the horizontal gray dashed line marking $R_{\rm vir}^{\rm host}$. \textbf{Bottom panels:} evolution of the alignment between $\vec{J}_{\rm sat}$ and  $\hat{e}_{\rm host}^{3}$, with the horizontal gray dashed line indicating the adopted co-orbitation threshold. In all panels, time is measured relative to infall for each satellite, marked by the vertical gray dashed line. Individual satellites are shown with different colors, and star markers indicate their values at $z=0$.}
   \label{fig:largest_prevspost}
\end{figure*}

To quantify the relative contribution of the two channels described above, we classify the KPP members according to whether their orbital angular-momentum poles are already aligned with the host equatorial plane at infall or become aligned only during their subsequent orbital evolution. Figure~\ref{fig:largest_prevspost} illustrates these two populations, as a representative example, for the Largest run. The left panels show satellites that are already aligned with the host equatorial plane at infall, which account for approximately 65\% of the KPP members in the Largest run. Their orbital poles remain aligned with $\hat{e}_{\rm host}^{3}$ throughout their subsequent evolution, on the stable equatorial plane. In contrast, the right panels show the remaining $\sim35\%$ of KPP satellites, which enter the halo on initially misaligned orbits around infall. As their orbits evolve and decay within the host halo, their orbital plane progressively aligns with the host equatorial plane, eventually reaching and remaining within it.  A similar contribution from post-infall reorientation is found in the other simulations hosting well-populated KPPs, representing 40\% (8/20) of the KPP members in the Fiducial run and 40\% (9/22) in the Larger run. These results emphasize that well-populated KPPs can arise from both preferential infall onto stable equatorial orbits (resulting from the alignment between the host halo and the cosmic web principal directions) and the post-infall reorientation of initially misaligned satellites.

In order to disentangle which mechanism dominates the post-infall alignment of satellite galaxies toward the equatorial plane -- between gravitational torques from the host potential and anisotropic dynamical friction -- we exploit the fact that the former should affect both DM particles and satellite galaxies, whereas the latter primarily affects satellite galaxies. To test this, we performed additional experiments using samples of DM particles selected to reproduce the same radial bias as our satellite population. We find that the impact on the orbital alignment of these DM particles is considerably weaker than that observed for satellite galaxies. This result suggests that anisotropic dynamical friction is the dominant mechanism driving the secular alignment of satellite galaxy orbits with the equatorial plane in flattened host DM halos, in agreement with the predictions of \cite{Binney_1977, Penarrubia_2002, Penarrubia_2004}.

Taken together, these results support two complementary channels linking sufficiently massive GSE-like mergers to the formation of KPPs. Such mergers leave behind a flattened and anisotropic host halo whose minor axis remains closely aligned with the preferred large-scale accretion direction:

\begin{enumerate}
    \item The close correspondence between $\hat{e}_{3}^{\mathrm{Host}}$ and $\hat{e}_{3}^{\mathrm{LV}}$causes satellite galaxies to be preferentially accreted through the equatorial plane of the host DM halo, so that a significant fraction of satellites already enter the halo on stable equatorial orbits.
    \item After infall, the orbital planes of satellite galaxies on initially non-equatorial orbits tend to progressively reorient toward the host DM halo equatorial plane, primarily as a result of anisotropic dynamical friction acting within the flattened halo potential.
\end{enumerate}

We note that the flattening of the LV is nearly identical across the simulation suite, as shown in Section~\ref{sec:lss}. Therefore, satellites are also preferentially accreted along a specific direction in the Smallest and Smaller simulations. However, in these runs the alignment between $\hat{e}_{3}^{\mathrm{Host}}$ and $\hat{e}_{3}^{\mathrm{LV}}$ is less effective and the host halos remain more spherical. Consequently, satellite galaxies that are accreted with orbital planes perpendicular to $\hat{e}_{3}^{\mathrm{LV}}$ do not enter the halo on stable equatorial orbits, which hinders their long-term persistence as kinematically coherent satellite structures and reduces the impact of anisotropic large-scale accretion. This contrasts with the Larger and Largest runs, where the halos are more flattened and more closely aligned with the large-scale accretion direction, allowing satellites accreted along the large-scale preferential direction to remain on stable equatorial orbits.

\section{Discussion}
\label{sec:discussion}

\subsection{Comparison with previous work}
\label{sec:disc_prev_work}

This work presents the first systematic study, using high-resolution cosmological simulations with genetically modified initial conditions, of the role played by the mass of a GSE-like merger in the formation of planes of satellites. This approach allows us to isolate the impact of the merger mass while keeping other relevant factors -- such as the large-scale structure environment -- effectively fixed.

The KPPs identified in several of our simulations reinforce the findings of \cite{santos_santos_2023} and \cite{Gamez_Marin_TNG_2025}, showing that kinematically coherent satellite planes can persist over extended timescales in $\Lambda$CDM simulations. To our knowledge, however, the Largest run provides the first example of a well-populated coherent plane ($N_{\rm sat}\sim25$, $f_{\rm sat}>50\%$) that maintains for more than 4~Gyr a thickness comparable to that of the on-plane VPOS satellites (Figure~\ref{fig:evolution_plane_multi}).

Furthermore, the instantaneous planarity of these long-lived structures can be enhanced by transient satellites, consistent with the scenario proposed for the MW by \citet{Sawala_Nature_2023}. In particular, the positional planes identified in the Larger and Largest simulations reproduce the structural properties of the VPOS for approximately 5\% ($\sim0.5$~Gyr) and 10\% ($\sim1$~Gyr) of their post-merger evolution, respectively. During these intervals, when considering almost the entire satellite population ($f_{\rm sat}=90\%$, $N_{\rm sat}\sim40$), the detected planes reach values as low as $c/a\sim0.24$ and $\Delta_{\rm RMS}\sim26.5$~kpc, while roughly 50\% of the satellites co-orbit around the plane normal (see Figure~\ref{fig:4GND_metrics_evolution}). These values are comparable to those measured for a similarly populated MW satellite sample (e.g., \citealt{Santos_santos_MW_2020,Pawlowski_2021}). Although the Larger and Largest runs also reproduce a substantial degree of orbital coherence in terms of co-orbitation (see Figure~\ref{fig:aitoff_angular}), they do not reach the very high corotation fraction inferred for the on-plane VPOS satellites, among which only 2 of 16 are counter-rotating (e.g., \citealt{Pawlowski_2021,Taibi_portrait_2024}). This supports the interpretation that the unusually high corotation fraction of the MW satellite system may be enhanced by the recent accretion of the LMC group \citep{santos_santos_2023,Martinez_Garcia_delpino_2025}.

In line with \cite{GamezMarin_2024}, we find that well-populated KPPs are strongly aligned with the axis of maximum compression of the large-scale structure, highlighting the connection between the anisotropic collapse of matter in $\Lambda$CDM and the formation of kinematically coherent planes of satellites. This is also consistent with previous studies proving preferential subhalo accretion along large-scale anisotropies \citep{Libeskind_Subhaloaccretion_2014,Dupuy_Accretion_2022}. However, our work provides an important additional perspective. Since the large-scale structure is fixed across our suite of genetically modified simulations (as shown in Section~\ref{sec:lss}), we find that simulations experiencing the same anisotropic collapse of the CW can nevertheless differ significantly in whether they develop well-populated KPPs. This indicates that, within the range of conditions explored here, anisotropies in the large-scale mass flows constitute a necessary but not sufficient condition for the formation of such structures. In particular, in our simulation suite, we find that well-populated KPPs form only when the host DM halo becomes sufficiently flattened and its minor axis is closely aligned with the axis of maximum compression of the LV.

Our results are consistent with those reported by \cite{Shao_2019, Shao_2021}, showing that kinematically co-orbiting planes of satellites tend to orbit around the equatorial plane of the host DM halo, that this alignment strengthens after satellite infall, and that the effect is more pronounced in more flattened host halos (i.e., with lower $c/a$). However, in contrast to these works, we propose that the mechanism primarily responsible for the post-infall reorientation of satellite orbits toward the host equatorial plane is anisotropic dynamical friction rather than tidal gravitational torques. This scenario is supported by the enhanced secular alignment observed in more flattened host DM halos with stronger velocity anisotropies, which is consistent with theoretical expectations for anisotropic dynamical friction \citep{Binney_1977, Penarrubia_2002, Penarrubia_2004}.

\citet{Gamez_Marin_2026} analyzed the formation of early KPPs in the TNG50-1 run using the TNG50 MW/M31 sample \citep{Pillepich_2024}. They found that, in well-populated systems hosting early-forming KPPs, the clustering of satellite orbital poles is typically established at early cosmic times ($T_{\rm uni}\sim4\,$Gyr), well before satellite infall. This disfavors post-infall orbital reshaping by host-halo anisotropies as the primary driver of early KPP formation in TNG50. Their results therefore highlight that, within a statistical sample of cosmological initial conditions, KPPs can form very early, driven by anisotropies in the LSS, without requiring a GSE-like merger history. This emphasizes the existence of multiple formation channels for planes of satellites. Nevertheless, such mergers and host DM halo effects may still play an important role in the formation of later-forming KPPs, as shown in our work, although the statistical significance of this pathway remains to be determined.

The strong impact of a GSE-like major merger on the formation of satellite planes identified in this work appears, however, to contrast with the results of \cite{Kanehisa_2023}. Using the IllustrisTNG large-volume simulations, they investigated the influence of major mergers (mass ratios above 1/3) on satellite plane properties and reported a negligible correlation between major mergers and the presence of such structures. A key difference is that our simulations isolate the effect of a GSE-like merger, whereas in large cosmological samples the impact of mergers may be entangled with other processes that also affect satellite distributions, such as group accretion or variations in the surrounding large-scale structure. Nevertheless, if the mechanism proposed here plays an important role, some level of correlation might still be expected in large statistical samples.

Another crucial difference between these works is numerical resolution. The analysis of \cite{Kanehisa_2023} relies on the hydrodynamical TNG100-1 run, whose mass resolution is substantially lower than that of our simulation suite -- approximately 40 times worse in DM particle mass and about 140 times worse in stellar particle mass. 
Such limited resolution may fail to adequately resolve the subhalo population in MW-mass halos and can lead to artificial disruption of satellites \citep{van_den_bosch_2018, Sawala_Nature_2023, Errani_2024, Santos_Santos_2025}, removing them before dynamical friction drives them into the inner regions of the host halo. Compounding this effect, if the analyses rely on standard halo finders and associated merger tree algorithms, this could further exacerbate the issue, as these are known to struggle when tracking subhalos through the dense central regions of their hosts \citep{Diemer_sparta_2024, Symfind_Mansfield_2024, Rodirguez_Cardoso_Agora_2025, Forouhar_Moreno_2025, Kong_Bloodhound_2026}, which can further contribute to an underrepresentation of inner satellites.

\begin{figure}
    \centering
    \includegraphics[width=0.9999\linewidth]{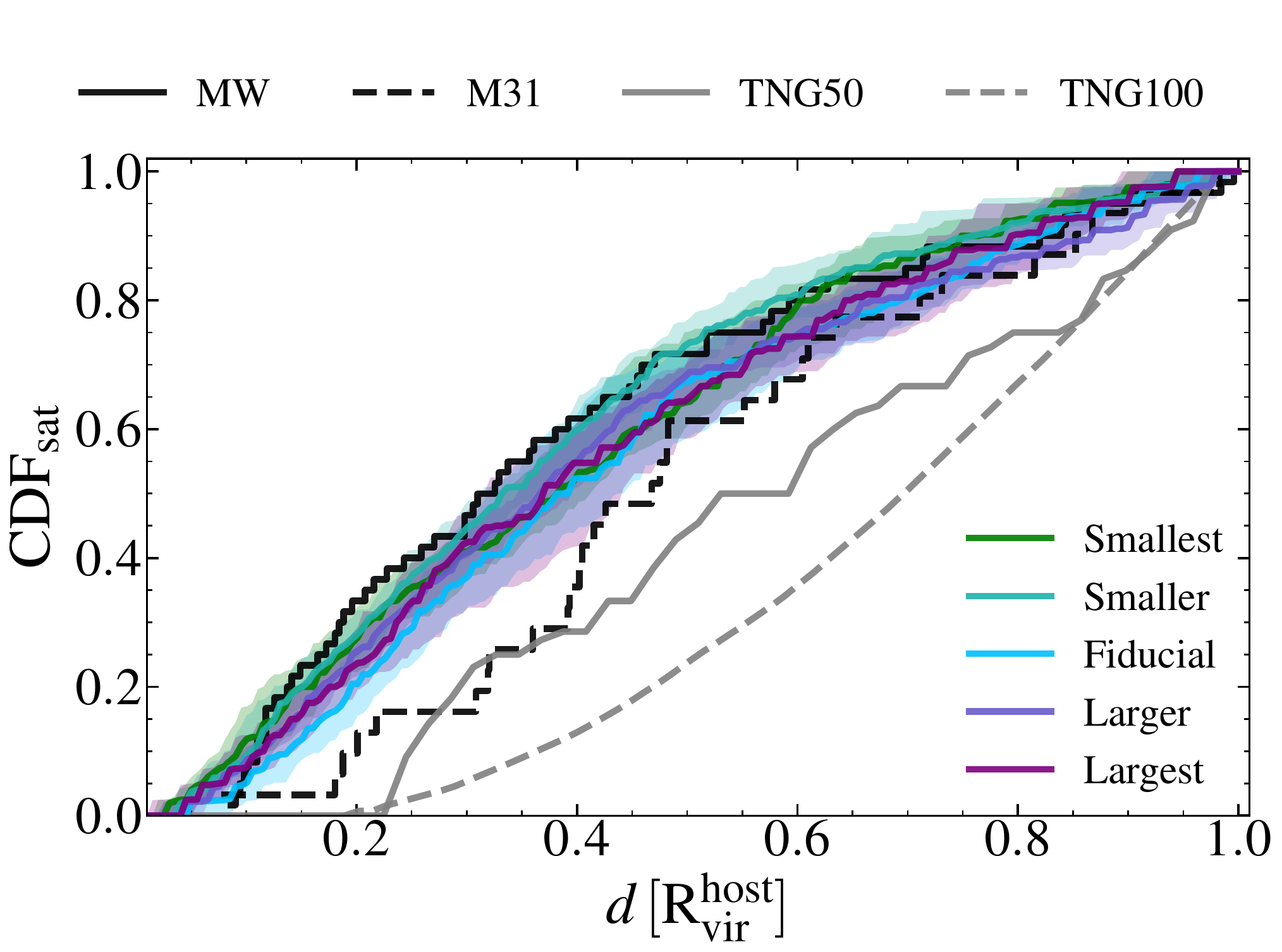}
  \caption{Satellite radial distributions normalized by host virial radius across our simulation suite. Solid colored lines show the median over 50 snapshots between $z=0.2$ and $z=0$, with shaded regions indicating the median absolute deviation. Observed distributions for the MW \citep{Drlica_Wagner_2020} and M31 \citep{Savino_2022} are shown as black solid and dashed lines, respectively. Gray dashed and solid lines show the median satellite radial distributions at $z=0$ reported for TNG100 \citep{Kanehisa_2023} and computed from the TNG50 MW/M31 sample catalogs \citep{Pillepich_2024}, respectively.}
              \label{fig:CDF_sats_rad_2}%
\end{figure}

In Figure~\ref{fig:CDF_sats_rad_2}, we compare the satellite radial distribution in our simulation suite  with the median value reported for the TNG100-1 sample employed in \cite{Kanehisa_2023}, and with the ones observed in MW \citep{Drlica_Wagner_2020} and M31 \citep{Savino_2022}, adopting $\rm{R}_{\rm vir}^{\rm MW} = 254$ kpc and $\rm{R}_{\rm vir}^{\rm M31} = 266$ kpc. For completeness, we also show the median value obtained from the TNG50 MW/M31  sample \citep{Pillepich_2024} which we have computed using the satellite positional data from the catalog\footnote{Publicly available at \url{https://www.tng-project.org/data/milkyway+andromeda/}}. The distribution reported for TNG100-1 run in \cite{Kanehisa_2023} differs significantly from that observed in the MW, particularly underpredicting the number of satellites in the inner halo. The median radial distribution obtained from the TNG50-1 sample is slightly more centrally concentrated than that of TNG100-1, as expected from its higher resolution, but it still underpredicts a considerable number of satellites in comparison with the MW, particularly in the inner regions. In contrast, our simulation suite recovers a significantly larger inner satellite population when applying the subhalo particle-tracking correction, yielding radial distributions consistent with those observed in the MW. Without this correction, the radial distributions obtained using standard RCT merger trees fail to recover the inner satellite population, highlighting the relevance of using accurate subhalo merger trees. The effectiveness of the subhalo particle-tracking method in recovering satellites that have decayed into the inner halo is illustrated in Appendix~\ref{app:rct_part_tracking} (Figures~\ref{fig:comparison_rct},~\ref{fig:comparison_rct_plot}, and \ref{fig:CDF_sats_rad}). These results highlight that, while a diversity of subhalo accretion histories can lead to a diversity of satellite radial distributions and concentrations, sufficiently high numerical resolution, combined with reliable subhalo tracking, produces more concentrated satellite radial distributions that are more compatible with that observed for the MW.

In the scenario proposed here, satellites progressively align with the host equatorial plane as their orbits decay toward the inner halo through anisotropic dynamical friction (see Figure~\ref{fig:largest_prevspost}). Detecting this process therefore requires sufficient resolution and robust tracking of satellites during the late stages of orbital decay. Premature subhalo disruption could consequently weaken or erase the correlation identified here, potentially contributing to the differences between our work and previous studies. Since our simulation suite does not allow us to determine the statistical prevalence of this mechanism, a more definitive assessment will require future studies based on large statistical samples with sufficiently high resolution and robust subhalo-tracking techniques such as those applied here.

\section{Conclusions}
\label{sec:conlcusions}

In this paper, we have investigated the formation of planes of satellite galaxies in the \textsc{Vintergatan-GM} suite of simulations, in which a GSE-like major merger at $z\sim2$ is systematically modified through genetic modifications of the ICs. Across the suite, the total (stellar) mass ratio of this merger is varied from 1:10 (1:25) to 1:2 (1:2), while the $z=0$ halo mass of the MW-halo and its surrounding large-scale environment are held fixed. Using these high-resolution zoom-in simulations, we examined whether planes of satellites comparable to that observed in the MW emerge, and assessed the role played by the mass ratio of a $z\sim2$ major merger in shaping kinematically coherent planar satellite structures. Our main findings can be summarized as follows:

\begin{enumerate}
    \item We applied the extended 4GND plot methodology \citep{Pawlowski_2013_4GND,Santos_santos_sim_2020,Santos_santos_MW_2020} to identify high-quality positional planes independently in each snapshot of every simulation. A consistent trend emerges across the simulation suite, as shown in Section~\ref{sec:positional_planes} and Figure~\ref{fig:4GND_metrics_evolution}. More massive GSE-like mergers are associated with: \textit{(i)} a higher degree of anisotropy in the distribution of 4-galaxy-normal vectors, \textit{(ii)} thinner satellite planes characterized by lower $c/a$ and $\Delta_{\rm RMS}$ values, and \textit{(iii)} an increased fraction of co-orbiting satellites aligned with the plane normal. In particular, the Larger and Largest runs exhibit plane properties comparable to those observed in the MW at certain epochs.

    \item The kinematic coherence of the satellite population shows the same systematic dependence on merger mass. The Fiducial, Larger, and Largest runs develop well-populated KPPs, containing $\gtrsim40\%$ of their satellites (as reflected in Figures~\ref{fig:aitoff_angular}~and~\ref{fig:f_sat_nkpp}). These KPPs provide a long-lived kinematic backbone around which thinner positional configurations can temporarily emerge as additional satellites contribute to the plane. In the Larger and Largest runs, coherent long-lived planes containing up to $N_{\rm sat}=25$ satellites maintain structural properties comparable to the MW on-plane VPOS population for $\sim4$~Gyr (see Figures~\ref{fig:evolution_plane_multi} and~\ref{fig:evolution_plane_multi_plot}). This suggests a transition toward VPOS-like configurations for GSE-like merger mass ratios around those of the Larger ($1{:}2.9$) run, corresponding to a progenitor mass of order $7\times10^{10}$--$10^{11}\,M_\odot$.
    
    \item We find in Section~\ref{sec:group_accretion} that the differences observed across the simulation suite are not driven by an increased contribution from group accretion during the GSE-like merger. The KPPs are composed almost entirely of satellites that were not originally associated with the GSE-like progenitor, and those satellites that were accreted with the progenitor are disrupted by $z=0$ or show no preferential alignment with the axis of maximum satellite co-orbitation.
    \item This work confirms that the dynamics of the cosmic web play a key role in driving the anisotropic accretion of satellites required for the formation of KPPs. Specifically, simulations that develop well-populated KPPs exhibit a clear alignment between the axis of maximum satellite co-orbitation and the local cosmic web principal direction of maximum collapse, in agreement with \cite{GamezMarin_2024}. However, in our specific simulation suite, this condition may be necessary but not sufficient: despite sharing an identical large-scale environment, some simulations develop well-populated KPPs, whereas others retain satellite angular momentum distributions that remain close to isotropic.
    \item The primary driver of the differences across the simulation suite are changes in the shape of the host DM halo and its alignment with the large-scale structure induced by the GSE-like major merger, as shown in Section~\ref{sec:host_halo_shape}. Increasing the mass of the merger leads to more flattened host halos (lower $c/a$) with more anisotropic velocity distributions (lower $\sigma_{\rm z}/\sigma_{\rm equat}$), which become more efficiently aligned with the large-scale direction of preferential subhalo accretion. As a result, well-populated KPPs in these runs form through two main mechanisms: \textit{(i)} the preferential accretion of subhalos through the host equatorial plane, where they remain on stable equatorial orbits; and \textit{(ii)} the secular alignment of initially non-equatorial satellite orbits toward the host equatorial plane through post-infall anisotropic dynamical friction, whose efficiency is enhanced in more flattened halos with stronger velocity anisotropies.
    \item Although we find that group satellite accretion occurring around 10 Gyr ago does not play a significant role in the formation of KPPs in our simulations, our results suggest that a more recent group accretion event -- such as the infall of the LMC in the MW -- may substantially enhance the population of kinematically coherent satellites. In particular, none of our simulations reproduces the high fraction of corotating, on-plane satellites observed in the MW, which likely reflects the recent accretion of the LMC and its associated satellite system sharing a common sense of rotation. Future simulations employing genetic modifications to introduce an LMC-like satellite at low redshift would therefore be valuable for isolating the impact of the LMC on the formation of the VPOS in the MW.
\end{enumerate}

\begin{acknowledgements}
    R.R.C., S.R.F., and J.G. acknowledge financial support from the Spanish Ministry of Science and Innovation through the research grants: PID2021-123417OBI00, funded by MCIN/AEI/10.13039/501100011033/FEDER, EU; PCI2022-135023-2, funded by MCIN/AEI/10.13039/501100011033 and the EU “NextGenerationEU” / PRTR; and PID2024-157374OBI00, funded by MICIU/AEI/10.13039/501100011033/FEDER, EU. Additionally, R.R.C. acknowledges financial support by IND2022/TIC-23643 project funded by Comunidad de Madrid. OA acknowledges support from the Knut and Alice Wallenberg Foundation, the Swedish Research Council (grant 2025-04892), the Swedish National Space Agency (SNSA Dnr 2023-00164), the LMK foundation, and eSSENCE, a Swedish strategic research programme in e-Science. JIR would like to acknowledge support from STFC grants ST/Y002865/1 and ST/Y002857/1. AP has been supported by funding from the European Research Council under the European Union’s Horizon 2020 research and innovation programmes (grant agreement no. 818085 GMGalaxies). I.S.S acknowledges support from the European Research Council (ERC) Advanced Investigator grant to C.S. Frenk, DMIDAS (GA 786910) and from the Science and Technology Facilities Council [ST/P000541/1] and [ST/X001075/1]. We acknowledge PRACE for awarding us access to Joliot-Curie at GENCI/CEA, France to perform the simulations presented in this work.  Computations presented in this work were in part performed on resources provided by the Swedish National Infrastructure for Computing (SNIC) at the Tetralith supercomputer, part of the National Supercomputer Centre, Linköping University. We also acknowledge EuroHPC Joint Undertaking for awarding us access to MareNostrum5 at BSC, Spain, under the Extreme Scale Access Mode project EHPC-EXT-2025E01-023, and to the Red Española de Supercómptuo (RES) to MareNostrum5 at BSC, Spain, with projects AECT-2024-2-0008, AECT-2024-3-0002, AECT-2025-1-0007, and AECT-2025-2-0005.
\end{acknowledgements}

%
%

\bibliography{aa59969-26}{}
\bibliographystyle{aa}

\newpage
\appendix
\section{Comparison between RCT and subhalo particle-tracking in VINTERGATAN-GM}
\label{app:rct_part_tracking}
As described in Section~\ref{sec:halofinder}, we post-processed the merger trees produced by \textsc{Rockstar+ConsistentTrees} (RCT) to follow subhalos during infall and characterize their orbital evolution. Specifically, we apply the particle-tracking algorithm introduced in \cite{Rodirguez_Cardoso_Agora_2025}, extensively tested across multiple numerical codes within the AGORA \texttt{CosmoRun} suite.

In this Appendix, we evaluate the performance of this particle-tracking approach relative to the standard RCT output for the VINTERGATAN-GM simulations, which have a higher mass resolution than the AGORA \texttt{CosmoRun} suite \citep{Roca_Fabrega_2021, roca_fabrega_2024}. For brevity, we present only the results for the Largest simulation, which is representative of the trends across the suite.

Figure~\ref{fig:comparison_rct} compares the evolution of five representative subhalos as identified by RCT and by the particle-tracking algorithm. The top, middle, and bottom rows show trajectories, total mass, and stellar mass (computed as in Section~\ref{sec:star_assigment}), respectively. The same systems are labeled in Figure~\ref{fig:comparison_rct_plot}. Their main evolutionary behaviors are summarized below:

\begin{figure*}
    \centering
    \includegraphics[width=0.99\linewidth]{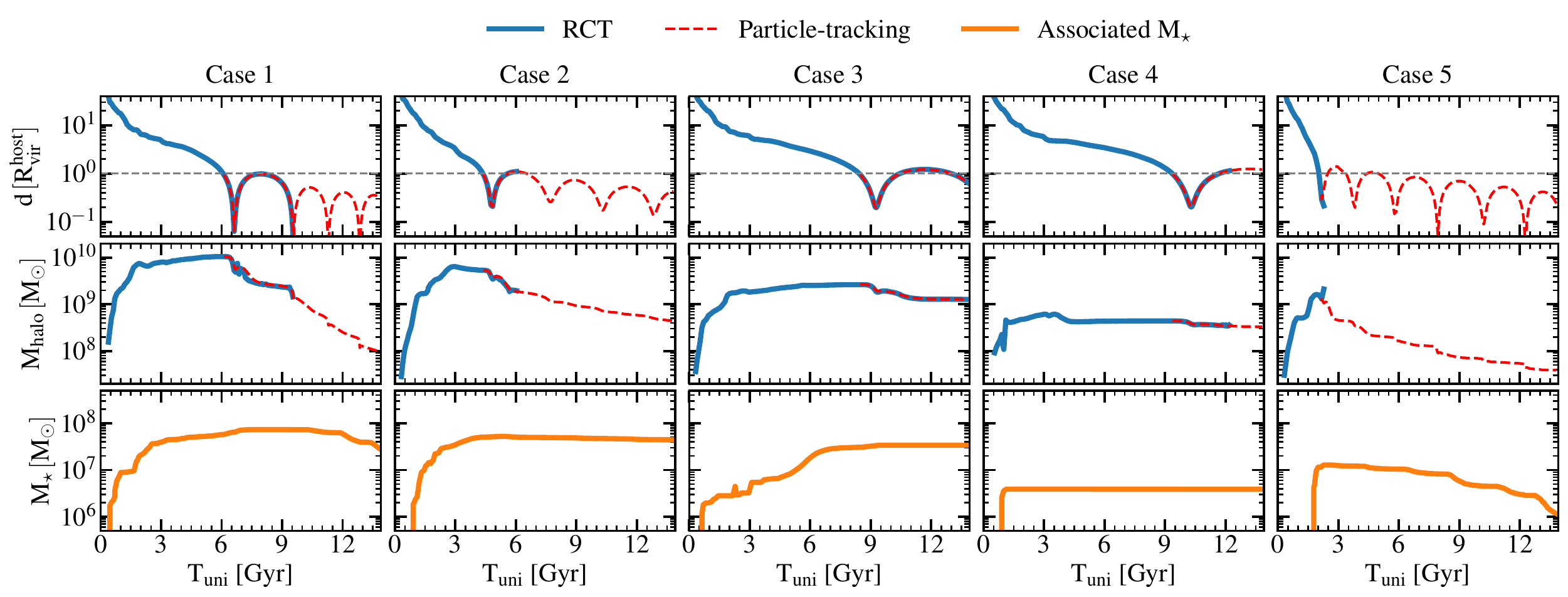}
    \caption{Comparison of the evolution of five representative subhalos in the Largest simulation, as measured by RCT (\textsc{Rockstar+ConsistentTrees}; solid blue lines) and the subhalo particle-tracking algorithm presented in \cite{Rodirguez_Cardoso_Agora_2025} and summarized in Section~\ref{sec:halofinder} (red dashed lines). \textbf{Top row:} subhalo's trajectory during its infall to the host halo. The host virial radius is indicated by a horizontal dashed line. \textbf{Central row:} Evolution of the subhalo mass. \textbf{Bottom row:} Evolution of the stellar mass associated with each subhalo, following the methodology described in Section~\ref{sec:star_assigment}. 
    }
              \label{fig:comparison_rct}%
\end{figure*}

\begin{figure*}
    \centering
    \includegraphics[width=0.99\linewidth]{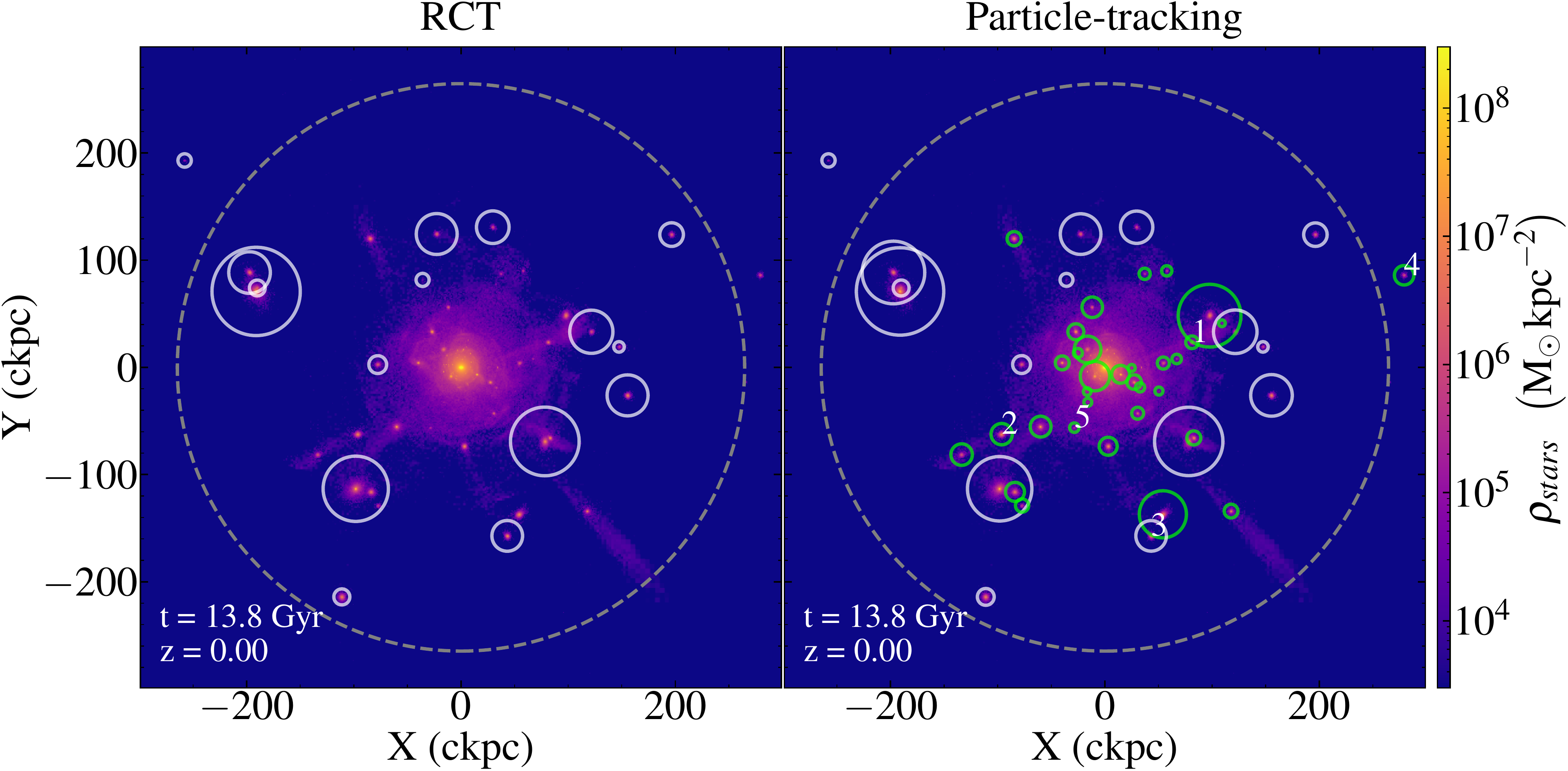}
    \caption{Projected stellar surface density for the Largest simulation at $z=0$, with the ``reliable'' (see Appendix~\ref{app:rct_part_tracking} for reliability definition) satellite galaxies detected by each method overplotted as solid circles representing half of the subhalo virial radius. \textbf{Left}: Reliable satellite galaxies identified by RCT. \textbf{Right}: Reliable satellite galaxies identified by the subhalo particle-tracking algorithm. Satellite galaxies also identified by RCT are shown as gray solid circles, while green solid circles indicate those classified as reliable exclusively by the subhalo particle-tracking algorithm. The representative subhalos shown in Figure~\ref{fig:comparison_rct} are highlighted and labeled by their case number in the upper-right corner of each subhalo. The great performance of the particle-tracking–corrected merger trees in recovering subhalos missed by RCT is clearly evident, particularly in the inner regions.
   }
              \label{fig:comparison_rct_plot}%
\end{figure*}

\begin{itemize}
    \item Case 1: This example illustrates one of the most common failure modes of RCT. The subhalo is tracked through one complete orbit after its first pericentric passage but is classified as merged at the second pericenter following substantial mass loss. In contrast, the particle-tracking algorithm continues to identify the system as an independent subhalo for more than two full orbits. At the time RCT declares the merger, the system still retains significant bound mass, ($M_{\rm halo}\sim10^9\,M_\odot$, $M_\star\sim5\times10^7\,M_\odot$). Tidal stripping initially affects the outer DM, while the stellar component remains largely intact, with significant stellar mass loss only after the second pericenter.
    \item Case 2: RCT tracks the subhalo for one orbit but loses it near the following apocenter after an interaction with another satellite. The particle-tracking algorithm instead follows it to $z=0$, capturing three additional orbits. After strong mass loss during the first orbit, stripping proceeds gradually, and the stellar component remains largely unaffected due to relatively shallow pericenters.

    \item Case 3: This case demonstrates excellent agreement between RCT and the particle-tracking algorithm throughout the entire evolution down to $z=0$. The correspondence between the two methods is remarkable, both in terms of the orbital trajectory and the mass evolution.

    \item Case 4: Similar to Case~2, RCT loses track of this subhalo near its first apocenter. However, in this instance the system has not undergone significant mass loss nor experienced a satellite--satellite interaction. At the time RCT classifies the subhalo as merged, it is located outside the host virial radius as a splashback galaxy, as shown in Figure~\ref{fig:comparison_rct_plot}. The particle-tracking algorithm naturally recovers and continues to follow this system.

    \item Case 5: This case illustrates a misidentification of distinct subhalos in RCT (e.g., \citealt{Symfind_Mansfield_2024}). During first pericenter, RCT incorrectly associates the merger-tree branch with an unrelated density peak, producing unphysical changes in mass and position. The particle-tracking method instead follows the correct subhalo for several additional orbits ($\sim10$~Gyr), smoothly capturing halo and stellar mass loss. Without post-processing, the RCT branch erroneously reaches its peak mass at its final, spurious snapshot.
\end{itemize}

Overall, the particle-tracking approach follows resolved subhalos until their physical disruption, whereas RCT often classifies them as merged even when they remain relatively massive ($N_{\rm DM} \geq 5000$; Cases~1, 2, and~5) or even far from the host halo center (Cases~2 and~4). Importantly, during the snapshots in which a subhalo is detected by both methods, they show excellent agreement in orbital trajectories and mass evolution, indicating convergence in the inferred subhalo properties.

In Figure~\ref{fig:comparison_rct_plot}, we compare the number of reliably detected satellite galaxies. A “reliable” satellite is defined as a subhalo that (i) formed outside the host virial radius, (ii) does not exhibit abrupt mass changes between snapshots ($\Delta M \geq 100\,M_{\rm sub}$), and (iii) has $M_\star > 10^5\,\rm{M_\odot}$. The figure shows the projected stellar surface density at $z=0$, with reliable satellites overplotted. RCT fails to trace a significant fraction of systems (green circles), which are recovered only by the particle-tracking algorithm. Although \textsc{Rockstar} may identify some of these subhalos at a given snapshot, \textsc{Consistent-Trees} can lose them at earlier times and misassociate their branches, preventing reliable reconstruction of their histories. In contrast, our method substantially increases the number of reliable satellites, particularly at low masses and in the inner regions. Overall, it recovers the full satellite population, whereas RCT loses subhalos that remain clearly identifiable even by eye, as shown in Figure~\ref{fig:comparison_rct_plot}.

\begin{figure*}
    \centering
    \includegraphics[width=0.99\linewidth]{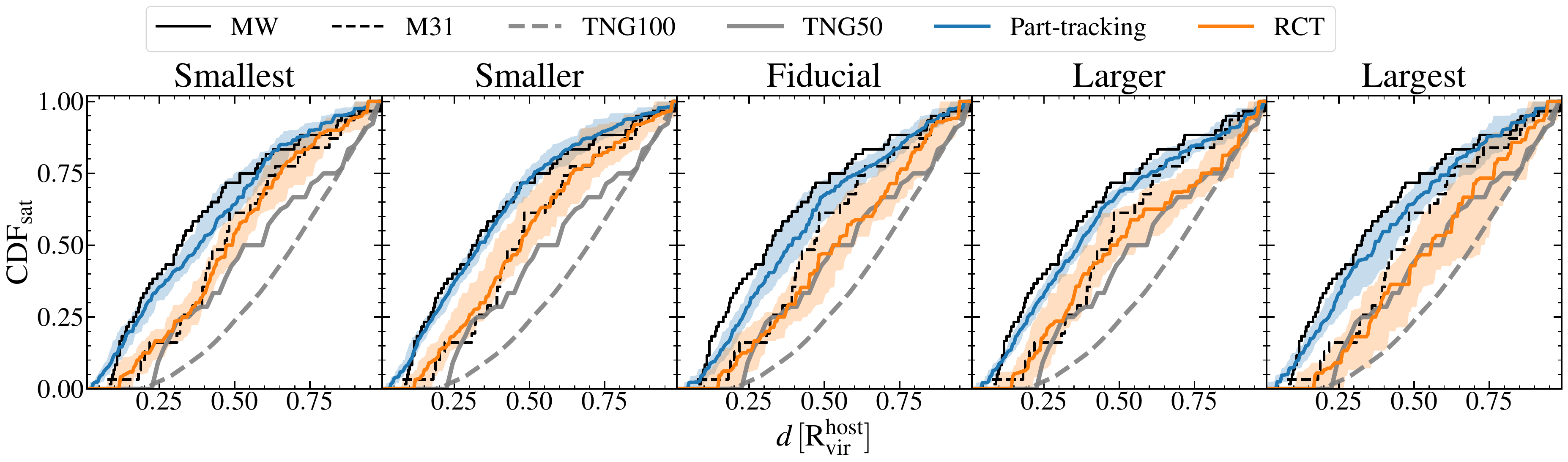}
  \caption{Satellite radial distributions normalized by the virial radius of their host halos across our simulation suite. Results are shown both using standard RCT merger trees and after applying the subhalo particle-tracking correction. Solid lines represent the median distributions computed from 50 snapshots between $z=0.2$ and $z=0$, while shaded regions indicate the median absolute deviation. For reference, the observed satellite radial distributions of the MW \citep{Drlica_Wagner_2020} and M31 \citep{Savino_2022} are shown as black solid and dashed lines, respectively. Gray dashed and solid lines show the median satellite radial distributions at $z=0$ reported for TNG100 \citep{Kanehisa_2023} and computed from the TNG50 MW/M31 sample catalogs \citep{Pillepich_2024}, respectively.}
              \label{fig:CDF_sats_rad}%
\end{figure*}

In Figure~\ref{fig:CDF_sats_rad}, we assess the impact of the subhalo particle-tracking correction on the satellite radial distribution of each simulation. As in Figure~\ref{fig:CDF_sats_rad_2}, we compare the simulated distributions with those observed in the MW, M31, as well as the TNG50-1 and TNG100-1 results, but now contrasting, for each simulation, the distributions obtained from the standard RCT merger trees with those recovered after applying the particle-tracking correction. This correction substantially increases the number of satellites identified in the inner halo -- as we show in Figure~\ref{fig:comparison_rct_plot} -- yielding a more centrally concentrated distribution that is much closer to that observed in the MW. In contrast, the standard RCT merger trees miss a significant fraction of inner satellites and therefore fail to reproduce the observed radial distribution. These results highlight that reliable subhalo tracking is crucial for recovering the underlying satellite radial distribution in our simulations.

A more detailed analysis of this particle-tracking approach will be presented in a future paper, where the algorithm will be made publicly available and integrated into \texttt{ytree} framework. Here we emphasize that this method enables robust subhalo tracking, which is essential for accurately studying the orbital evolution of satellite galaxies within their host halos.

\section{Sensitivity of 4GND metrics to $f_{\rm sat}$.}
\label{app:collection_planes}
As described in step (6) of Section~\ref{sec:extension}, using the extended 4GND plot methodology \citep{Santos_santos_sim_2020, Santos_santos_MW_2020}, a collection of planes with different values of $N_{\rm sat}$ (or equivalently $f_{\rm sat}$) is constructed for each density peak in the 4GND plot (see e.g., Figure~4 in \citealt{Santos_santos_MW_2020} for the different properties of the MW plane when including a larger number of satellites).

In the second row of Figure~\ref{fig:4GND_metrics_evolution}, these collections of planes are used to show the time evolution of the $c/a$ ratio when considering $f_{\rm sat} = 90\%$ at each timestep, thereby characterizing the planarity of nearly the entire satellite population. The same plane defined by $f_{\rm sat} = 90\%$ is used to determine $\Delta_{\rm RMS}$ and the plane normal employed to compute $f_{\rm sat}^{\rm coorbit}$, shown in the third and fourth rows of Figure~\ref{fig:4GND_metrics_evolution}.

\begin{figure*}
    \centering
    \begin{subfigure}{0.99\textwidth}
    \includegraphics[width=\textwidth]{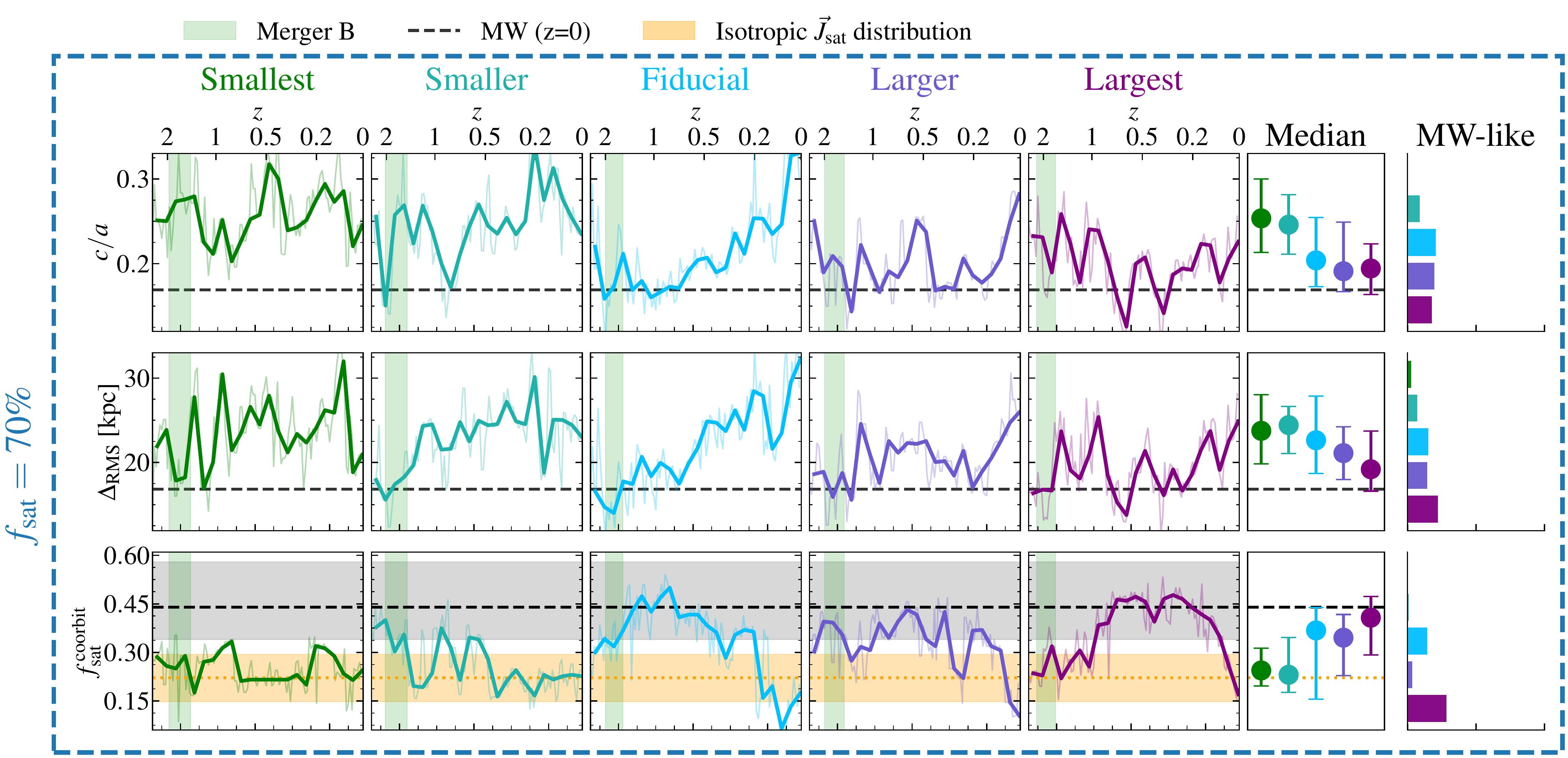}
    \end{subfigure}
    \begin{subfigure}{0.99\textwidth}
    \includegraphics[width=\textwidth]{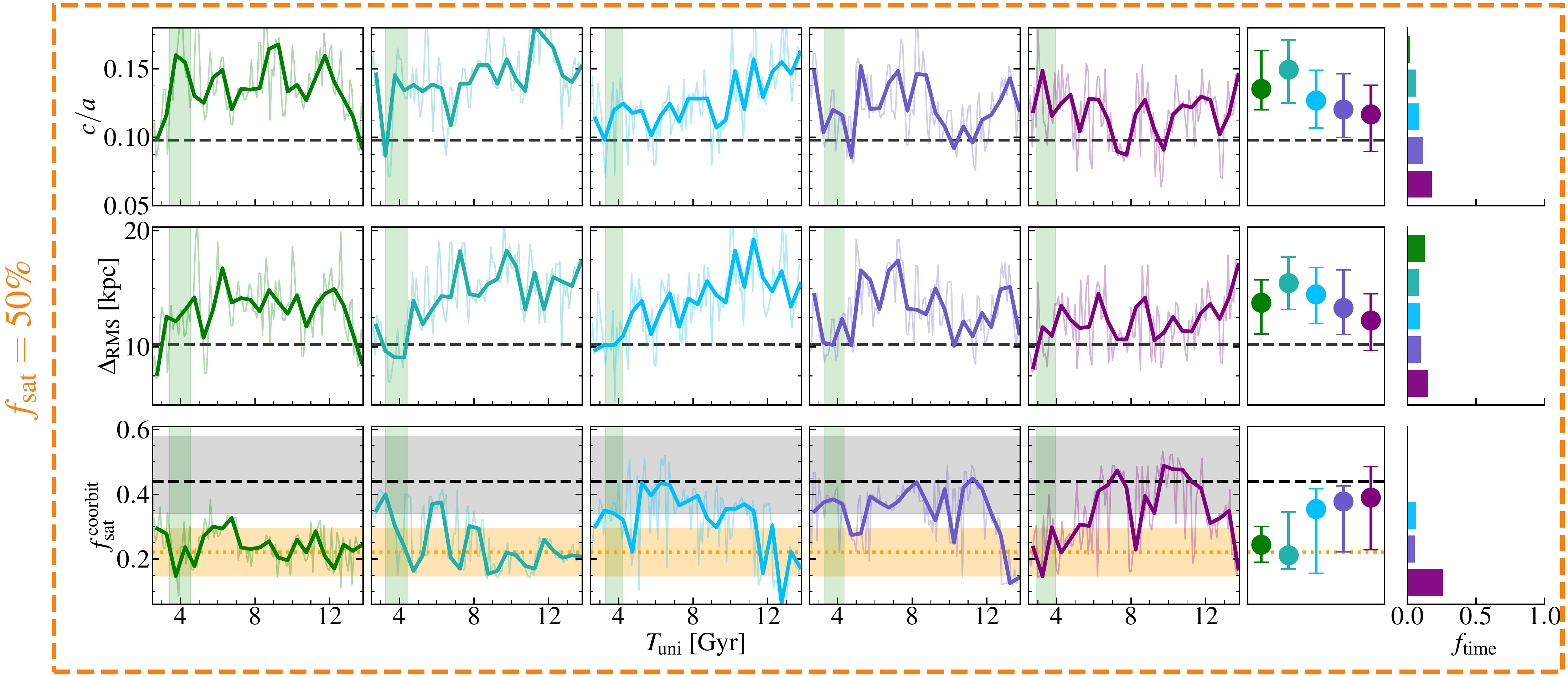}
    \end{subfigure}
    \caption{Same as the three bottom rows of Figure~\ref{fig:4GND_metrics_evolution}, but using $f_{\rm sat} = 70\%$ (\textbf{top panels}) and $f_{\rm sat} = 50\%$ (\textbf{bottom panels}) to determine the temporal evolution of the metrics.}
   \label{fig:f_sat_70_50}
\end{figure*}

The evolution of these metrics when considering lower values of $f_{\rm sat}$, such as $70\%$ and $50\%$, is shown in Figure~\ref{fig:f_sat_70_50}. The axis ranges are adjusted separately for each $f_{\rm sat}$ to emphasize the relative differences across the simulation suite, since our aim here is to test whether the trends identified for $f_{\rm sat}=90\%$ persist when varying the satellite fraction, rather than to directly compare the absolute metric values between different $f_{\rm sat}$. As the fraction of satellites considered decreases, both $c/a$ and $\Delta_{\rm RMS}$ naturally decrease for all simulations. Differences between simulations also become smaller for lower $f_{\rm sat}$, reflecting the fact that it is generally easier to identify a thin plane when considering only a subset of satellites, even if the overall satellite population is not distributed in a planar configuration. Nevertheless, the relative trends in planarity across simulations, although less pronounced, remain visible when considering $f_{\rm sat} = 70\%$ and $50\%$. Regarding $f_{\rm sat}^{\rm coorbit}$, it becomes systematically noisier and slightly lower as $f_{\rm sat}$ decreases, indicating that when only a small subset of satellites is considered it is more likely to find as the best plane a random alignment of satellites that is not kinematically supported. Interestingly, the fraction of time $f_{\rm time}$ during which the simulations reproduce structures with planarity comparable to or greater than that observed in the MW remains similar when decreasing $f_{\rm sat}$.

\section{Persistent and transient population in positional planes}
\label{app:persvstrans}
As described in Section~\ref{sec:positional_planes}, positional planes identified through the 4GND plot methodology vary their orientation and membership with time (see Figure~\ref{fig:4GND_metrics_evolution}). In order to investigate how persistent is the membership of positional planes, we focus on a well-defined positional configuration in each simulation. Specifically, we select the snapshot at which the $f_{\rm sat}=90\%$ plane reaches its minimum $c/a$, corresponding to the thinnest configuration involving almost the entire satellite population. These minima occur at $T_{\rm uni}=7.2$, $6.5$, $6.5$, $6.9$, and $9.5$ Gyr in the Smallest, Smaller, Fiducial, Larger, and Largest simulations, respectively. We labeled the normal to this thinnest plane $\hat{n}_{\rm best}$. We then selected the $f_{\rm sat}=90\%$ satellites with the largest contributions to the corresponding 4GND peak and followed their orbital evolution since the GSE-merger event until $z=0$. For each satellite and at every available snapshot in this interval, we computed its orbital angular-momentum direction, $\hat{J}_{\rm sat}$, and its angular separation from the fixed direction $\hat{n}_{\rm best}$.  We classified a satellite as a persistent member when its median angular separation over the considered interval satisfies $\alpha_{\rm med}<\alpha_{\rm coorbit}=36.87^\circ$, and as a transient member otherwise. In this way, ``persistent'' refers to satellites whose orbital planes remain preferentially aligned with the selected positional plane, whereas ``transient'' satellites contribute to the thin positional configuration at the selected snapshot but do not maintain an orbit aligned with it.

\begin{figure*}
    \centering
    \includegraphics[width=0.99\linewidth]{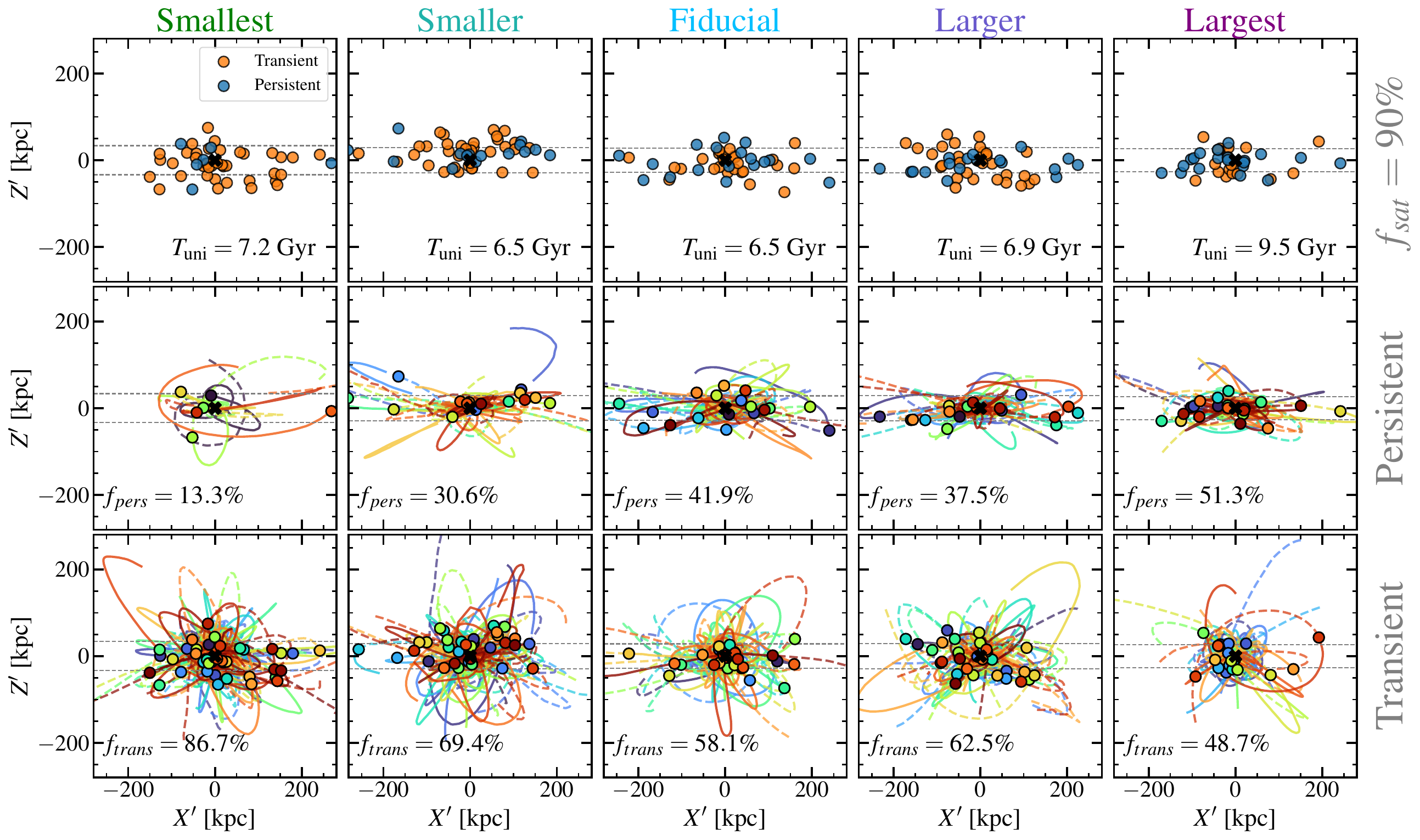}
   \caption{Persistent and transient members of the thinnest $f_{\rm sat}=90\%$ positional plane in each simulation. Each column corresponds to one simulation and the selected cosmic time $T_{\rm uni}$ is indicated in the top row. \textbf{Top row:} projected $X'-Z'$ distribution of the satellites defining the $f_{\rm sat}=90\%$ plane at the snapshot at which its axis ratio $c/a$ reaches its minimum. The coordinate system is rotated such that $Z'$ is parallel to the normal of the selected positional plane, $\hat{n}_{\rm best}$. Satellites classified as persistent and transient are shown in blue and orange, respectively. \textbf{Middle row:} orbital trajectories of the persistent satellites, followed from 2 Gyr before to 4 Gyr after the selected snapshot. Dashed and solid segments indicate the evolution before and after the selected time, respectively. Each satellite and its corresponding trajectory are shown with the same color. \textbf{Bottom row:} same as the middle row, but for transient satellites. A satellite is classified as persistent when the median axial angular separation between its orbital angular-momentum direction and $\hat{n}_{\rm pos}$ over the considered interval is smaller than $\alpha_{\rm coorbit}=36.87^\circ$, and as transient otherwise. Horizontal gray dashed lines indicate $\pm\Delta_{\rm RMS}$ of the selected positional plane. The corresponding fractions of persistent and transient satellites, relative to the $f_{\rm sat}=90\%$ sample, are indicated in the middle and bottom panels.}
              \label{fig:persvstrans}%
\end{figure*}

The result is shown in Figure~\ref{fig:persvstrans}. The persistent fractions are $f_{\rm pers}=13.3\%$, $30.6\%$, $41.9\%$, $37.5\%$, and $51.3\%$ from the Smallest to the Largest simulation, respectively, with the complementary transient fractions being $86.7\%$, $69.4\%$, $58.1\%$, $62.5\%$, and $48.7\%$. Thus, although the trend is not strictly monotonic, the fraction of satellites that persistently support the thinnest positional planes increases substantially toward the simulations with more massive GSE-like mergers. In the Smallest run, the thinnest positional configuration is overwhelmingly produced by satellites that only temporarily cross the plane, whereas in the Largest run approximately half of the satellites contributing to the plane follow orbits persistently aligned through their previous and subsequent evolution.

\begin{figure}[h!]
    \centering
    \includegraphics[width=0.99\linewidth]{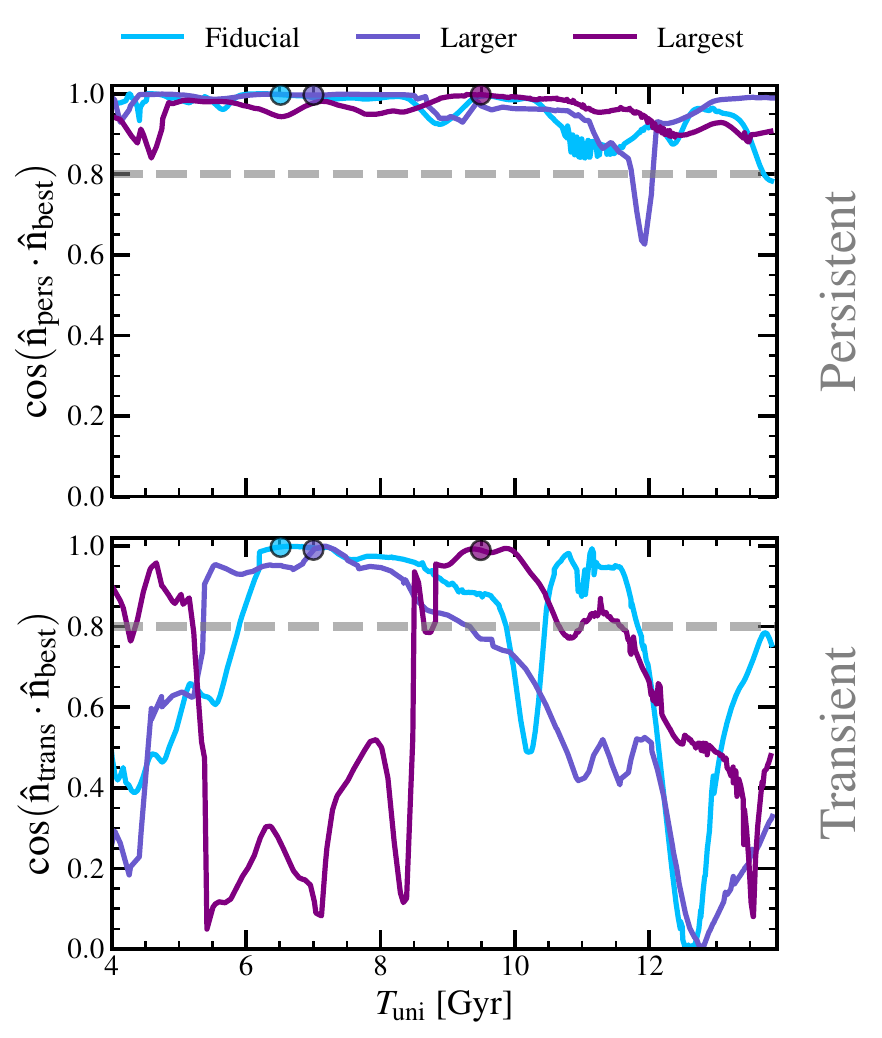}
   \caption{Time evolution of the alignment between the normal to the reference positional plane, $\hat{n}_{\rm best}$, shown in Figure~\ref{fig:persvstrans}, and the normal to the best-fitting plane defined at each snapshot by the persistent ($\hat{n}_{\rm pers}(z)$, top panel) and transient ($\hat{n}_{\rm trans}(z)$, bottom panel) satellite populations. The colored circles mark the snapshot at which the thinnest plane with the minimum $c/a$ is identified, while the horizontal gray dashed line indicates $|\cos\alpha|=0.8$, corresponding to $\alpha_{\rm coorbit}=36.87^\circ$.}
              \label{fig:aligment_perstrans}%
\end{figure}

Having defined the persistent and transient populations relative to a fixed reference plane, we can further investigate the origin of the decrease in planarity and anisotropy observed at $T_{\rm uni}\gtrsim12$ Gyr in the Fiducial, Larger, and Largest runs. For this purpose, at each snapshot we independently compute the best-fitting plane defined by the persistent and transient populations and measure the alignment of its normal with $\hat{n}_{\rm best}$. The results are shown in Figure~\ref{fig:aligment_perstrans}, with the top and bottom panels showing the evolution of the alignment for the persistent and transient populations, respectively. Colored circles indicate the time at which the reference $f_{\rm sat}=90\%$ plane reaches its minimum $c/a$.

The plane defined by the persistent satellites remains strongly aligned with $\hat{n}_{\rm best}$ for most of the subsequent evolution, supporting the interpretation that this population defines a stable underlying planar structure. In contrast, the orientation of the plane defined by the transient population varies substantially with time. At some epochs it becomes closely aligned with the persistent structure, whereas at others it is strongly misaligned. Notably, the thinnest positional configurations occur when the transient population is also aligned with $\hat{n}_{\rm best}$, temporarily reinforcing the planar distribution defined by the persistent satellites. At $T_{\rm uni}\gtrsim12$ Gyr, however, the transient population becomes substantially misaligned in the Fiducial, Larger, and Largest runs, thereby reducing the overall anisotropy and planarity of the full satellite distribution seen in Figures~1 and 2 of the manuscript.

In addition, these runs experience a late accretion event at around $T_{\rm uni}\sim12$ Gyr that introduces approximately 6--7 satellites, depending on the simulation, and the spatial distribution of these recently accreted satellites is generally misaligned with the reference plane. Their accretion therefore further reduces the anisotropy and planarity of the overall satellite population at late times.

Taken together, these results suggest that the persistent satellite population provides the long-lived backbone of the planar configuration, while transient satellites can temporarily enhance its apparent planarity when their instantaneous spatial distribution becomes aligned with this structure. Conversely, the misalignment of the transient population, together with the late accretion of additional satellites along a different direction, can substantially weaken the positional plane. Thus, although the underlying persistent structure can remain stable over several Gyr, the exceptionally thin positional configurations identified at individual snapshots are more transient and sensitive to the instantaneous and variable distribution of the remaining satellite population.

\section{Projected evolution of long-lived satellite planes}

\begin{figure*}
    \centering

    \includegraphics[width=0.99\linewidth]{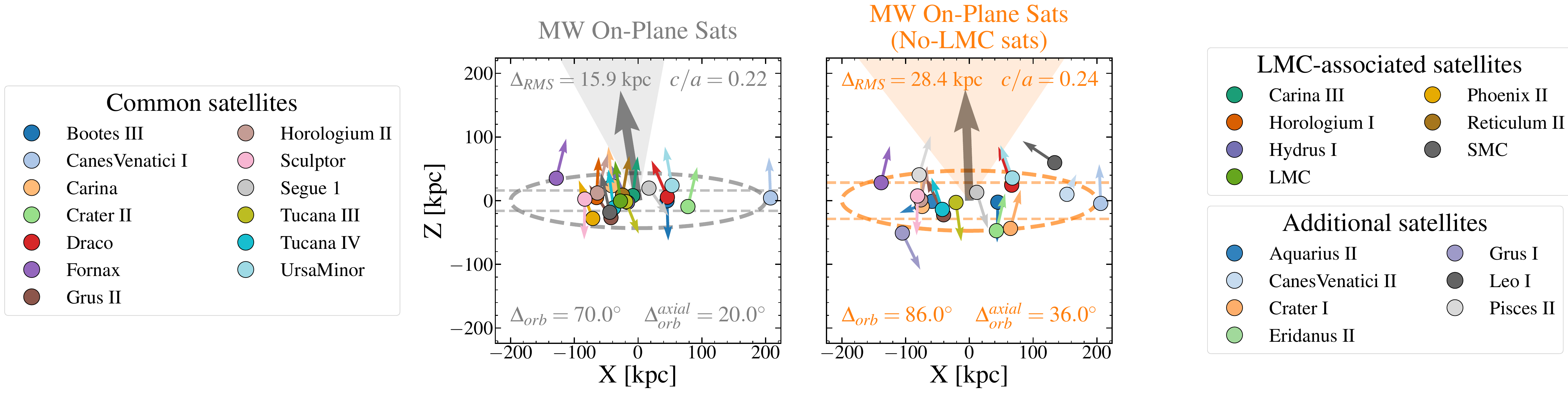}

    \vspace{0.4em} 

    \includegraphics[width=\linewidth]{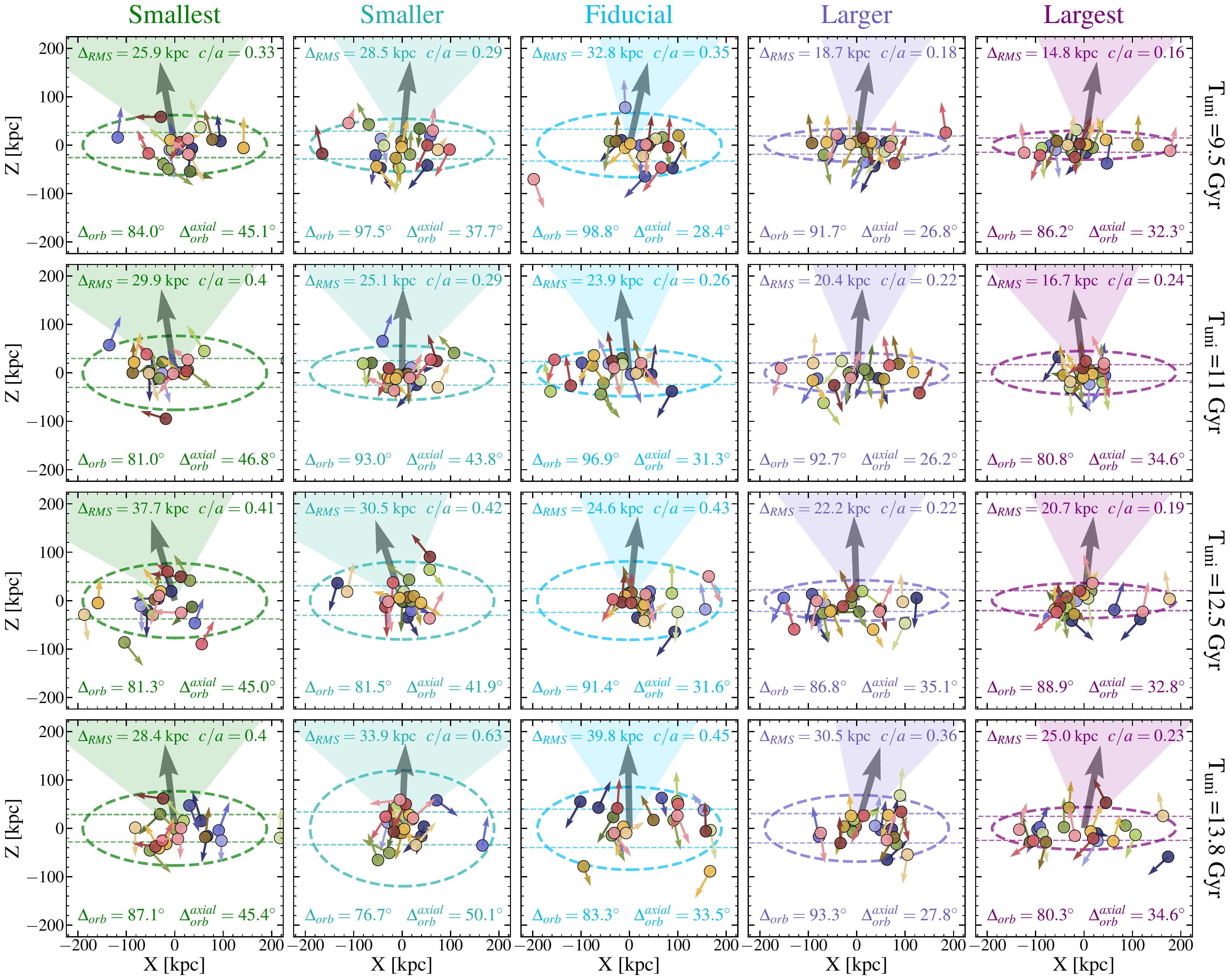}

    \caption{Comparison between the projected distribution of MW on-plane satellites and the satellites defining long-lived planes used in Figure~\ref{fig:evolution_plane_multi} for $N_{\rm sat}=20$ across the simulation suite. The two top panels show the plane properties for the 20 satellites most kinematically coherent with the VPOS, including (left) and excluding (right) satellites associated with the LMC. The four rows below present the time evolution of the satellites belonging to long-lived ($\geq 4$~Gyr) kinematically coherent planes identified in each simulation, with rows corresponding to cosmic times $T_{\rm uni}=9.5$, 11, 12.5, and 13.8~Gyr, and columns to different simulations. In each panel, satellite positions and projected angular-momentum poles are shown in a reference frame where the normal to the best-fitting plane is aligned with the positive $z$-axis. For the simulations, the same set of $N_{\rm sat}=20$ satellites is tracked over time, with consistent colors indicating identical satellite IDs. The plane thickness is indicated by colored horizontal dashed lines corresponding to the RMS height, $\Delta_{\rm RMS}$, while the minor-to-major axis ratio, $c/a$, is represented by an ellipse with the corresponding axial ratio. The mean axial angular-momentum direction, $\vec{J}_{\rm mean}^{\rm axial}$, and the axial orbital-pole dispersion, $\Delta_{\rm orb}^{\rm axial}$, are shown as a gray arrow and a colored shaded cone, respectively. All plane metrics are explicitly labeled in each panel. Details of the satellite selection procedure in both the MW and the simulations are provided in Section~\ref{sec:long_lived}. The maintained planarity of the long-lived satellite plane in the Largest run is visually evident, reproducing the level of planarity observed for the MW on-plane satellites for more than 4 Gyr.}

    \label{fig:evolution_plane_multi_plot}
\end{figure*}

This appendix provides a visual counterpart to the quantitative analysis of long-lived satellite planes presented in Section~\ref{sec:long_lived}. Figure~\ref{fig:evolution_plane_multi_plot} shows the projected evolution of the $N_{\rm sat}=20$ long-lived planes analyzed in Figure~\ref{fig:evolution_plane_multi}. The two upper panels show the projected distribution of the 20 MW satellites most kinematically aligned with the VPOS, including and excluding LMC-associated satellites, while the panels below follow the corresponding $N_{\rm sat}=20$ satellite populations across the simulation suite.

For the simulations, we show four approximately equally spaced snapshots over the final $\sim4$~Gyr of evolution, after all satellites belonging to the selected long-lived planes have been accreted. The figure visually illustrates the trends quantified in Figure~\ref{fig:evolution_plane_multi}. In the Smallest and Smaller runs, the $N_{\rm sat}=20$ selections extend well beyond the relatively small KPP populations, and the different orbital orientations of the additional satellites lead to comparatively thick and rapidly varying configurations. In contrast, the Larger and especially the Largest simulations contain sufficiently populated and kinematically coherent KPPs to sustain thin, flattened configurations over several Gyr. This is also evident from the stronger alignment of the satellites' angular-momentum poles around the normal to the plane. The Fiducial run shows an intermediate behavior. In the Largest run, the plane remains visually comparable in thickness and flattening to the MW on-plane VPOS satellite sample throughout much of the analyzed interval. This provides a direct visual illustration of how increasing KPP richness translates into the ability to sustain well-populated, long-lived satellite planes.

\section{LV shape evolution plots}
\label{app:LV}

\begin{figure*}
    \centering
    \includegraphics[width=0.99\linewidth]{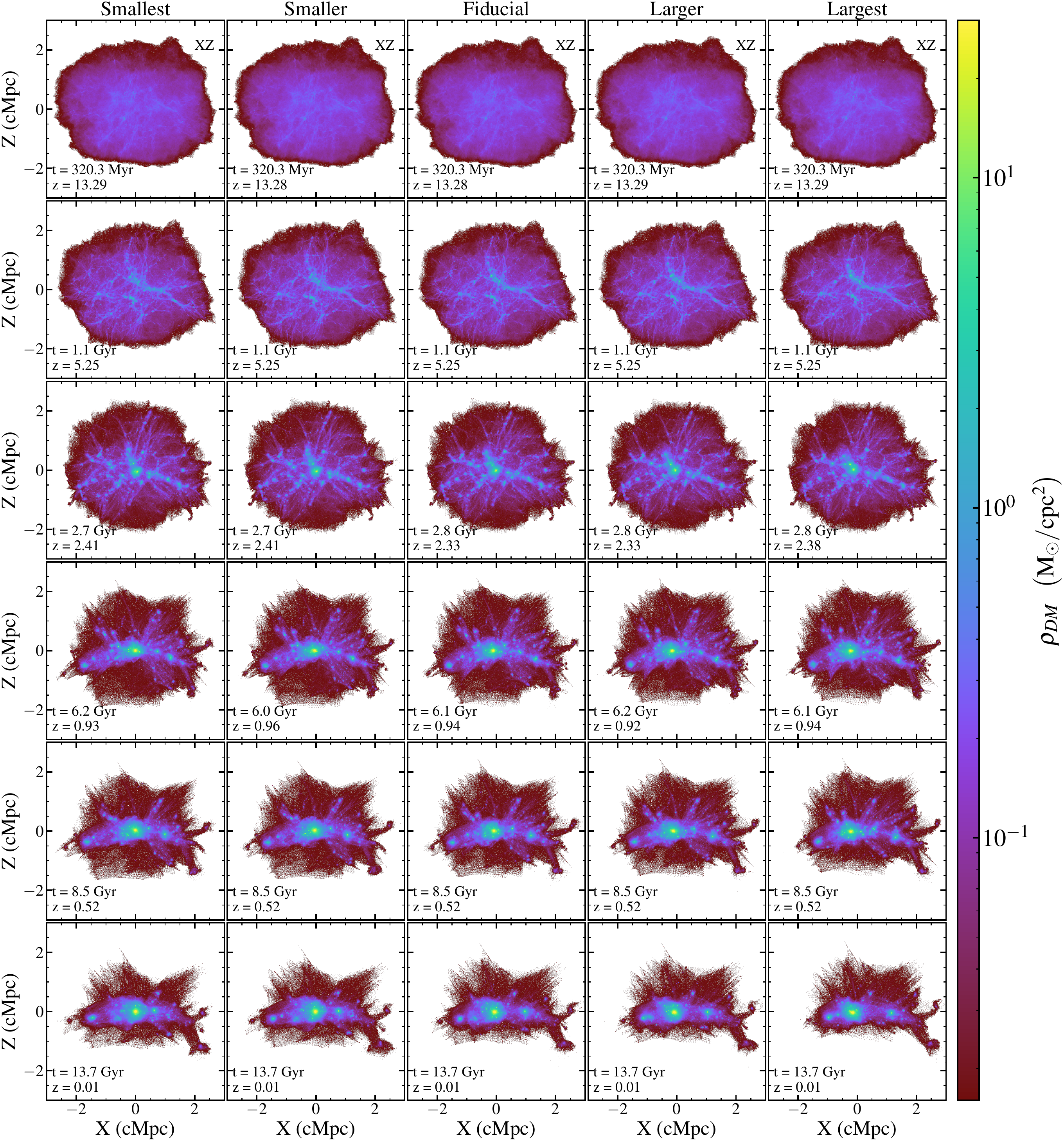}
   \caption{Shape evolution of the LV for each simulation. Each row corresponds to a different timestep from $z \sim 13$ to $z = 0$, and each column to a different simulation. Each panel shows the projected DM density of LV particles onto the $X$-$Z$ plane, projecting the normal plane to the direction defined by $\hat{e}_2^{LV}$ at $z=0$. The figure illustrates how the initially spherical mass distribution gradually flattens over time: by $z \sim 0.9$, a wall-like structure has emerged, and by $z \sim 0.5$, most of the mass has concentrated into a dominant filament within the $X$–$Z$ plane. However, since the large-scale structure was defined to be identical across simulations by construction, the LV shape remains nearly unchanged across the suite, ensuring that these anisotropies are consistent between simulations.}
              \label{fig:lv_evolution}%
\end{figure*}

In Section~\ref{sec:lss}, we display the results of the evolution of LV deformations based on the approach described in \cite{Robles_LV_2015} and \cite{GamezMarin_2024}. Here, we provide a description of this approach. In order to characterize the evolution of the CW, we define the LV by first selecting all DM particles located within the host virial radius at $z=0$, $R_{\rm vir,\,z=0}$, and tracing them back to high redshift $z_{\rm high} \sim 13$ (our first stored snapshot). Using their positions at $z_{\rm high}$, we compute a new center of mass, $\vec{r}_{\rm c}$, and select all particles enclosed within a sphere of radius $R_{\rm LV} = K\times R_{vir,\,z=0}/(1+z_{\rm high})$, with $K=15$ around their respective centers $\vec{r_c}$ (see first row of Figure~\ref{fig:lv_evolution}). These particles sample the mass flows that shape the elements of the CW as the Universe evolves. They initially follow geodesic trajectories until they become trapped within, or accreted onto, CW structural elements. Our interest here focuses on the global deformations of LVs as an average of the fate of their constituent particles. We therefore track the evolution of these particles from $z_{\rm high}$ to $z=0$.  A visual representation of the shape evolution of the LV for each simulation is shown in Appendix~\ref{app:LV} in Figure~\ref{fig:lv_evolution}.

To quantify the local LV transformations across our simulations, we compute the reduced inertia tensor, $I^{\,\rm r}_{ij}$, of each LV with respect to its center of mass at different redshifts:
\begin{equation}
   I_{ij}^{\rm r} =\sum_{n}m_n\frac{(\delta_{ij}r_{n}^2 - r_{i,n}r_{j,n})}{r_{n}^2}, \hspace{1cm} n=1, ..., N
   \label{reducedI}
\end{equation}
where $r_{i,n}$ and $r_{j,n}$ are the $i$-th and $j$-th components of the position vector of the $n$-th LV particle relative to the LV center of mass, $r_n$ is its distance to the center of mass, $m_n$ is the particle mass, and $N$ is the total number of particles in the LV. To isolate the evolution of the LV itself, and minimize the influence of the host halo shape\footnote{An analysis of the evolution of the host halo shape and its impact is presented in Section~\ref{sec:host_halo_shape}.}, we exclude the host halo particles at each snapshot when computing $I^{\,\rm r}_{ij}$. We adopt the reduced inertia tensor rather than the standard (non-reduced) form to suppress the influence of substructure in the outer regions of the LV \citep{Gerhard_1983, Bailin_Steinmetz_2005, Robles_LV_2015, GamezMarin_2024}. In addition, the reduced inertia tensor is invariant under purely radial mass rearrangements with respect to the LV center of mass, making it particularly well suited for characterizing anisotropic mass deformations, such as those predicted by nonlinear models of gravitational CW evolution (e.g., \citealt{Gurbatov_2012} and references therein). 

In order to measure the LV shape evolution, we calculated the principal axes of the inertia ellipsoid, $a,\,b$ and $c$, derived from the eigenvalues ($\lambda_i$, with $\lambda_1\leq \lambda_2 \leq \lambda_3$) of the $I^{\,\rm r}_{ij}$ tensor, so that $a\geq b \geq c$ \citep{Gonzalez_Garcia_2009, Robles_LV_2015, GamezMarin_2024},
\begin{equation}
    a = \sqrt{\frac{\lambda_2-\lambda_1+\lambda_3}{2M}}, \, b = \sqrt{\frac{\lambda_3-\lambda_2+\lambda_1}{2M}}, \, c = \sqrt{\frac{\lambda_1-\lambda_3+\lambda_2}{2M}},
\end{equation}

where $M$ is the total mass of a given LV. We denote the directions of the principal axes of inertia by $\hat{e}_i^{LV},\,i=1,2,3$, where $\hat{e}_1^{LV}$ corresponds to the major axis, $\hat{e}_2^{LV}$ to the intermediate axis and $\hat{e}_3^{LV}$ to the minor axis.

Figure~\ref{fig:lv_evolution} shows the time evolution of the projected DM density of the LV particles. As already noted in the analysis of Figure~\ref{fig:evolution_eigenvectors}, the LV evolution is nearly identical across the simulation suite, reflecting that the large-scale structure surrounding the host halo is fixed by construction in the genetically modified simulations. 

In the third row of the figure, at $z \sim 2.4$, small-scale differences associated with the infall of the GSE-like major merger are visible, with the mass of this overdensity systematically increasing from the Smallest to the Largest run. These localized differences introduce minor discrepancies in the relative centering of the host halo with respect to the LV from the merger to $z=0$. Nevertheless, the surrounding large-scale structure remains nearly identical across simulations, in agreement with the analysis presented in Section~\ref{sec:lss}. Overall, the LV evolves from an initially spherical configuration into a flattened structure preferentially distributed in the $X$–$Y$ plane, perpendicular to the $\hat{e}_3^{LV}$ direction.

\end{document}